\documentclass[iop,revtex4]{emulateapj}

\usepackage{natbib}
\usepackage{hyperref}
\usepackage{booktabs}
\usepackage{amsmath}
\usepackage[dvipsnames]{xcolor}
\usepackage{lscape}
\usepackage{subfigure}

\shorttitle{UV Spectrum of H$_2$}
\shortauthors{Jaeggli, Judge, \& Daw}

\newcommand{\hh}
{\ifmmode{\rm H_ 2}\else${\rm H_2}$\fi}

\newcommand{\bstate}
{$B\,^1\Sigma^+_u$}
\newcommand{\cstate}
{$C\,^1\Pi_u$}

\newcommand{\level}[3]{$\mathcal{J}_{\rm {#3}}={#1}, v_{\rm {#3}}={#2}$}
\newcommand{\upjv}[2]{$\mathcal{J}_{\rm u},v_{\rm u}={#1},{#2}$}

\newcommand{\re}[1]{{\color{Black} #1}}

\begin{document}

\title{Formation of the UV Spectrum of Molecular Hydrogen in the Sun}

\author{S.~A. Jaeggli}
\affil{National Solar Observatory, 8 Kiopa$'$a Street, Suite 201, Pukalani, HI 96768, USA}
\email{sjaeggli@nso.edu}

\author{P.~G. Judge}
\affil{High Altitude Observatory, National Center for Atmospheric Research, P.O. Box 3000, Boulder CO 80307-3000, USA}
\email{judge@ucar.edu}

\author{A.~N. Daw}
\affil{NASA Goddard Space Flight Center, Solar Physics Laboratory, Code 671, Greenbelt, MD 20771, USA}
\email{adrian.daw@nasa.gov}

\begin{abstract}
Ultraviolet lines of molecular hydrogen have been observed in solar spectra for almost four decades, but the behavior of the molecular spectrum and its implications for solar atmospheric structure are not fully understood.  Data from the HRTS instrument revealed that H$_2$ emission forms in particular regions, selectively excited by bright UV transition region and chromospheric lines.  We test the conditions under which H$_2$ emission can originate by studying non-LTE models sampling a broad range of temperature stratifications and radiation conditions.  Stratification plays the dominant role in determining the population densities of H$_2$, which forms in greatest abundance near the continuum photosphere.  However, opacity due to photoionization of silicon and other neutrals determines the depth to which UV radiation can penetrate to excite the H$_2$.  Thus the majority of H$_2$ emission forms in a narrow region, at about 650 km in standard 1D models of the quiet-Sun, near the $\tau$=1 opacity surface for the exciting UV radiation, generally coming from above.  When irradiated from above using observed intensities of bright UV emission lines, detailed non-LTE calculations show that the spectrum of H$_2$ seen in the quiet-Sun SUMER atlas spectrum and HRTS light bridge spectrum can be satisfactorily \re{reproduced} in 1D stratified atmospheres, without including 3D or time dependent thermal structures.  A detailed comparison to observations from 1205 to 1550 \AA\ is presented, and the success of this 1D approach to modeling solar UV H$_2$ emission is illustrated by the identification of previously unidentified lines and upper levels in HRTS spectra.
\end{abstract}



\section{Introduction}
The archetypal homonuclear \hh{} molecule has no electric dipole moment in its lowest electronic state $X\,^1\Sigma^+_g$ and can therefore support no vibrational-rotational electric dipole (E1) radiative transitions, but radiative E1 transitions do occur between the B\,$^1\Sigma^+_u$ (Lyman) and C\,$^1\Pi_u$ (Werner) states to the ground state.  Given solar elemental abundances and \re{typical temperatures and densities}, molecular hydrogen (\hh) should be the most abundant molecule in the \re{photosphere and low chromosphere.}  Yet molecular hydrogen was not identified in solar spectra until 1977 when high spectral and spatial resolution UV data became available.  \citet{jordan77} reported the first observations of lines from the Lyman band of \hh{} made using the High Resolution Telescope and Spectrograph \citep[HRTS,][]{bartoe75}. Lines from the Werner and Lyman bands of \hh{} were reported by \citet{bartoe79}. These authors showed that the UV emission spectrum of \hh{} must originate from a process that selects particular upper levels.

Earlier, \citet{krishna75} predicted intensities for the UV \hh\ lines based on LTE populations and found that only weak emission and absorption might occur.  Noting particular patterns in the observed emission spectra, \citet{jordan78} and \citet{bartoe79} showed that the observed UV lines of \hh\ result from {\it fluorescent excitation by strong UV lines}, including \re{\ion{H}{1} Ly $\alpha$ 1215.67 \AA}, \ion{C}{2} 1334.53 and 1335.70 \AA, and \ion{Si}{4} 1393.76 and 1402.77 \AA.  Accidental wavelength coincidences arise between the densely packed vibration-rotation substructure of the electronic Lyman and Werner transitions and these strong solar UV emission lines.  Photo-excitation of selected \hh{} levels is explicitly a non-LTE process.  The authors calculated the non-LTE fluorescent spectrum using a quiet-Sun model provided by Avrett including the depth dependent intensities for \ion{H}{1} Ly$\alpha$, and using the intensities of bright UV lines also observed by HRTS. The calculated \hh{} line intensities were in reasonable agreement with observations, and \hh{} line ratios generally followed the branching ratios, suggesting the \hh{} lines have modest optical depths.  They also found that \hh{} emission tends to appear under particular conditions within active regions: above sunspots, in light bridges, and in flare ribbons.  Additional identifications were added by \citet{sandlin86} based on the wavelengths and relative intensities of \re{lines with common upper levels}, however the excitation pathways for the new lines were not clear as they were for the lines identified by Jordan and Bartoe.

In other relevant work, large thermal inhomogeneities were needed to simultaneously explain infrared spectra of the CO molecule, whose abundance strongly decreases with \re{increasing} temperature, and the emission cores of \ion{Ca}{2}, which strongly increase in brightness with \re{increasing} temperature.  In static models horizontal temperature fluctuations are needed to explain these effects \citep[e.g.][]{ayres86, avrett95}.  In dynamical models strong thermal fluctuations occur naturally in the mid- to upper-chromosphere during the course of evolution \citep{carlsson95, leenaarts11}.  While \citet{leenaarts11} emphasize these inhomogeneities, the largest inhomogeneities are always seen in the most tenuous parts of the chromosphere where the \hh{} abundance is very small anyway.  These inhomogeneities may not have a significant effect on deeper, denser layers of the solar atmosphere where \hh{} is most abundant.

Are strong thermal inhomogeneities necessary for the UV \hh{} emission to appear, or does it merely require a source of UV photons within the context of simple models?  Our approach is to explore the simplest models first, trying to identify the most important physical processes needed to understand \hh{} observations and the limits of such models.  We study conditions in typical 1D models to see how the \hh{} UV spectra might form in both cool and warm atmospheric models (warm compared to the kinds of model proposed by \citealp{ayres86}, for example).  We study models of increasing complexity to find the simplest physical picture compatible with atlas spectra from SUMER and HRTS.  Using these 1D models, incorporating updated molecular data available from \cite{abgrall00} and references therein, we are able to largely reproduce the observed distribution of \hh\ lines, including 54 previously unidentified lines in the atlas spectra that we identify in Section 4.

Our purpose is to lay the groundwork for future analysis of many \hh{} spectra acquired under a broad variety of conditions by the Interface Region Imaging Spectrograph \citep[IRIS,][]{depontieu14}. IRIS covers a more limited wavelength range from 1332-1358 and 1389-1406 \AA\ in the far ultraviolet.  However, this region targets some of the major excitation sources (\ion{C}{2} and \ion{Si}{4}) for \hh\ and includes many previously identified \hh\ lines, with superior spatial, spectral, and temporal resolution as compared to HRTS or SUMER.  First we review earlier relevant results.

\begin{figure}
 	\begin{center}
		\includegraphics[width=3in]{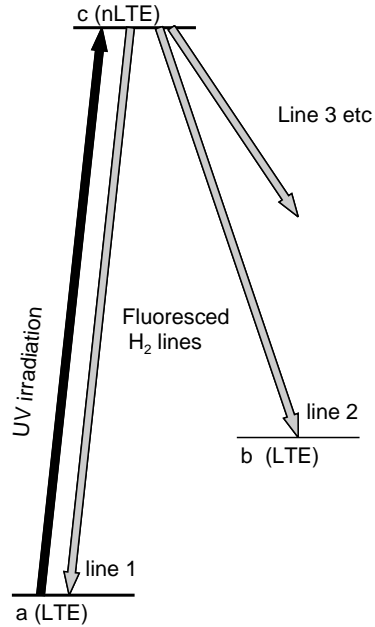}
	\end{center}
	\caption{An illustration of the essential processes for \hh\ fluorescence. \hh\ energy levels are labeled $a$, $b$, $c$ with radiative transitions labeled line 1, 2, 3 etc.  A transition occurs upwards via the absorption of intense UV radiation by line 1, downward transitions via line 1, 2, 3 etc. are spontaneous emissions.  The lower levels $a$ and $b$, containing the bulk of the \hh\ population, belong to the lowest electronic state, and they are assumed to be strongly coupled collisionally.  Their relative populations are therefore close to LTE, unlike level $c$.  Continuum transitions from level $c$ to a dissociated state are treated the same way as the transitions for radiative emission.}
	\label{fig:3l}
\end{figure}

\section{Reconciling earlier work}\label{earlier}
Strong UV emission from \hh\ in the solar atmosphere arises from fluorescence, a process illustrated schematically in Figure \ref{fig:3l} with representative energy levels (labeled $a$, $b$, $c$).  In this example a Lyman or Werner transition, from level $a$ in the ground electronic state to level $c$ in an excited electronic state, occurs via absorption of external UV radiation of the same wavelength.  Based on the selection rules and the relative probability for each transition, spontaneous emission occurs back down to many different possible levels ($a$, $b$, etc.) in the ground state or, with a typically much lower probability, into a continuum state of two dissociated atoms.  The energy levels of the ground state are collisionally coupled (i.e. in LTE), so level populations are quickly redistributed after fluorescence occurs.  The fluorescently excited upper level has a lifetime that is very short compared to the collisional timescale\re{, which is characteristic of a non-LTE process.}  In practice a full multi-level model, containing all necessary levels for both Lyman and Werner bands, is adopted for the calculations made below.

The process that produces fluorescence is physically simple, but determining the molecular populations and solving the equations of radiative transfer may be complicated if the thermally inhomogeneous and time-varying structure of the chromosphere must be taken into account.  A full solution to the time-dependent non-LTE equations for the non-equilibrium populations of the most abundant molecular species has been performed by \citet{leenaarts11} in 2D.  These authors emphasize the highly structured energy balance of the acoustically-heated chromosphere and its relationship to latent heat of dissociation of the most abundant molecules of \hh\ and CO.  One might infer that a dynamical approach is therefore required to understand the excitation of \hh\ molecules by UV photons.

\begin{figure*}
\epsscale{1.2}
\plotone{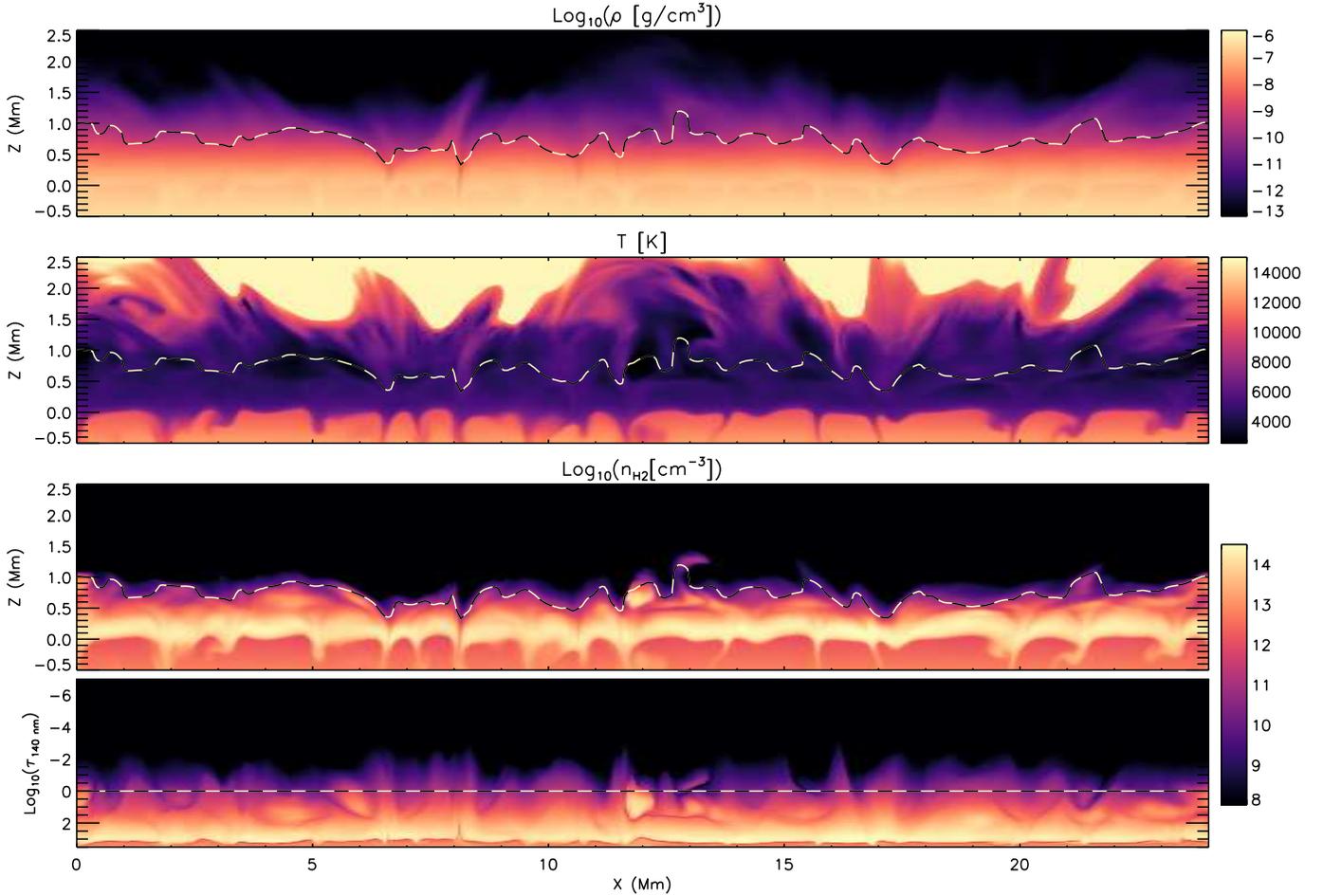}
	\caption{A vertical slice through time step 385 of the 3D simulation {\tt en024048\_hion} from {\tt http://sdc.uio.no/vol/simulations} made with the {\it Bifrost} code is used to show mass density, temperature, and number density of \hh\ molecules calculated in dissociative equilibrium (top three panels, respectively) in a region containing photospheric magnetic field concentrations. The bottom of the four panels is the same \hh\ number density calculation as in the third panel, but with a vertical axis that reflects the vertical opacity rather than the height, specifically, the logarithm of continuum optical depth at 140 nm. The dashed line in all panels shows the height of $\tau_{140\rm nm}=1$ where most of the \hh{} UV emission originates.}
	\label{fig:bifrost}
\end{figure*}

In apparent contrast, much earlier work successfully \re{reproduced} \hh\ intensities working with a 1D stratified model atmosphere \citep{jordan78,bartoe79}.  Using a model for the fluorescence containing 17 \hh\ lines (see their equation 1), \citet{bartoe79} derived an \hh\ mean column density of $\approx 1.0\times10^{18}$ cm$^{-2}$.  Using a density scale height for \hh\ of $\mathcal{H}/2 \approx 60$ km (where $\mathcal{H}$ is the density scale height of the gas), typical \hh\ number densities at the region of emission are then $\approx 10^{11}$ cm$^{-3}$.  This places the molecular emission close to a height of 600 km above the continuum photosphere in typical models.  The mass column density $m$ of the plasma within this preferred \hh\ emission region lies close to $\mathcal{H} \times \rho$ with $\rho \approx 10^{-9}$ to $10^{-10}$ g~cm$^{-3}$, so that $m \approx 0.001-0.01$ g~cm$^{-2}$.

Can bright UV line radiation (from H L$\alpha$, \ion{C}{2}, \ion{Si}{4}, \ion{O}{4}, see Table~\protect\ref{tbl:pumps}) that excites \hh{} \re{fluorescence} penetrate down from the tenuous hotter regions (middle panel of Figure~\protect\ref{fig:bifrost}) to these \re{cooler} layers?  Indeed it can, even in 1D models.  \re{As noted by \citet{jordan78}, transitions to the Lyman and Werner bands occur from vibrationally excited levels in the ground state which are only significantly populated in relatively warm gas.  As a result the \hh\ molecules producing fluorescence lie above the temperature minimum where UV photons can reach them.}  Between 911 and 1520 \AA, the opacity is dominated by photoionization of neutral C and Si (\citealt{vernazza76}, \citealt{socas15}).  Continuum optical depths become unity between $m=$0.002 and 0.01 g cm$^{-2}$.  The fact that the locations of the observed \hh\ emission and continuum optical depth unity coincide implies that molecules deeper in the atmosphere are shielded from UV photons by the \re{continuum} opacity.

The apparent contradiction between these two approaches is resolved when one looks carefully at conditions in the low chromosphere where most \hh\ molecules must reside owing simply to the dominance of density and temperature stratification.  In regions less than $\approx 1$ Mm above the continuum photosphere, horizontal deviations from hydrostatic equilibrium are small \citep[e.g.][see his Figure 1]{judge15}.

We investigate this physical picture using a 2D vertical cut from a 3D MHD simulation of the Sun's atmosphere, \verb+en024048_hion+ from \citet{carlsson16}, constructed using the {\it Bifrost} code \citep{Gudiksen2011}.  Some properties of this slice are shown in Figure~\protect\ref{fig:bifrost}.  The top three panels in the figure show the mass density $\log_{10}(\rho)$, the temperature, and the equilibrium \hh\ number density $\log_{10}(n_{H_2})$, all with the same vertical scale in height.  We have calculated the equilibrium population of \hh, given the plasma temperature and the population density of neutral hydrogen $n({\rm H})$, using the (Saha) equilibrium equation:
\begin{equation}\label{eqeq}
n(\hh)= n^2({\rm H}) \left(\frac {h^2}{\pi m_{\rm H} k T} \right )^{3/2}\frac{Z_{\hh}(T)}  {4}  e^{D/kT}
\end{equation}
where $D =4.478$ eV is the dissociation energy of \hh\ \citep{liu09} and $Z_{H_2}(T)$ is the temperature-dependent partition function of \hh\ \re{\citep[we use the tabulated function of][]{barklem16}}.  Note that because of the three-body formation process, \hh\ densities vary with about 1/2 the vertical scale height of the gas densities.

\re{The continuum opacity at 140 nm is vastly dominated by photoionization of neutral Si.  For the purposes of this illustration, the opacity was calculated using the Si photoionization cross sections of \citep{nahar00}.  Thompson scattering was included to prevent unrealistically low coronal opacities when setting the vertical scale to match that of the corresponding 1-D geometrically averaged atmosphere.  The Si population was calculated both in LTE and by including Si explicitly in a non-LTE calculation of statistical equilibrium using the Rybiki-Hummer radiative transfer and chemical equilibrium code \citep{uitenbroek00}}.  Both methods yield the same results for the purposes of this illustration.  The dashed line in all four panels of Figure \ref{fig:bifrost} shows the height of $\tau_{140\rm nm}=1$.  The fourth (bottom) panel also shows the equilibrium \hh\ number density, but with the vertical scale in terms of the optical depth at 140 nm.

\re{Obvious multi-dimensional structure, including intense magnetic flux concentrations in the photosphere and complex temperature structure higher up, are present.  It is important to note that the authors of this simulation caution that the lowest temperatures found in the simulation are unrealistic} and that a higher, more uniform temperature structure in the mid-chromosphere is required to explain the appearance of the \ion{Mg}{2} lines in the vicinity of strong magnetic concentrations \citep{carlsson15}.  However, this simulation still serves to demonstrate an important point.  In those layers where \hh\ molecules are abundant, the iso-surfaces of molecular density remain closer to horizontal than vertical.  \re{That said}, the silicon opacity shields the regions where \hh\ molecules are most abundant from the incoming UV radiation, so the observed \hh\ emission does not originate from the regions of highest \hh\ density - it originates near the $\tau=1$ surface of the incoming/outgoing radiation (which are typically close in wavelength). 

The stratification of plasma and \hh\ serves to demonstrate that the process of radiative transfer occurs predominantly in the vertical direction.  Therefore, one dimensional models in principle might be suitable for studying the excitation of fluoresced \hh\ UV line emission in spite of the fact that the sources of UV radiation can cause (and originate from) highly inhomogeneous heating (middle panel of Figure~\protect\ref{fig:bifrost}).  The surest measure of success is provided by compelling comparison with observations of the real Sun.  Therefore, it is important to explore the simplest models first.

Recent work with the IRIS instrument has highlighted new properties of the \hh\ UV spectrum observed under more unusual circumstances. \cite{schmit14} reported \hh\ absorption features on top of bright continuum and highly variable emission from lines of \ion{C}{2} and \ion{Si}{4} usually associated with the chromosphere-corona transition region.  These observations show that hotter ($10^4-10^5$ K) plasma can occasionally exist deeper than cooler, molecule-bearing gas, a situation absent in static 1D models.  Further departures from the 1D picture are discussed in Section 5, where we argue that the vertical scale height is still an important aide to our understanding.

\section{Non-LTE Radiative Transfer Calculations}
Based on the arguments in Section \ref{earlier}, we adopt a 1D, stratified model of \hh\ fluorescence containing the physics described by \citet{jordan78} and \citet{bartoe79}, including the vertical transfer of downward propagating UV radiation from the solar transition region.  The observed \hh\ lines are orders of magnitude less intense than the transition region lines that excite them, and their line ratios are well matched to the theoretical branching ratios \citep{bartoe79}, implying that the optical depths of the \hh\ lines are at most unity.  Therefore, their impact on the radiative transfer is small.  This allows us to decouple the calculation of detailed radiative transfer in the model atmosphere from the calculation of the thousands of \hh\ line intensities and carry them out in two separate pieces which are each described in the following subsections.

Observations from IRIS and previous missions show \hh\ fluorescence occurs in a variety of different regions in response to bright emission from nearby \re{hot} sources.  As input to these calculations we use semi-empirical 1D models and external radiative input, which is scaled from its quiet-Sun values and added through the top of the atmosphere, to simulate a range of thermal and radiative conditions from cool atmospheres with quiet-Sun intensities to hot atmospheres with bright flare radiation.  The transition region line intensities given in Table \ref{tbl:pumps} are used as the input spectrum.

\begin{deluxetable}{lrcc}
\tabletypesize{\scriptsize}
\tablecaption{Integrated line intensities of strong UV lines\label{tbl:pumps}}
\tablehead{\colhead{Transition} & \colhead{ $\lambda$ } & \colhead{Line intensity} & \colhead{Linewidth }\\
\colhead{} & \colhead{[\AA]} & \colhead{[erg~cm$^{-2}$s$^{-1}$sr$^{-1}$]} & \colhead{units of 10$^{10}$ Hz} } 
\startdata
  \ion{H}{1} Ly $\gamma$ &    972.5 &     129 &      62 \\
          \ion{C}{3} &    977.0 &     630 &      33 \\
   \ion{H}{1} Ly $\beta$ &   1025.7 &     561 &      58 \\
  \ion{H}{1} Ly $\alpha$ &   1215.6 &   45042 &      49 \\
           \ion{C}{2} &   1334.5 &     593 &      13 \\
           \ion{C}{2} &   1335.7 &     709 &      13 \\
           \ion{O}{5} &   1371.3 &      13 &      18 \\
           \ion{Si}{4} &   1393.8 &     226 &      17 \\
           \ion{O}{4} &   1397.2 &       9 &      18 \\
           \ion{O}{4} &   1399.8 &      13 &      18 \\
           \ion{O}{4} &   1401.2 &      34 &      18 \\
           \ion{Si}{4} &   1402.8 &     105 &      17 \\
      \ion{S}{4}/\ion{O}{4} &   1404.8 &      24 &      18
\enddata
\tablecomments{Intensities were integrated over wavelength, minus the background continuum, from data for the SUMER quiet-Sun atlas spectrum of \citet{curdt01}.}
\end{deluxetable}

We have selected three diverse 1D, semi-empirical atmospheric models for the calculations:  the average quiet-Sun model atmosphere ``C'' from \citet{fontenla93} (henceforth ``FALC''), the cool model atmosphere COX from \citet{avrett95}, and the F2 flare model atmosphere from \citet{machado80}.  FALC is a ``warm'' quiet-Sun atmosphere calculated based on the average intensity of UV emission lines observed in quiet-Sun regions.  The photosphere and low chromosphere from this model are warmer than the earlier model ``C'' of \citet{vernazza81}, following the work of \citet{maltby86} and \citet{lemaire81} respectively.  FALC is far warmer than models previously developed to understand the CO spectrum, including the COX model.  COX was developed by \citet{avrett95} (where it is called $M_{CO}$) to explain the average appearance of strong vibration-rotation lines of the CO molecule observed in the infrared, however this atmosphere does not correctly reproduce many other properties of average chromospheric observations.  The COX model reaches a lower minimum temperature much higher in the atmosphere as compared to FALC.  The F2 atmospheric model was produced using, what was at the time, the most comprehensive set of chromospheric observables obtained during flares.  The F2 model was specifically intended to reproduce the strong continuum and line emission of \ion{H}{1}, \ion{Si}{1}, \ion{C}{1}, \ion{Ca}{2}, and \ion{Mg}{2} seen in bright flares.  This model is characterized by a warm chromosphere and a low-lying, steep transition region.

\begin{figure*}
	\begin{center}
		\includegraphics[width=6in]{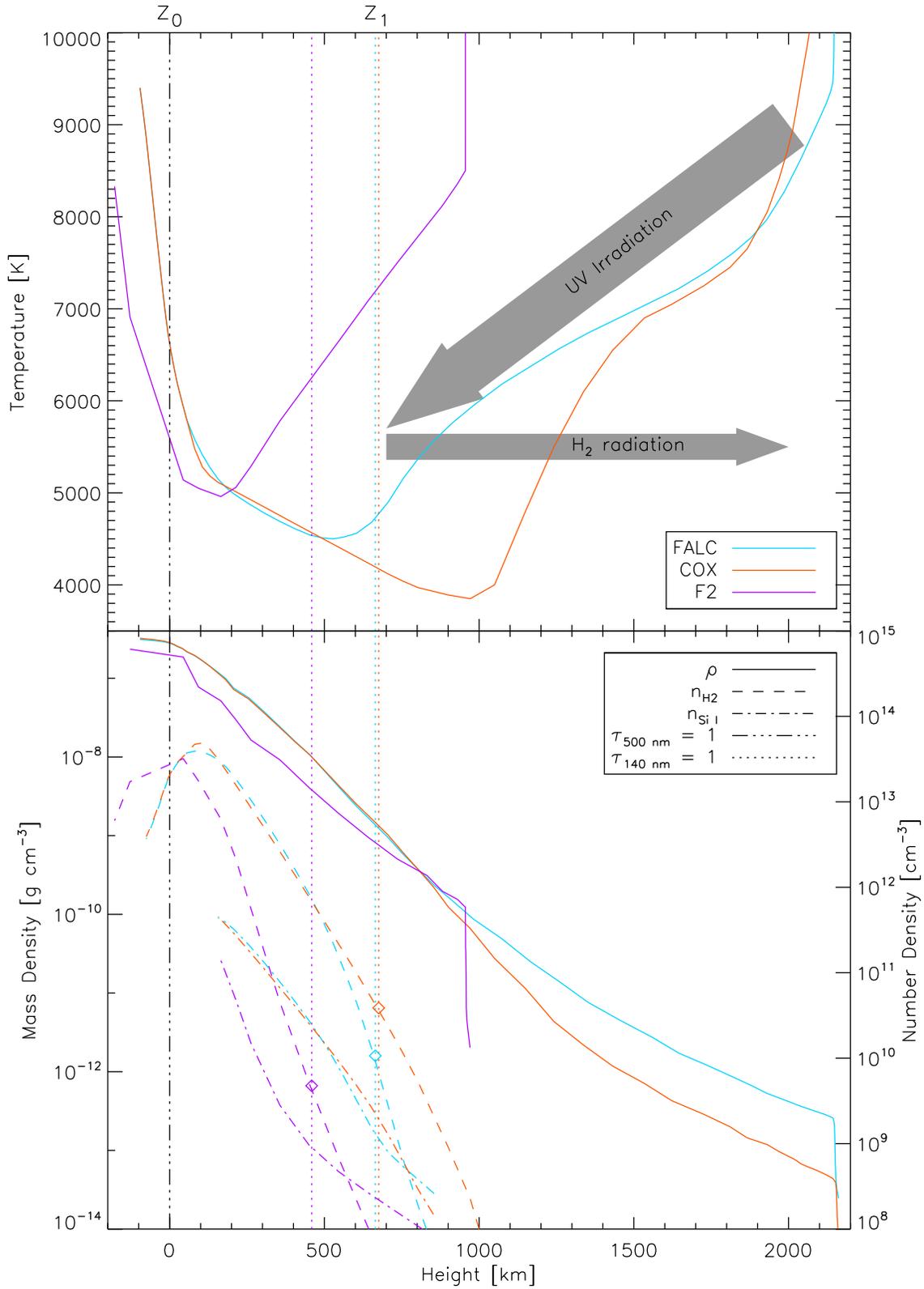}
	\end{center}
	\caption{Electron temperature (top panel) and the number densities of hydrogen and \hh\ (bottom panel) are shown as a function of height for the FALC, COX, and F2 model atmospheres.  The $\tau$ = 1 depth computed at 1400 \AA\ (indicated by the vertical dotted line at z$_1$) is dominated by the \ion{Si}{1} photoionization continuum, which was treated in non-LTE and is almost identical in FALC and COX.  The number densities of neutral silicon are also shown (dot-dashed lines). In such a model the UV irradiation from above (diagonal arrow) can penetrate only to the depth $z_1$ and drops rapidly below this height.}
	\label{fig:atmos}
\end{figure*}

The properties of these models can be seen in Figure \ref{fig:atmos}, which shows the temperature (top panel), and the total mass density and the number density of \hh\ (bottom panel) as a function of height.  We also show the \ion{Si}{1} number density, which is representative of of the UV opacity.  The UV continuum optical depth ($\tau_{140\rm nm}=1$) is shown for each model by the dotted lines in the figure.  The properties of FALC and COX are quite similar in this region, making the $\tau=1$ depth nearly the same.  The F2 atmosphere is significantly warmer, which pushes $\tau=1$ to greater depth.  

The background continuum opacities, due to ionization of \ion{Si}{1}, drop exponentially with a scale height of $\mathcal{H}$, while \hh\ densities fall off more steeply with height as $\mathcal{H}/2$.  There is therefore a ``sweet spot'' in depth for each atmosphere, close to continuum optical depth $\tau_\kappa = 1$ where \hh\ molecules are abundant enough, and continuum optical depths are small enough for photo-excitation to occur.  As direct photo-dissociation of ground state \hh\ is a forbidden transition, photo-dissociation of \hh\ occurs via electronic excitation followed by radiative dissociation and therefore only for photons shorter than 1108 \AA\ \citep{heays17, vandishoeck11}.  Since the UV opacity increases steeply with decreasing wavelength, very few dissociating photons can find their way down to the sweet spot for excitation of \hh\.  This narrow formation range for \hh\ fluorescence lends itself to a zero-dimensional approximation which is explored in the final subsection.

\subsection{Detailed Radiative Transfer Calculations}
We have solved the equations of radiative transfer and non-LTE statistical equilibrium for atoms of H, C, O, Si, and Fe using the RH code mentioned earlier.  We have included N, Al, S, Na, Mg, and Ca in the calculations assuming LTE.  Selected atomic transitions were actively calculated.  In the wavelength range of interest these include: \ion{H}{1} Lyman $\alpha$ 1215.6 \AA; \ion{O}{1} 1302.2, 1304.9, and 1306.0 \AA; and \ion{C}{2} 1334.5 and 1335.7 \AA.  RH gives self-consistent sets of atomic and molecular populations, pressures, opacities, and spectral intensities as functions of depth and angle relative to the Sun's local vertical.

Two cases for the radiative upper boundary conditions were run: zero incoming radiation, and radiation set by line emission from the transition region lying immediately above.  \re{We assume that the transition region emission is optically thin and radiates isotropically, covering $\omega = 2\pi$ steradians of the chromosphere.  The angle-averaged mean intensity, $\bar{J}$, seen by the chromosphere is therefore ${1\over2}I$.}  In the calculations we used multiples ($m$) of the measured UV transition region lines from the SUMER atlas of \citet{curdt01} shown in Table \ref{tbl:pumps}.  \re{The angle-averaged intensities are then ${1\over2}mI$.}  \re{In the case of Ly $\alpha$ and \ion{C}{2}, radiation was added to the line radiation calculated by RH.}  For multipliers $m=$30 and greater, the lines were also made twice as broad, consistent with typical behavior of very bright radiation observed during flares.
  
\begin{figure}
	\begin{center}
		\includegraphics[width=3.5in]{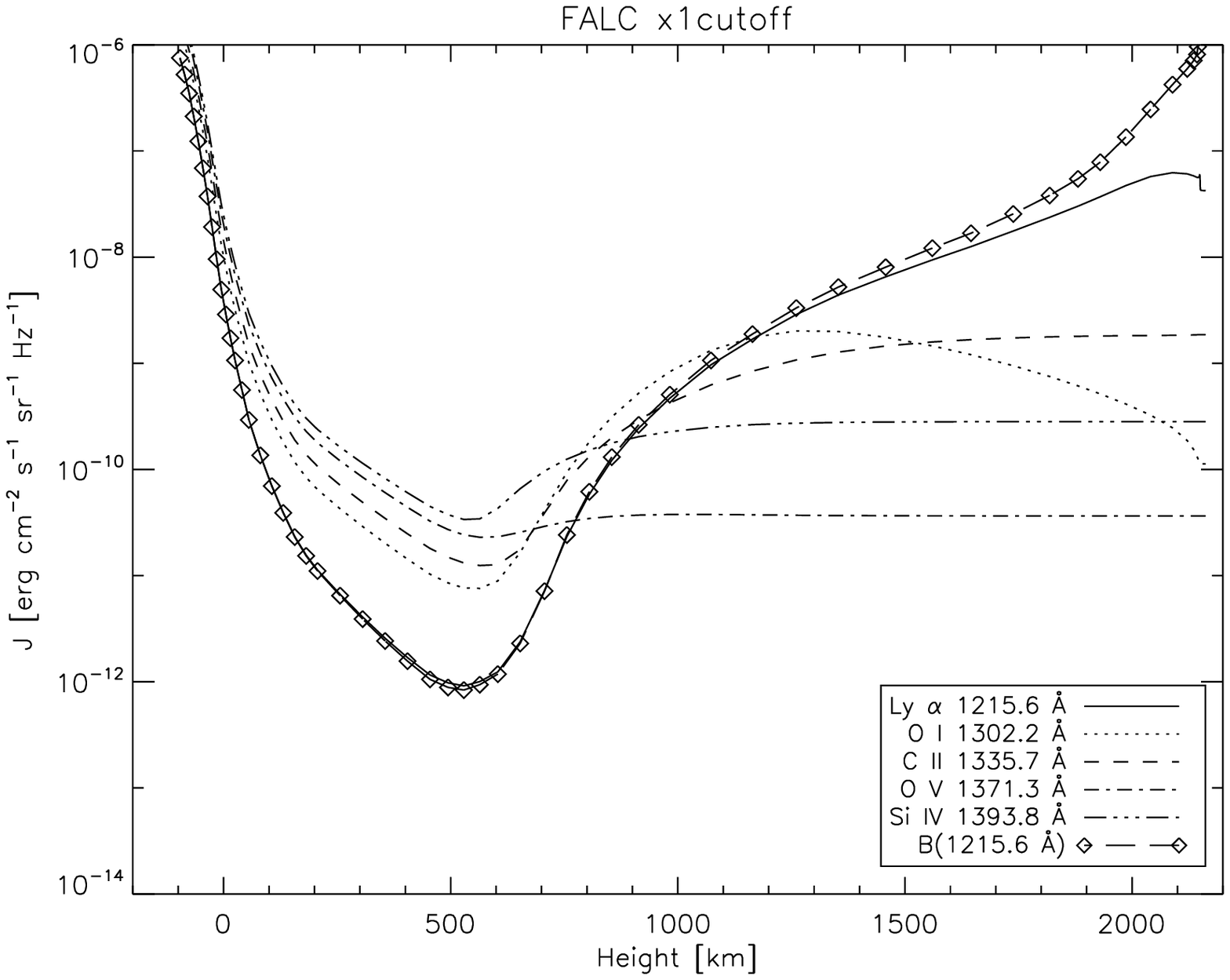}
		\includegraphics[width=3.5in]{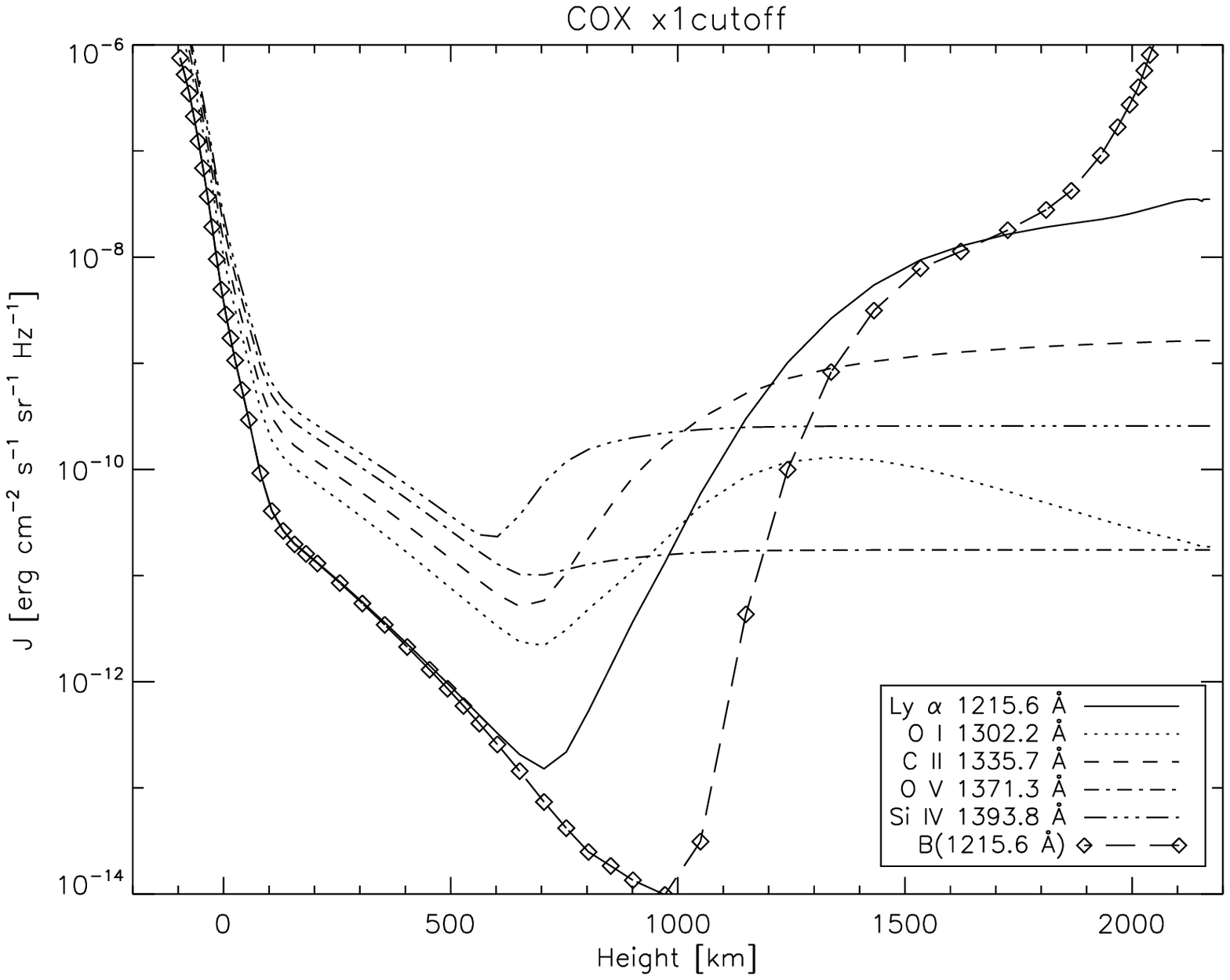}
		\includegraphics[width=3.5in]{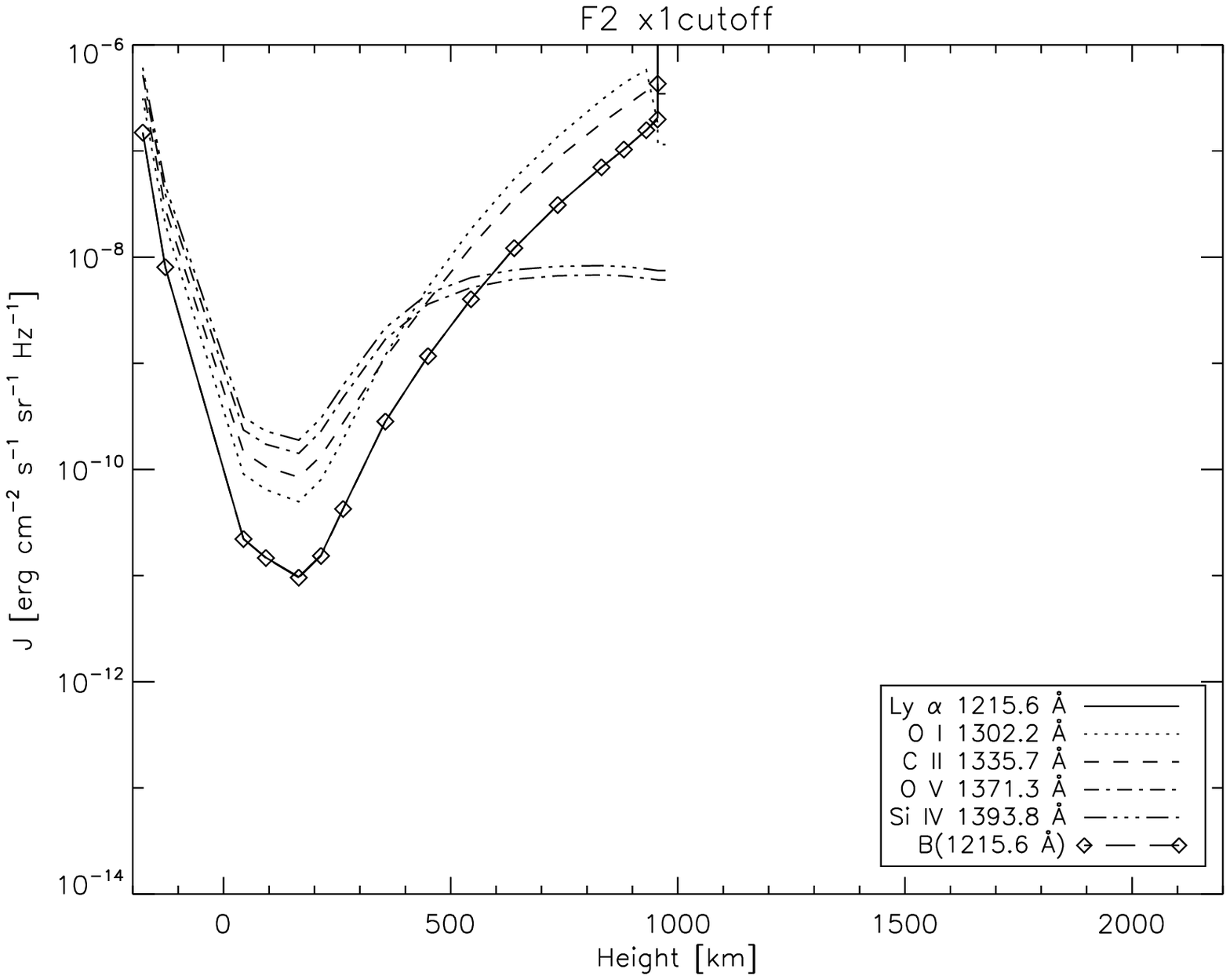}
	\end{center}
	\caption{Angle-averaged intensities at the centers of strong UV lines, and the blackbody intensity at the location of Ly $\alpha$ are shown as a function of height in models FALC, COX, and F2.  Downward directed line intensities were added at $\times$1 those of the typical quiet-Sun (with the exception of \ion{C}{1} where no additional intensity was added).}
	\label{fig:j}
\end{figure}

The mean angle-averaged intensities computed from the models using RH were saved to a file and then used as input to excite the \hh{} lines in a separate calculation, described below (Section \ref{sec:art}).  Figure \ref{fig:j} shows $\bar{J}$ as a function of height computed at the central wavelengths of several strong UV emission lines  in models FALC, COX, and F2 for the $m=$1 case of down-going radiation.  The blackbody intensity at the location of Ly $\alpha$ is also shown for comparison.  These intensities are leading terms in determining the intensities of the \hh{} fluorescent lines (see eq. \protect\ref{eqn2} below).  Two classes of behavior are seen in the Figure.  The lines of \ion{H}{1} Ly $\alpha$ 1215.6 \AA, \ion{O}{1} 1302.2 \AA, and \ion{C}{2} 1335.7 \AA, which are actively calculated by RH, have significant optical depths throughout the chromosphere.  The behavior of $J$ in these chromospheric lines is controlled by the fully coupled radiation and non-LTE atomic populations (source functions), and by the thermalization of the source functions ($S_\nu$ approaching $B_\nu$) deep in the atmosphere.  In contrast, the transition region lines of Si IV 1393.8 \AA\ and O V 1371.3 \AA, which were not explicitly calculated by RH, have essentially no opacity across the chromosphere.  The $J$ values show a simple $\exp{(-\tau)}$ variation where down-going radiation is attenuated by this term only deep in the atmosphere (near 600-700 km), where $\tau_k \rightarrow 1$.  The mean intensities for transition region lines throughout the chromosphere are dominated by the UV radiation coming down from the upper boundary.  All models show this simple behavior with some variation in the height where the greatest attenuation is achieved.

\subsection{Radiative Transfer Calculations for \hh \label{sec:art}}
In the calculations of \hh\ fluorescent intensity we use the wavelengths, energy levels, and transition probability data for \re{27,981} lines in the \hh\ Lyman and Werner bands from the semi-empirical calculations of \citet{abgrall93a, abgrall93b} \re{and \citet{abgrall00} taken from the their on-line database at \url{http://sesam.obspm.fr/}}.  The ground state energy levels include rotational quantum number up to $\mathcal{J}$=26 and vibrational quantum number up to $v$=14. The \bstate\ and \cstate\ levels include up to $\mathcal{J}$=25, and up to $v=37$ and 13, respectively.  Selection rules for allowed radiative transitions in the Lyman band are $\Delta\mathcal{J}=\pm 1$ and for the Werner bands $\Delta\mathcal{J}=0, \pm 1$.  As a homonuclear molecule, \hh\ supports both ortho- (singlet) and para- (triplet) nuclear spin states.  In the ground and Werner electronic states ortho-states have odd $\mathcal{J}$, while para-states have even $\mathcal{J}$.  In the Lyman band this is reversed, ortho-states have even $\mathcal{J}$ while para-states have odd $\mathcal{J}$.  Therefore based on the selection rules, transitions to and from the Lyman band preserve the nuclear spin state of the molecule, while transitions from the Werner band may not.

Radiative dissociation producing continuum emission does sometimes occur for these fluorescently excited \hh\ levels.  The probabilities of dissociative continuum transitions are small, occurring less than 1\% of the time for the majority of the brightest lines in our calculations, which are Lyman transitions from excited levels in the \bstate\ state with low vibrational and rotational quantum numbers. We use the total continuum emission probabilities of \cite{abgrall00}, which go up to $\mathcal{J}$=10, and integrate the continuum emission probability densities of \cite{abgrall97} for $\mathcal{J}>$10. We do not calculate the continuum emission, as its spectral irradiance is extremely weak.  Based on the dissociative probabilities and the upper level populations for all the transitions in our calculation, dissociation occurs for approximately 10\% of transitions at the heights important for \hh\ emission.   The effect of this dissociation on the target \hh\ population is not significant, and for the purposes of our calculation the largest effect of dissociation resulting from these transitions is a modest decrease in some of the \hh\ line intensities.

Assuming that the \hh\ lines do not affect the intensities computed from RH, we calculated the fluorescent spectrum of \hh\ using the number densities and intensities from the non-LTE radiative transfer calculations.  This is an excellent assumption for the strong atomic lines of the chromosphere since the opacities of those lines are orders of magnitude larger than those of \hh, and for the transition region lines it is justified by the success of the calculation in \citet{bartoe79}.  Optically thin line formation was assumed by \citet{jordan78}, and is justified by the \re{matching} of the \hh\ line intensity ratios, however we treat the line opacities approximately using escape probabilities as shown below.

For simplicity of exposition here, we show an example where absorption by only one line (line ``1'' excited from level $a$) contributes to the \hh\ emission from a single upper level.  All thermal and radiation variables ($n, \rho, T, S, J, \tau$) are only functions of height $z$ in the 1D stratified atmosphere.  The statistical equilibrium equation for a typical level $c$ reduces to:
\begin{equation}\label{eqn1}
n_{a} B_1 \overline{J_1} = n_c(A_1 + A_2 + A_3 + \ldots)
\end{equation}
where levels $a$ and $c$ are representative levels of the lower and upper electronic states (Figure~\protect\ref{fig:3l}).  We have assumed that the de-populating process is spontaneous decay through bound-bound transitions to the lower energy level or \re{bound-to-dissociated transitions}.  $A_1$ and $B_1$ are the Einstein coefficients for spontaneous emission and absorption respectively.  $J_1(z)$ is the angle-averaged intensity of the irradiating source, integrated over the absorption profile of the \hh\ molecules, at each depth $z$.  This absorption profile is implicitly assumed to be narrow compared with the widths of the irradiating chromospheric and transition region lines.

The lower level populations $n_a$ are assumed to be thermally populated and are therefore represented by a Boltzmann distribution:
\begin{equation}\label{eqn1a}
n_a = \frac{n_{\rm H_2} g_a}{Z_{\rm H_2}(T)} \exp{(-E_a/kT)}
\end{equation}
where $n_{\rm H_2}$ is the number density of molecular hydrogen, $g_a$ is the statistical weight of the level, $Z_{\rm H_2}(T)$ is the partition function, and $E_a$ is the energy of the level.  Again, we use the tabulated partition function of \citet{barklem16} in our calculations.  The statistical weight $g$ of a given level is nominally $2\mathcal{J}+1$, where $\mathcal{J}$ is the rotational quantum number, but because \hh\ is homonuclear this must be multiplied by an additional term of 3/4 for ortho-states or 1/4 for para- states for collisional equilibrium \citep{irwin87}.  In the ground state, ortho-states have odd $\mathcal{J}$, while para-states have even $\mathcal{J}$, therefore:
\begin{equation}\label{eqn1b}
g_a=(2\mathcal{J}_a+1) \times \begin{cases} 1/4, & \text{$\mathcal{J}_a$ even} \\ 3/4, & \text{$\mathcal{J}_a$ odd}. \end{cases}
\end{equation}
As determined by the selection rules, radiative transitions between ortho- and para- states are not allowed for Lyman band transitions, but do occur for Werner band transitions, however the molecules can also change spin states through collisions while in the ground state.  In a high collision environment such as the Sun, the states rapidly distribute to their 1/4 para-, 3/4 ortho- distribution, but this is often not the case in rarefied astrophysical regimes \citep{shaw05}.

When the fluorescent emission in \hh\ is optically thin, the excess emission over any background continuum, integrated over the line ``2'' profile is simply
\begin{equation}\label{eqn2}
I_2 ={h\nu_2 \over 4\pi} \int_{\Delta z} n_c A_2 dz \ \ \ {\rm erg \ cm^{-2} \ sr^{-1} \ s^{-1}}
\end{equation}
where $\Delta z$ is the height range over which the \hh\ lines are formed and $\nu_2 = c/\lambda_2$ is the laboratory frequency of transition ``2.'' Let
\begin{equation}\label{eqn3}
b_2 ={A_2 \over A_1 + A_2 + A_3 + \ldots}
\end{equation}
be the radiative branching ratio for line ``2,'' including all line and continuum (dissociating) transition probabilities in the denominator, then
\begin{equation}\label{eqn4}
I_2 = {h \nu_2 \over 4\pi} \int_{\delta z} n_a  B_1 \overline{J_1} b_2 dz,
\end{equation}
an  expression identical to Equation 3 of \citet{jordan78} when we replace $\overline{J_1}$ with $I_\nu$ and allow for a sum over all lower levels $a$ with a radiative transition to the upper level $c$.

We allow approximately for photon trapping by using escape probabilities. The above equations are used but with the replacement:
\begin{equation}\label{eqn5}
A_2 \rightarrow A_2 p_2(\tau_2, \tau_\kappa), \ \ \ p(\tau,\tau_\kappa) ={1 \over 1 + \tau / \sqrt{\pi}} \exp{(-\tau_\kappa)}
\end{equation}
where $p(\tau, \tau_\kappa)$ is an approximation to the frequency averaged escape probability of \hh\ line photons from a region \re{where the line center optical depth is $\tau$ and the background continuum optical depth is $\tau_k$} \citep[e.g.,][]{rybicki84}.  The term $\exp{(-\tau_\kappa)}$ allows for extinction by the background continuum.  Since values of $\tau_2$ are determined by the \hh{} LTE populations of the lower levels in the ground state, they are essentially independent of the far smaller population of the upper level $n_{c}$, and no iterations are needed between the statistical equilibrium equation (\ref{eqn1}), and the radiation transfer represented by equations (\ref{eqn2}) and (\ref{eqn5}). Note that $b_2$ and $n_c$ depend on $\tau_2$ and $\tau_\kappa$ through $p_2$.

There are many sources of line and continuum opacity at longer wavelengths that are not fully taken into account by the radiative transfer calculation performed using RH.  This leads to the unrealistic pumping of lines long-ward of about 1520 \AA\ where the \ion{Si}{1} opacity edge resides.  The continuum opacities are highly sensitive to temperature gradients.  To prevent excess pumping of \hh{} by the long wavelength continuum we enforce a cutoff at 1520 \AA, only radiation shorter than is is allowed to pump the \hh{} lines.

\subsection{A zero-dimensional approximation}
These 1D calculations are amenable to reduction to a one-point (``zero dimensional'', 0D) model that further illuminates the essential processes.  In section \protect\ref{sec:art} (especially Figure \protect\ref{fig:j}) we showed that two fundamentally different kinds of lines excite the \hh\ lines: those that form throughout the chromosphere whose transfer is dominated by line opacity of the transition itself (\ion{H}{1} Ly $\alpha$, \ion{C}{2} lines), and the other lines (canonical ``transition region lines'' like \ion{Si}{4}) whose emission usually occurs above the chromosphere and is then attenuated only by the background continuum opacity.  When $\overline{J}$ is dominated by irradiation attenuated only by continuum opacity, we have at each depth
\begin{equation}\label{eqn6}
\overline{J} \approx \frac{\omega}{ 4\pi} I_{\overline{\nu_1}} \exp{(-\tau_\kappa)} \ \ \ {\rm erg \ cm^{-2} \ sr^{-1} \ s^{-1} \ Hz^{-1}}
\end{equation}
where $\omega$ is the angular size of the downward radiation source as seen from the \hh\ molecules, $I_{\overline{\nu_1}}$ is the intensity of the downward irradiation\re{, and $\tau_k$ is the continuum optical depth.}

To a rough approximation, the exponential can be replaced  by 1 when $\tau_\kappa < 1$ and 0 otherwise.  Then, the integral  equation (6) reduces to an integral over depth $z$ of $n_a(z)$ from the origin of the irradiation down to the point $z_1$ where the continuum optical depth is unity.  Since $n_a(z) \propto exp(- 2z/\mathcal{H})$, we find
\begin{equation}\label{eqn7}
I_2 \approx {h\nu_2 \over 4\pi} n_a(z_1) B_1  {\omega \over 4\pi} I_{\overline{\nu_1}} b_2 \frac{\mathcal{H}}{2}.
\end{equation}
$I_{\overline{\nu_1}}$ is in units of erg cm$^{-2}$ sr$^{-1}$ s$^{-1}$ Hz$^{-1}$.  This is related to the frequency-integrated intensity of the illuminating line, $I_{ILL}$ by
\begin{equation}\label{eq9}
I_{\overline{\nu_1}} \Delta \nu_1 = I_{ILL}
\end{equation}
where $\Delta \nu_1$ is the Doppler width of the exciting line in Hertz.  Then, since we can set $I_{ILL} = I_{obs}$ (i.e., the photo-excited \hh{} molecules see the same radiation we observe because $\tau_\kappa < 1$), and with the Einstein relation $$\frac{2h \nu_1^3}{c^2} g_aB_1 = g_c A_1$$ we have
\begin{multline}\label{eqn10}
{I_2 \over I_{ILL}} \approx \frac{c^2}{8 \pi}  \left( {g_c \over g_a} {\nu_2 \over  \nu_1^3} A_1 b_2 \right)
\left( {\omega \over 4\pi} {1 \over \Delta \nu_1} n_a(z_1) {\mathcal{H} \over 2} \right).
\end{multline}
Equation (\ref{eqn10}) relates the ratio of two observables, the frequency-integrated \hh{} and the irradiating line intensities, in terms of atomic parameters (first bracket) and solar parameters (second bracket):  the solid angle of irradiation, Doppler widths, number densities, and scale heights.  Equation~(\protect\ref{eqn1a}) gives $n_a(z_1)$ in terms of atomic and molecular data and $T(z_1)$ and $n_{\rm H_2}(z_1)$.  

The results from this zero-dimensional approximation are compared with the more detailed calculations below.

\begin{figure*}
	\begin{center}
		\includegraphics[width=7in]{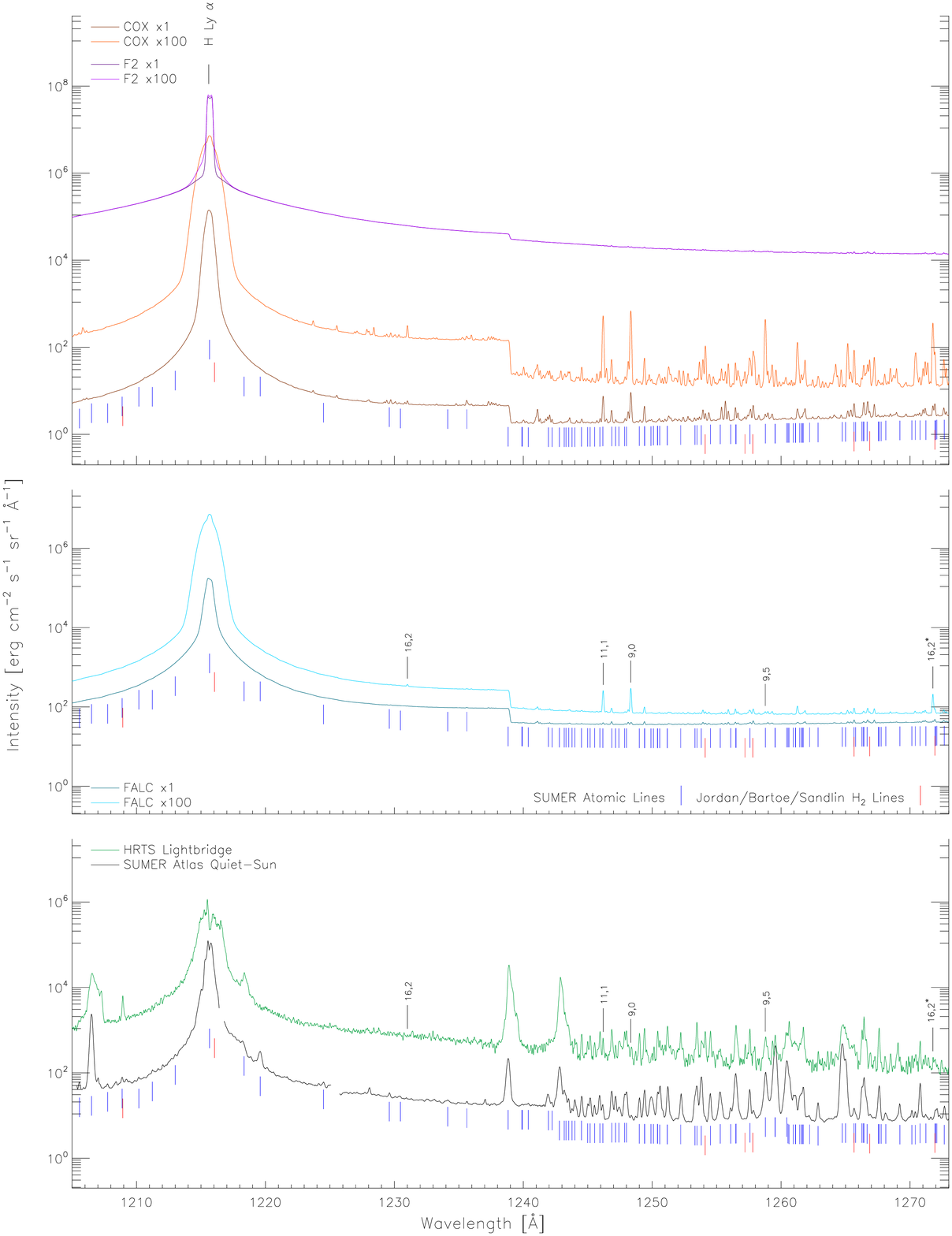}
	\end{center}
	\caption{(Top and middle) Synthetic \hh\ fluorescence spectra generated with the FALC, COX, and F2 atmospheres.  For each model we show the spectra resulting from multiplying the down-going radiation $\times1$ and $\times100$ its quiet-Sun value.  (Bottom) Observed spectral intensities from the SUMER quiet-Sun atlas spectrum \citep{curdt01} and the HRTS light bridge atlas spectrum \cite{brekke91} for the range 1205-1550 \AA.  Please refer to Section \ref{sec:spectra} for a detailed description of the figure annotations.}
	\label{fig:obs}
\end{figure*}

\addtocounter{figure}{-1}
\begin{figure*}
	\begin{center}
		\includegraphics[width=7in]{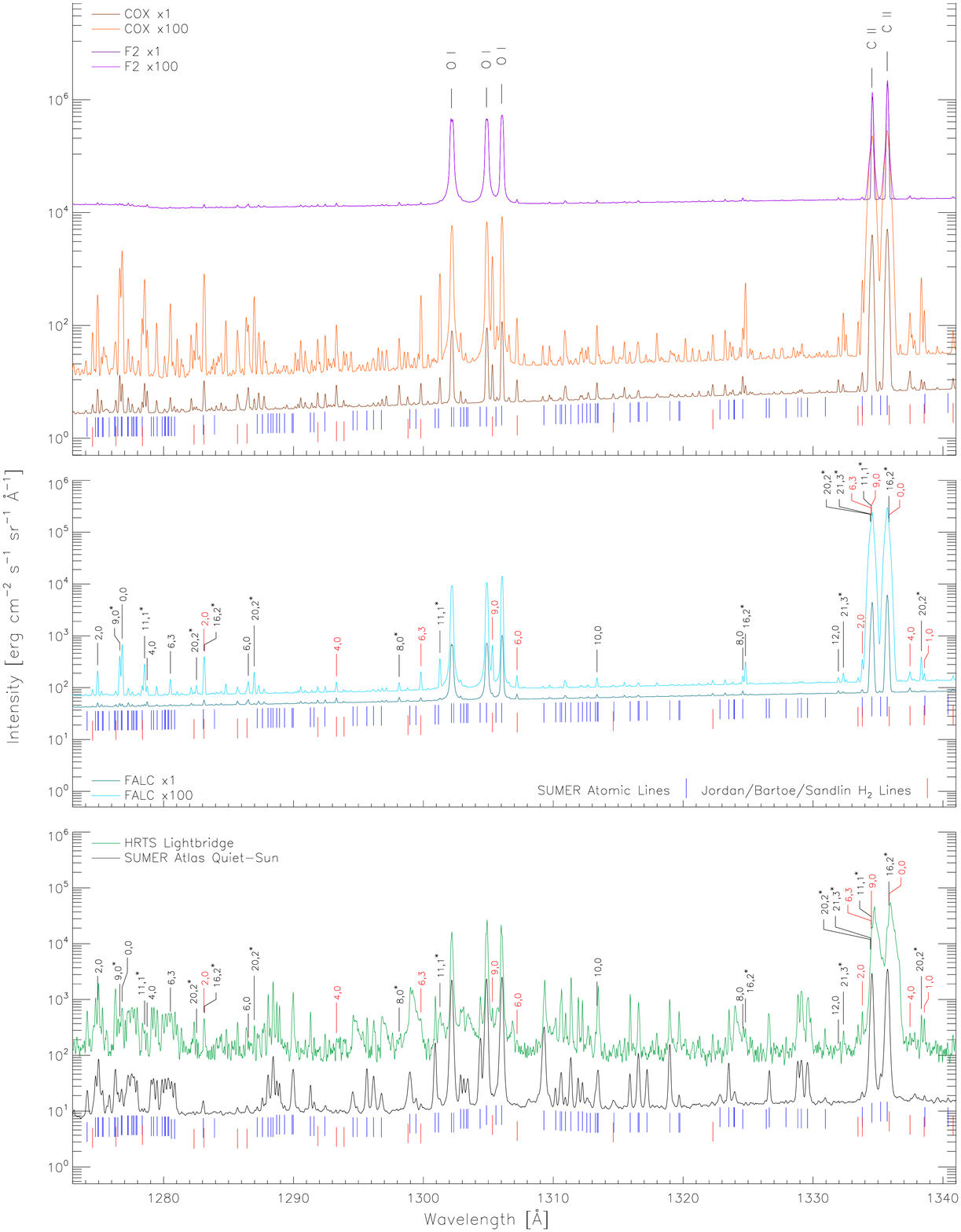}
	\end{center}
	\caption{{\it Continued.}}
\end{figure*}

\addtocounter{figure}{-1}
\begin{figure*}
	\begin{center}
		\includegraphics[width=7in]{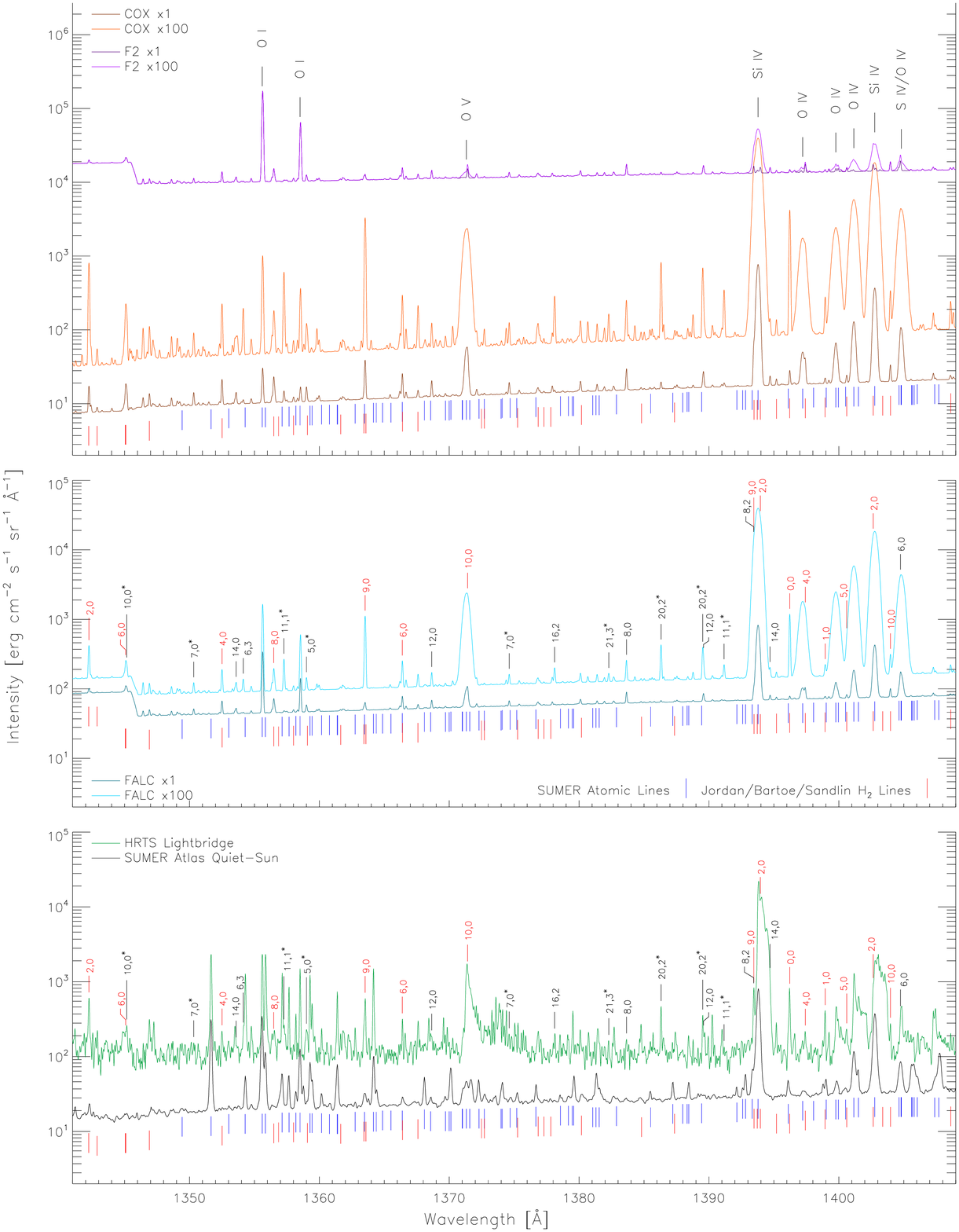}
	\end{center}
	\caption{{\it Continued.}}
\end{figure*}

\addtocounter{figure}{-1}
\begin{figure*}
	\begin{center}
		\includegraphics[width=7in]{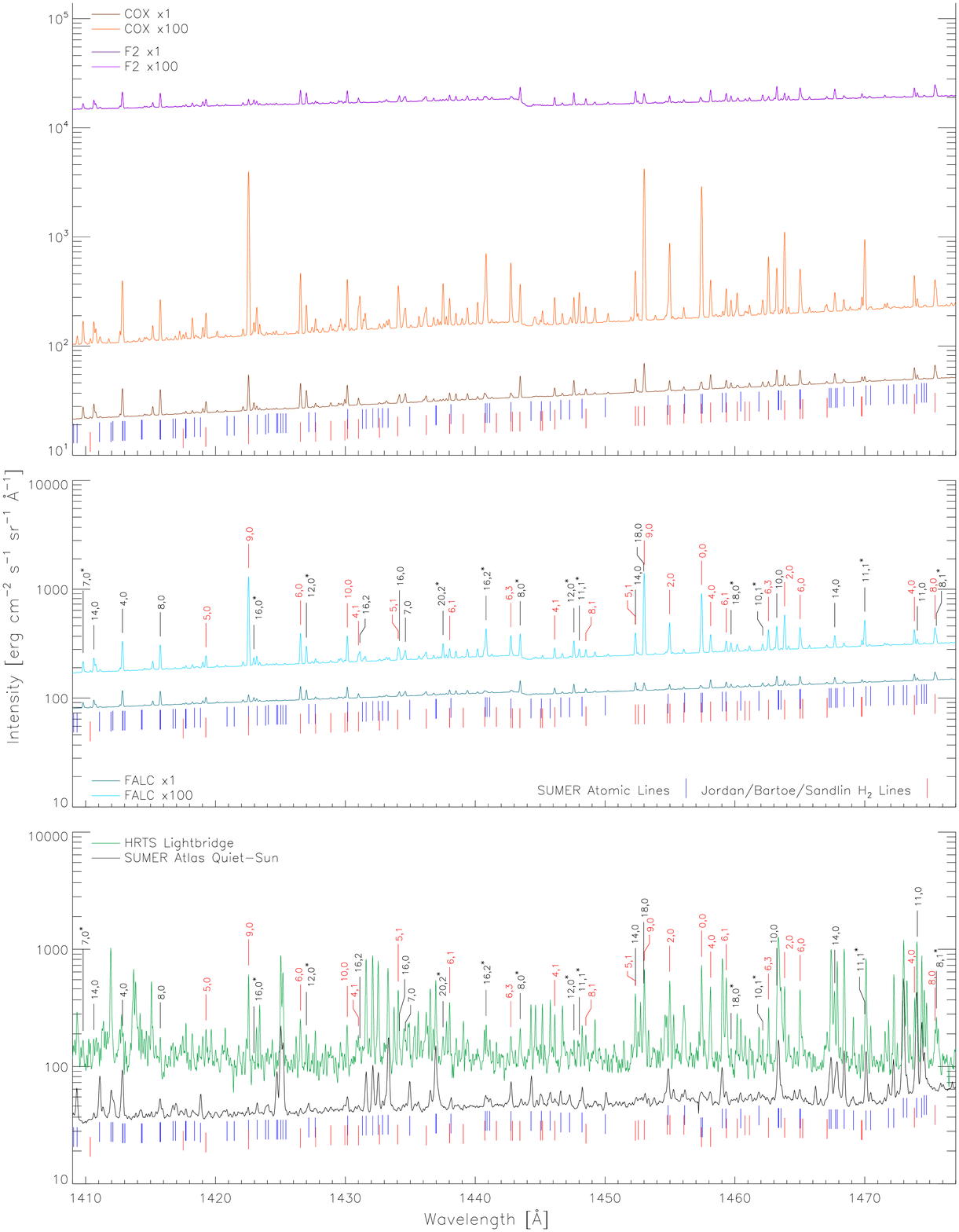}
	\end{center}
	\caption{{\it Continued.}}
\end{figure*}

\addtocounter{figure}{-1}
\begin{figure*}
	\begin{center}
		\includegraphics[width=7in]{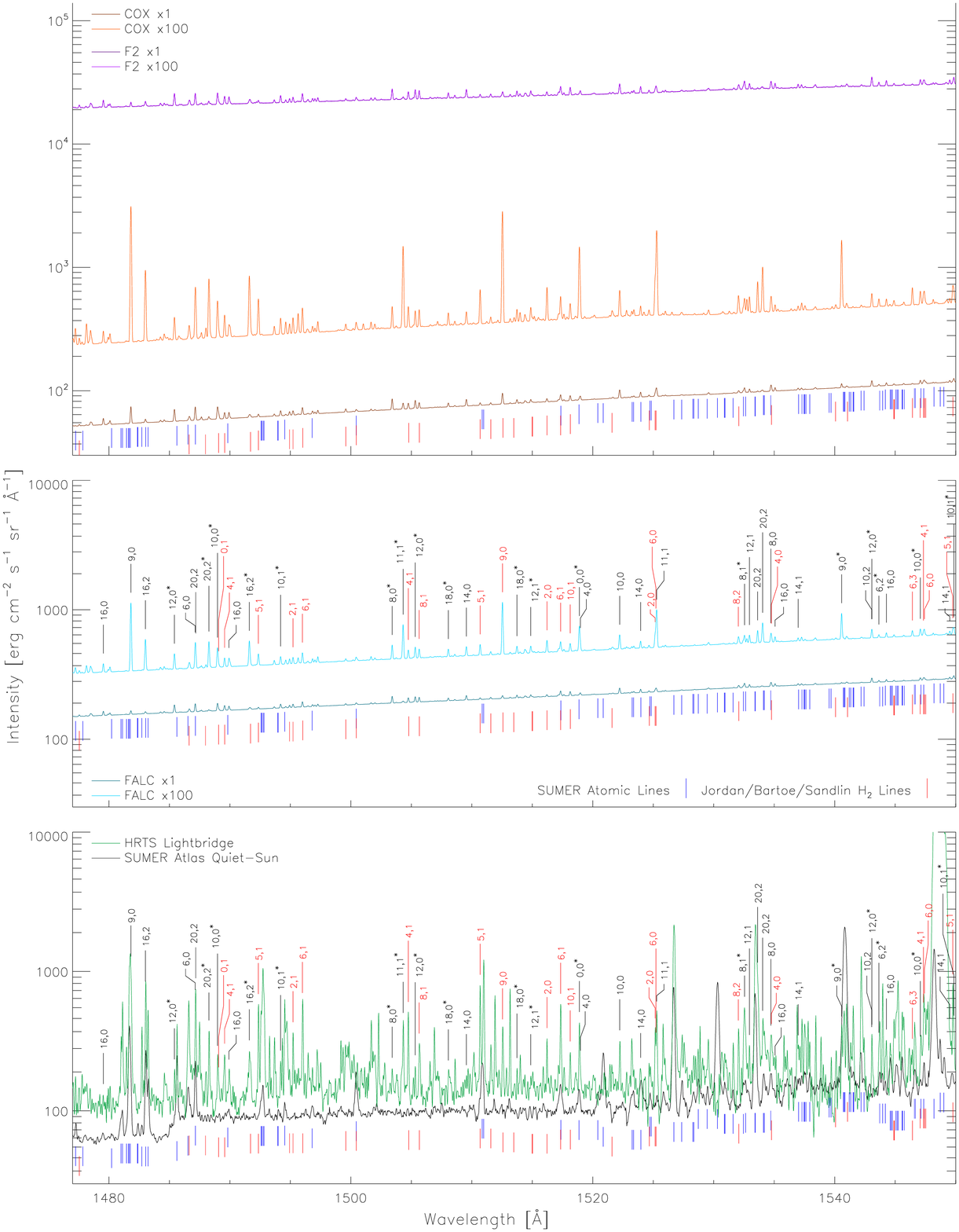}
	\end{center}
	\caption{{\it Continued.}}
\end{figure*}

\section{Results}
In this section we show the resulting UV spectrum of \hh, and look at the behavior of \hh\ lines as a function of the \re{down-going intensity} multiplier for the different models we have used.  We also dig deeper into the intermediate results of the \hh\ fluorescence calculations to determine where \hh\ fluorescence originates and how different radiation sources contribute to the population of the upper energy levels.  We demonstrate the agreement between the synthetic results and observed spectra by doing a detailed comparison for the strongest lines predicted by our calculations.  Finally, we look in depth at the excitation sources for three particular families of lines, where each family of lines shares common upper level, to see if radiation from the \hh\ lines and from the excitation sources is consistent in the observations and synthesis.

\subsection{Synthetic Spectra\label{sec:spectra}}
The top two panels in Figure \ref{fig:obs} demonstrate the diversity of possible synthetic \hh\ spectra over the wavelength range 1205-1550 \AA\ by showing the computed intensities resulting from the FALC, COX, and F2 atmospheres for multipliers of 1 and 100 applied to the quiet-Sun values of the down-going radiation.  The synthetic spectra are composed of the continuum and line intensities calculated with RH, including the down-going radiation we have added through the top of the atmosphere and the calculated intensities for the fluoresced \hh\ lines.  The positions of all atomic lines calculated by RH or added as down-going radiation are marked in the top panel so that they can be easily distinguished from the \hh\ lines.  The lines of \ion{H}{1} Lyman $\alpha$ and \ion{C}{2} show line profiles with a component calculated internally by RH, with broad damping wings and a narrow line core which shows self-absorption, and a Gaussian component from our added radiation spectrum.  The continuum calculated by RH generally increases smoothly in intensity with wavelength, with small jumps at 1240, 1345, and 1443 \AA, due to the photoionization threshold of the metastable neutrals $^1D_2$ C, $^1D_2$ S, and $^1S_0$ C, respectively.  Whether these continuum edges appear in absorption or emission is very sensitive to temperature gradients, and these jumps appear larger or smaller in the different atmospheric models.  See \citet{socas15} for a good illustration of the UV continuum opacity and its sources.

The remaining spectral features in the top two panels, consisting of narrow peaks, are \emph{all} due to \hh.  The \hh\ lines shown were computed using the escape probability method (Equation \ref{eqn5}).  The stronger \hh\ lines in this wavelength range have optical depths of order unity.  Thus, optically thin calculations (not shown) produce \hh\ lines in this region that are between 1 and 2 times brighter than those shown, as \hh\ photons are mildly trapped at these wavelengths leading to increased photon pumping and escape in transitions sharing a common upper level with lines above the \ion{Si}{1} edge at 1520 \AA.  The effects of photon trapping in the \hh\ transitions, less than factors of two, are therefore relatively minor.  This behavior is characteristic of all calculations done here.  We assume a 5 km s$^{-1}$ Doppler width for the \hh\ lines, which is typical of neutrals in the low chromosphere.

The bottom panel of Figure \ref{fig:obs} shows the SUMER quiet-Sun atlas spectrum of \citet{curdt01}, and the HRTS light bridge spectrum of \cite{brekke91}.  This light bridge is the same one analyzed by \citet{bartoe79} observed during the second flight of HRTS.  For comparison, the synthetic spectra have been convolved with a Gaussian profile of width 0.04 \AA\ which is equivalent to the average spectral resolution of the SUMER spectrograph operating in first order \citep{curdt01}.

To provide a representative sample of \hh\ lines at wavelengths between 1205 and 1550 \AA\ for a variety of solar conditions, we compiled a composite list of the 100 strongest lines from all calculations, which comprises \re{170} lines.  We add three lines to this list that did not meet the ``top 100'' selection criteria because they are new identifications particularly pertinent to that discussion, and thus, the line list includes \re{173} lines.  The line parameters for this list (from \citet{abgrall93a,abgrall93b}) and the line intensities and wavelengths of the major pumps predicted by our calculations are presented in Table \ref{tbl:h2lines}.  The locations of these strongest \hh\ lines are marked above the spectra in the bottom two panels of Figure \ref{fig:obs} and labeled with the $\mathcal{J},v$ of the upper level.  The 67 lines in this list that were previously identified by \citet{jordan77,jordan78} and \citet{bartoe79} are colored red, while lines not previously identified are in black, and the 54 new lines identified by this work have labels which are marked with an asterisk.  To help with comparison, and show where blends may be an issue, we also show previously identified atomic lines from \citet{curdt01} (blue tic marks) and all \hh\ lines from \citet{jordan78}, \citet{bartoe79}, and \citet{sandlin86} (red tic marks) underneath the spectra in all panels of the figures.  The line parameters of all \re{27,981} \hh\ lines, calculated intensities, and synthetic spectra, for all of the models and intensity multipliers listed here, are available as ASCII tables in the on-line material.

The distribution of computed line emission from \hh\ in the synthetic spectra is in reasonable agreement with observed spectra, including the previously identified \hh\ lines, and many lines that have not been previously identified.  Of these new lines, 16 are close matches to unidentified features reported by \citet{sandlin86} and are labeled with a $u$ in Table \ref{tbl:h2lines}.  Two `new' identifications are re-assignments of previously identified solar \hh\ lines, specifically, our calculations indicate that the \re{1443.465} and \re{1547.081} \AA\ lines from the \upjv{8}{0} and \upjv{10}{0} upper levels, respectively, are responsible for the corresponding features in solar spectra, rather than the \upjv{2}{5} and \upjv{2}{3} levels previously thought.

We do not label new identifications where the observed and predicted features have blend issues with other species or \hh, with three notable exceptions: the \re{1283.143} \AA\ line from the \upjv{16}{2} upper level is included, as the calculations and observations presented here demonstrate that it is an equal partner to the previously identified line at \re{1283.110} \AA\ from \upjv{2}{0}; and the \re{1532.546} and \re{1543.652} \AA\ lines, as possible blends with \ion{P}{2} are precluded by the absence of enough emission for the other lines of the \ion{P}{2} $3s^23p^2$ $^3P$ - $3s3p^3$ $^3D^o$ multiplet.  A detailed comparison of many of the strong lines in the synthetic spectra and the atlas spectra is given in Section~\ref{comparison}.

While the method of spectral synthesis presented here is clearly a valuable tool for identification of solar UV \hh\ lines, it is also not the only tool required when blends with other species are present.  Therefore, the identifications here are intended to be definitive but not exhaustive.
Detailed identification and analysis of the \hh\ lines in high resolution IRIS spectra will be carried out in a subsequent paper.

\begin{figure}
	\begin{center}
		\includegraphics[width=3.5in]{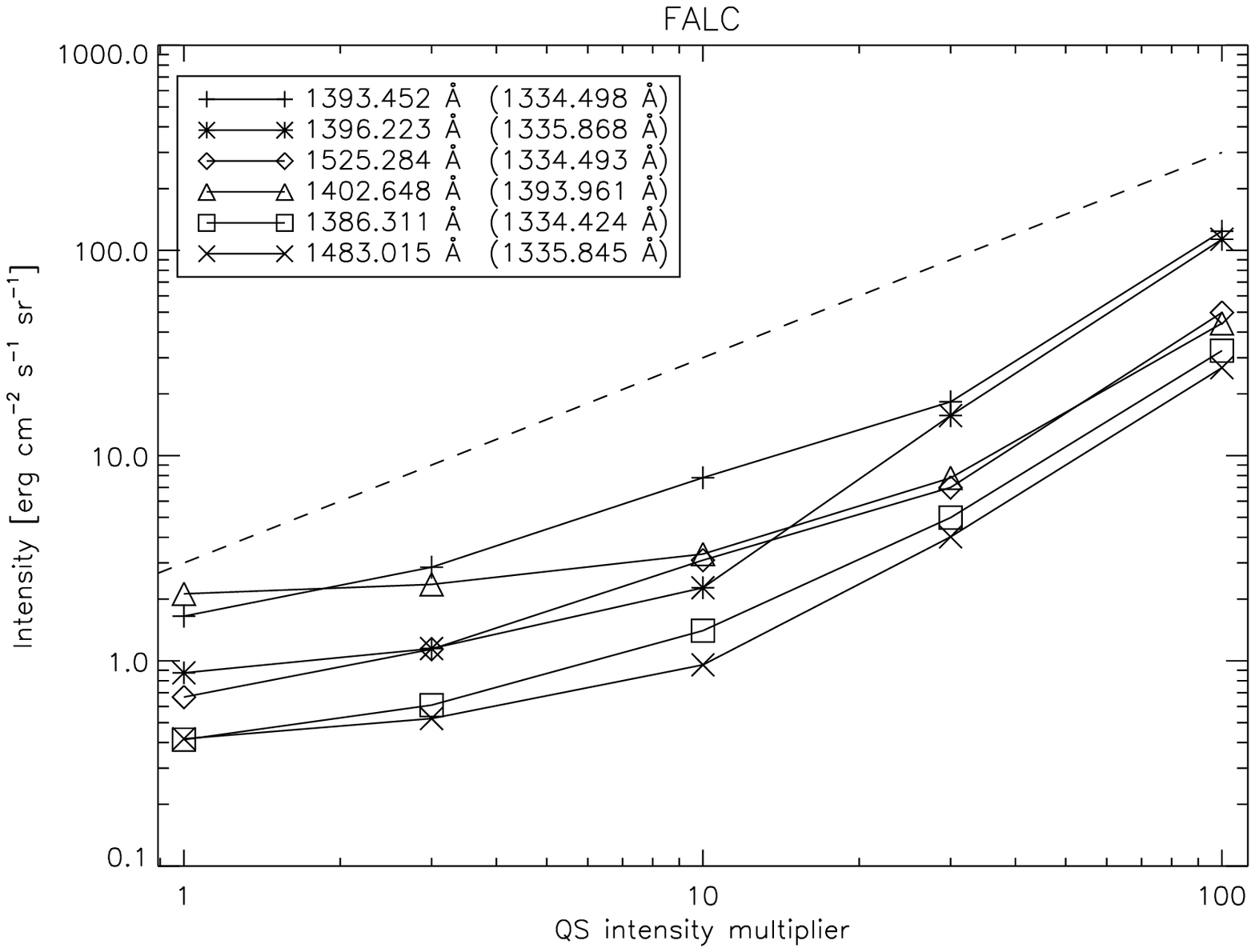}
		\includegraphics[width=3.5in]{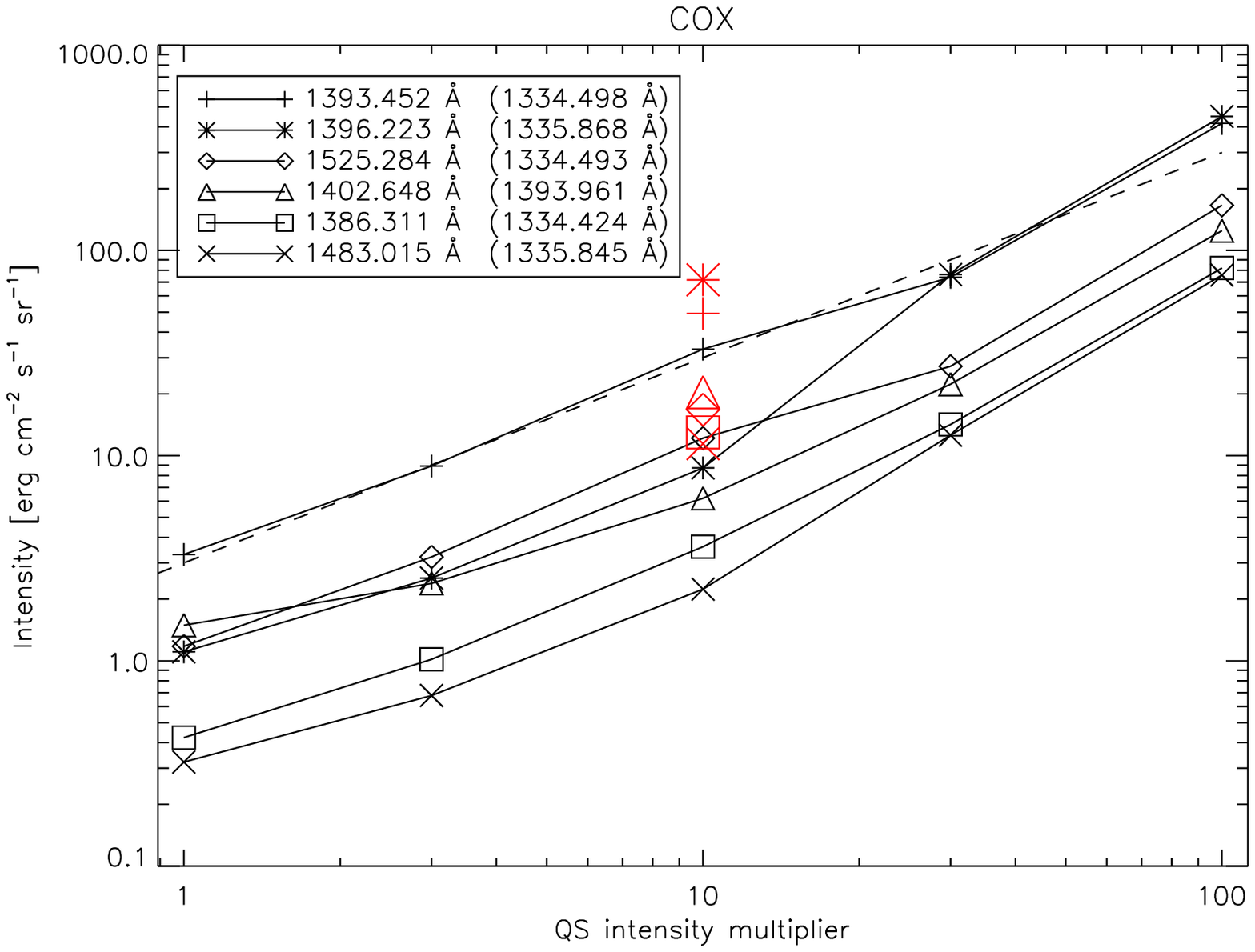}
		\includegraphics[width=3.5in]{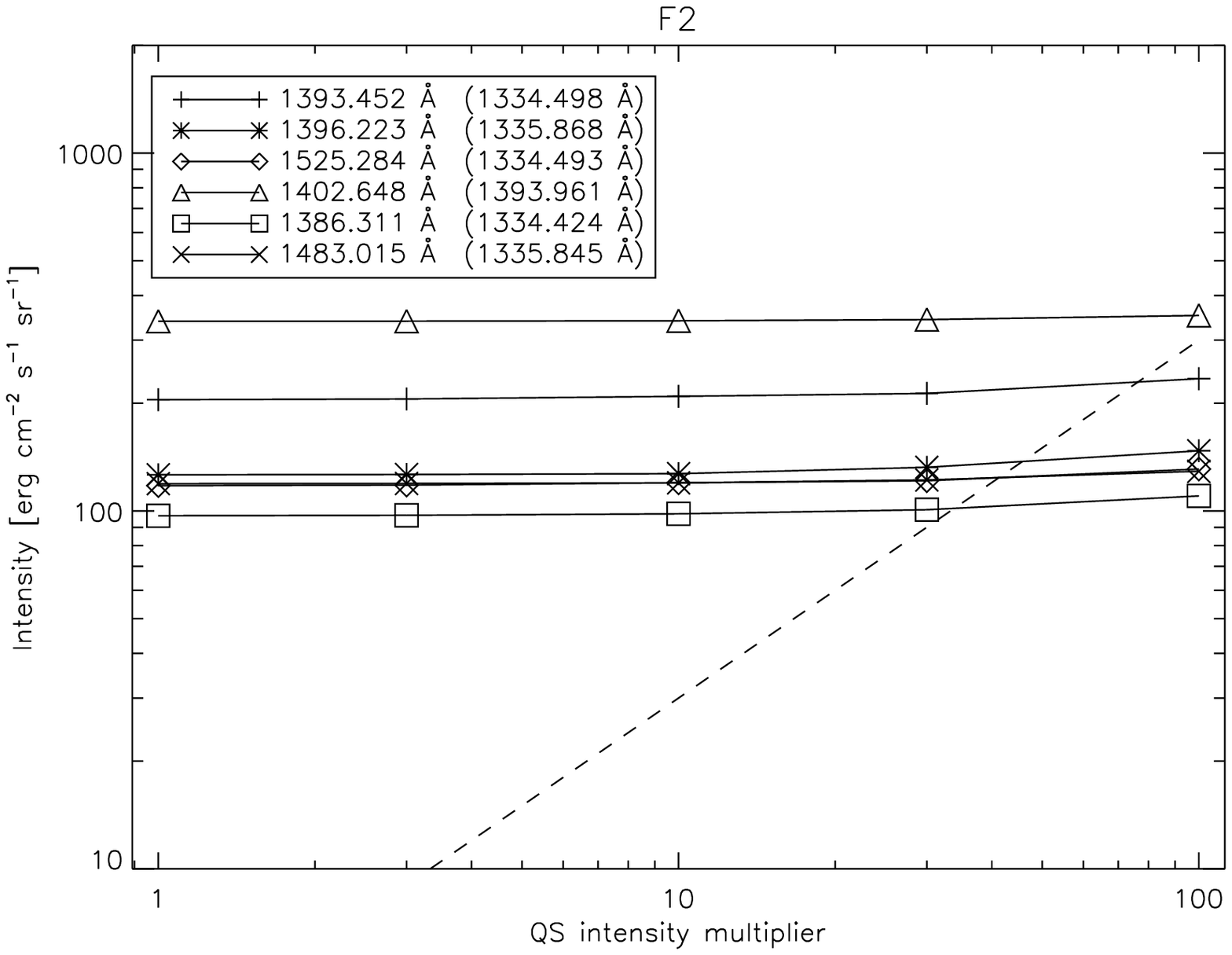}
	\end{center}
	\caption{Wavelength-integrated line intensities of the brightest \hh\ lines from 6 distinct upper levels are shown for FALC, COX, and F2 as a function of the factor by which the quiet-Sun intensities were multiplied in the downward irradiation of the chromosphere in lines of \ion{H}{1}, \ion{C}{2}, \ion{C}{3}, \ion{Si}{4}, \ion{O}{4}, and \ion{O}{5}.  The legend gives the wavelength of the strongest emission line for each upper level, and the wavelength of the most significant pump is given in parentheses.  The dashed line shows y = 3x.  The large red symbols show calculations using the approximate formula (Equation \ref{eqn10}) for the multiplier value of 10.  This approximation yields intensities of \hh\ lines which are linear in the multiplier on the abscissa.}
	\label{fig:summary}
\end{figure}

The three atmospheric models show very different radiation properties and the response of \hh\ to the radiation field is correspondingly different.  The large continuum level in the F2 atmosphere arises from the much higher temperature of the low chromosphere, and even large multiples of the quiet-Sun radiation scarcely have an impact on the thermal and radiative state of the atmosphere.  Meanwhile, the continuum of the COX atmosphere jumps up nearly two orders of magnitude with the added radiation.  The response of the \hh\ lines to these different radiation conditions can be seen in Figure \ref{fig:summary}, where we show the integrated \hh\ line intensities for six of the strongest lines between 1205 and 1550 \AA\ from different upper levels as a function of the quiet-Sun intensity multiplier for the three models, FALC (top), COX (middle), and F2 (bottom).  The \hh\ lines from different upper levels show different relative strengths based on whether they are pumped by a strong continuum (as for the F2 atmosphere and multipliers thereof) or strong line emission (as for COX and higher multipliers of the FALC model).  Similar to the continuum response noted above, added radiation to the F2 model scarcely impacts the \hh\ intensities, while for the case of the COX model they essentially scale linearly with the multiplier.  The behavior of \hh\ lines in the FALC atmosphere appears more continuum dominated for low multipliers and more line dominated for high multipliers.  It is also worth noting, that despite the low \hh\ densities present in the very warm F2 atmosphere, the \hh\ lines are much stronger than those produced by the other atmospheres.  This is due to the massive amount of continuum radiation produced by this model and the lowered opacity of the high temperature atmosphere.

The large red symbols in the middle panel of Figure \ref{fig:summary} show the line intensities computed using the zero-dimensional approximation of Equation (\ref{eqn10}) for downward-directed irradiances that are ten times the quiet-Sun value.  The zero-dimensional calculation only considers line radiation, therefore we only compare it with the COX model, which has a very low continuum level.  We used conditions at a height of $z_1 = 650$ km in the COX model, $T = 4220$ K, $n_{H_2} = 5 \times 10^{10}$ and $n_H = 46.7 \times 10^{14}$ cm$^{-3}$.  All intensities are within a factor of 10 of the more detailed calculations, but better than that for most cases.  These differences are not surprising given the crude approximation of the integrals and very crude treatment of radiative transfer in this model.  The ``0D'' approximation tends to over-estimate the intensities in most lines by a factor of several.  Thus, while the detailed calculations are to be preferred, the ``0D'' approximation serves to demonstrate the essential physics of the generation of \hh\ emission from a stratified atmosphere.

\subsection{Line and Population Contributions}
\begin{figure}
	\begin{center}
		\includegraphics[width=3.5in]{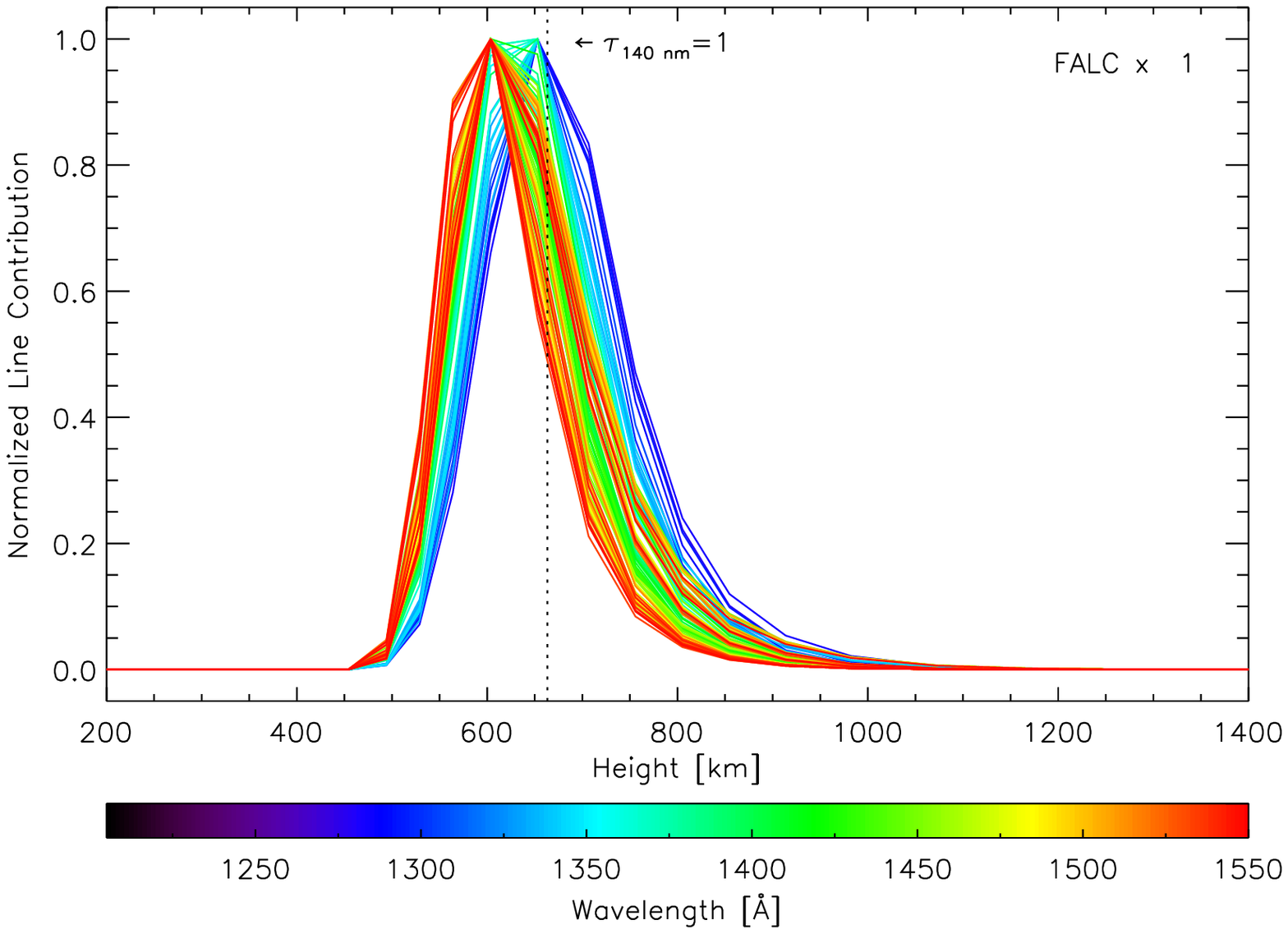}
	\end{center}
	\caption{The normalized line contribution as a function of height for the 100 brightest \hh\ lines between 1205 and 1550 \AA\ from the FALC$\times1$ calculation.  In general the lines originate in a 150 km thick region centered at 650 km.  Red lines originate slightly deeper than blue lines in agreement with the wavelength dependence of opacity, but some lines originate higher due to their excitation source.}
	\label{fig:contrib1}
\end{figure}

The calculations give us the opportunity to investigate at what height \hh\ fluorescence occurs in the model atmospheres and which radiation sources contribute the most to pumping the upper levels.  To demonstrate where the \hh\ fluorescent emission originates,  in Figure \ref{fig:contrib1} we show the normalized contribution to the \hh\ line intensity from the FALC model for down-going radiation $\times1$ as a function of height for the 100 strongest \hh\ lines between 1205 and 1550 \AA.  Coloration from violet to red indicates the wavelength of the line from shorter to longer wavelength respectively.  The peak contribution for FALC comes from a region tightly centered between 550 and 600 km above the photosphere near the height of $\tau_{140\rm nm}=1$.   As the chromosphere becomes less opaque at longer wavelengths, longer wavelength \hh\ lines originate from deeper in the atmosphere and shorter wavelength lines originate higher.  Lines pumped by exciters at significantly different wavelengths contribute to scatter in the distribution.

\begin{figure}
	\begin{center}
		\includegraphics[width=3.5in]{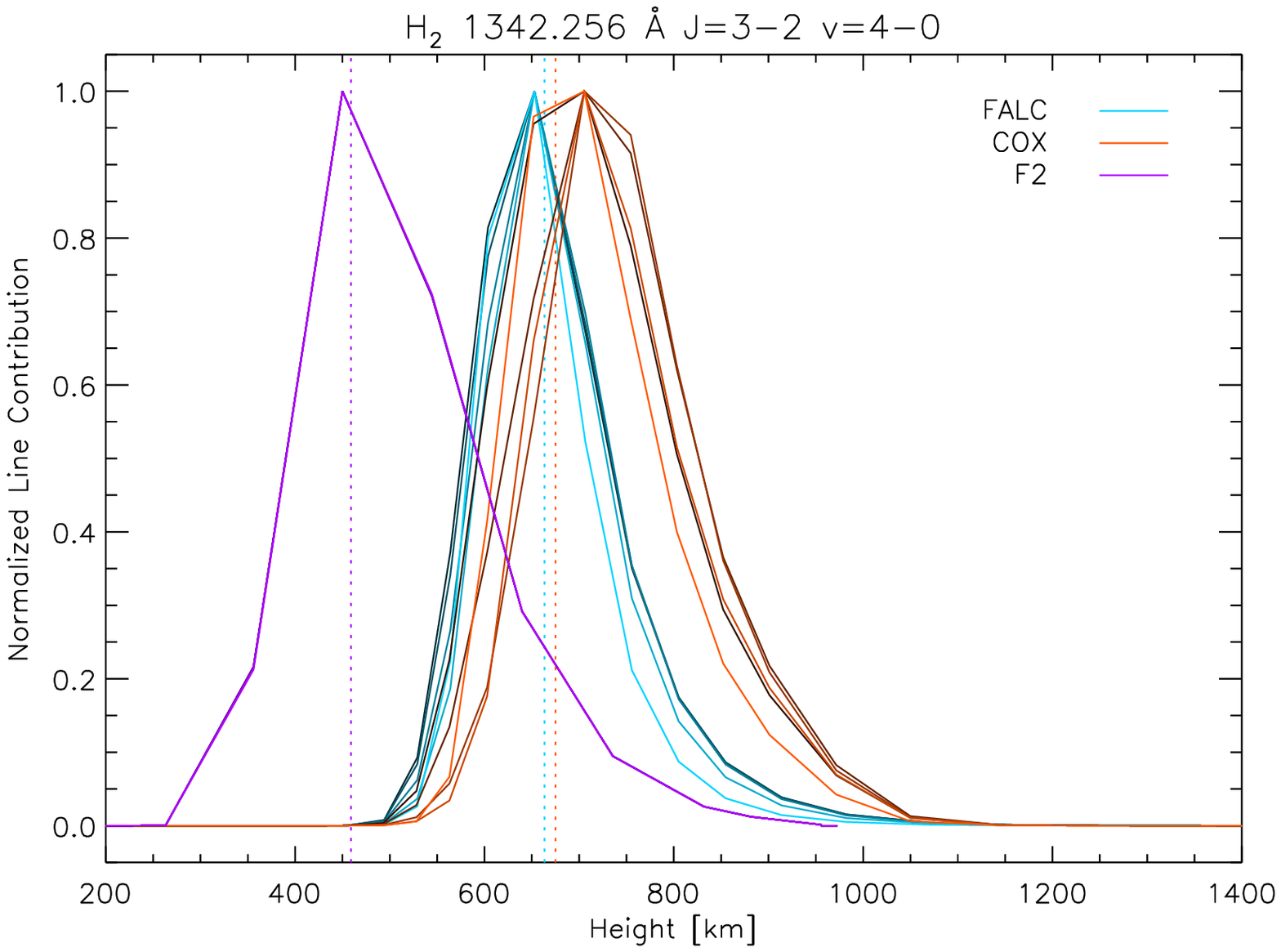}
	\end{center}
	\caption{The normalized line contribution function for the \re{1342.256} \AA\ \hh\ line shown for all atmospheric models (color hue) and radiation multipliers (color intensity).  The peak of the contribution function occurs near the height of $\tau=1$ in the ultraviolet.}
	\label{fig:contrib2}
\end{figure}

Figure \ref{fig:contrib2} shows the intensity contribution function of just one line, the 1342.256 \AA\ line belonging to the \level{2}{0}{u} upper level, for each of the different models as a function of height.  The three models are shown with different hues and each multiplier of that model is indicated by a different saturation of that hue, where the lightest colors indicate the highest multipliers.  The peak line contribution comes from a height near $\tau_{140 nm}=1$, shown by the dotted lines, at heights of approximately 650, 700, and 450 km above $\tau_{500{\rm nm}}=1$ in the photosphere for the FALC, COX, and F2 models respectively, and changes only slightly with multiplier.

To further investigate the properties of \hh\ fluorescence we focus on the six upper levels that produce some of the strongest \hh\ lines of the Lyman band and the most emission between 1205 and 1550 \AA , specifically, the levels \level{0,2,9}{0}{u}; \level{11}{1}{u}; and \level{16,20}{2}{u}, shown previously in Figure \ref{fig:summary}.  For large values of the radiation multiplier these six levels produce over half the \hh\ emission in this wavelength range for the FALC and COX atmospheres.

\begin{figure*}
	\begin{center}
		\includegraphics[width=7in]{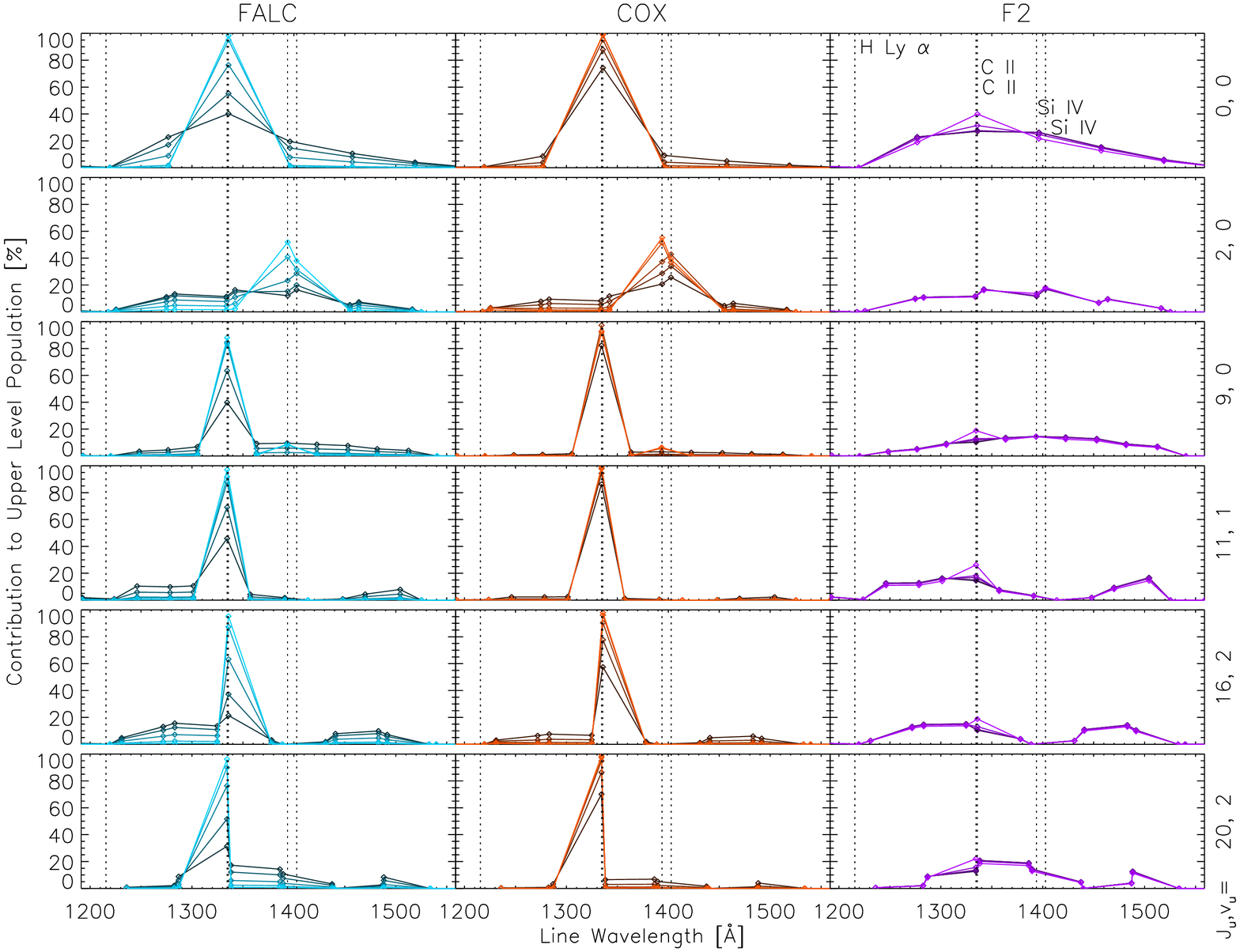}
	\end{center}
	\caption{The contribution to the upper level populations for lines between 1205 and 1550 \AA\ with the shared upper levels \level{0,2,9}{0}{u}; \level{11}{1}{u}; and \level{16,20}{2}{u}, integrated over height for the case of different atmospheric models ({\it left}: FALC, {\it middle}: COX, {\it right}: F2) and radiation multipliers (color intensity).  The wavelengths of the brightest atomic lines of the transition region and chromosphere are marked by the vertical dotted lines and labeled in the lower left panel.  In continuum dominated cases (F2 and low multipliers of FALC) many \hh\ lines contribute to the upper level population, while in line dominated cases (COX and high multipliers of FALC) one or two \hh\ lines are pumped primarily by the \ion{C}{2} lines (\level{0,9}{0}{u}; \level{11}{1}{u}; and \level{16,20}{2}{u}) or by the \ion{Si}{4} lines (\level{2}{0}{u}).}
	\label{fig:pumps}
\end{figure*}

The source of pumping radiation is a crucial piece of information in determining the behavior of \hh{} fluorescence.  As noted in Section 4.1, continuum sources and line sources can change the relative strength of \hh\ lines that result from different upper levels.  For these six upper levels, Figure \ref{fig:pumps} shows the fractional contribution to the upper level population by each transition for each model and each multiplier of that model using the same color scheme as the previous figure.  In cases with strong continuum, as for the entire set of F2 models (right column), the continuum contribution accounts for more than 50\% of the \hh\ upper level population for each of these three upper levels.  Although \hh{} transitions from the \level{0}{0}{u} and \level{2}{0}{u} upper levels fall near the \ion{H}{1} Ly $\alpha$ line core, these lines are not pumped significantly.  The large opacity in Ly $\alpha$ prevents illumination of \hh{} and the escape of \hh{} photons.  The \ion{C}{2} line 1335.7 \AA\ is the single most important pump for the \level{0}{0}{u} and \level{16}{2}{u} upper levels, while the \ion{C}{2} line at 1334.5 \AA\ is the most important for the \level{20}{2}{u}; \level{9}{0}{u}; and \level{11}{1}{u} upper levels.  For these levels, the \ion{C}{2} lines are significantly more important than any other sources except for cases where there is a very strong continuum.  Both lines of \ion{Si}{4} are important pumps for the \level{2}{0}{u} upper level.  It is important to consider both \ion{Si}{4} lines, and possibly \ion{C}{2} when it becomes broad, when evaluating the radiation sources for this family of \hh\ lines.

\subsection{Detailed Comparison with Atlas Spectra}\label{comparison}
\begin{figure*}
	\begin{center}
		\includegraphics[height=3.75in]{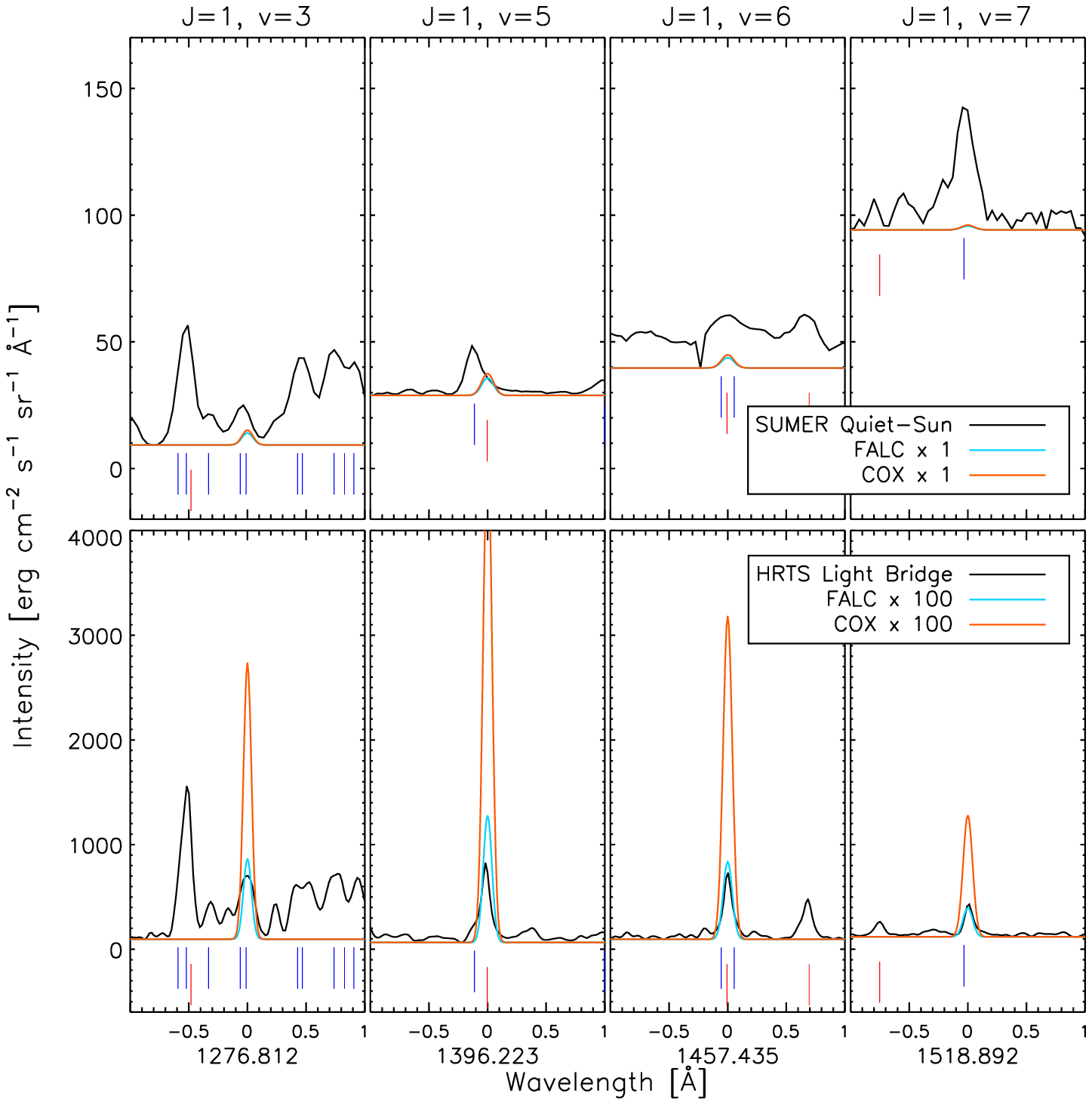}
	\end{center}
	\caption{Spectra from regions around selected \hh{} lines from the \level{0}{0}{u} upper level.  The top panels show the SUMER quiet-Sun atlas spectrum (black) alongside the \hh\ line profile predicted by FALC (blue) and COX (red) radiation $\times$1 cases, which nearly overlay one another.  The bottom panels show the HRTS light bridge atlas spectrum (black) alongside the FALC (blue) and COX (red) radiation $\times$100 cases.  As in Figure \ref{fig:obs} the blue and red tic marks below the spectra indicate the wavelengths of previously identified atomic and \hh\ lines respectively.}
	\label{fig:obszoom1}
\end{figure*}

\begin{figure*}
	\begin{center}
		\includegraphics[height=3.75in]{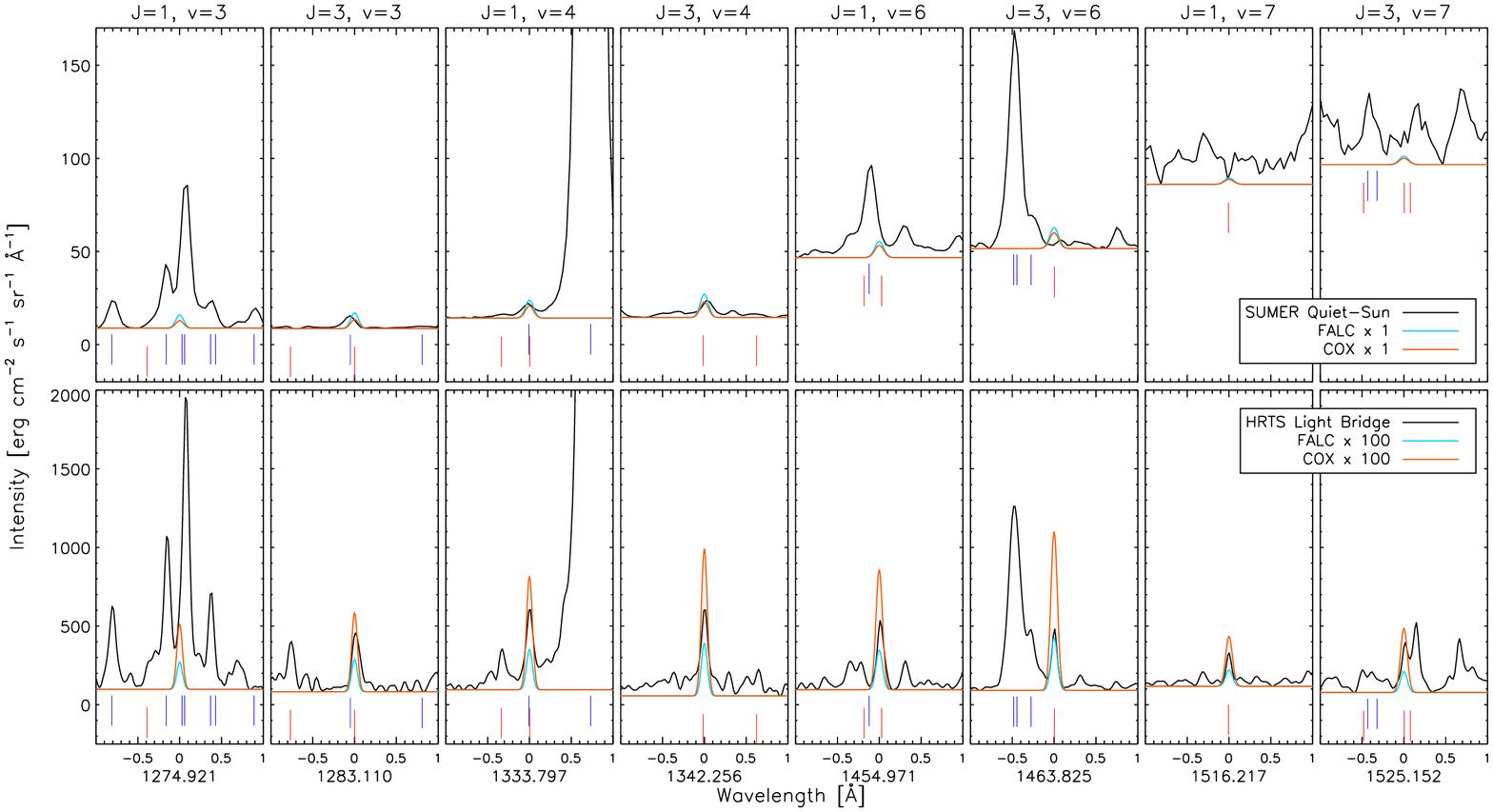}
	\end{center}
	\caption{Spectra from regions around selected \hh{} lines from the \level{2}{0}{u} upper level.  The top panels show the SUMER quiet-Sun atlas spectrum (black) alongside the \hh\ line profile predicted by FALC (blue) and COX (red) radiation $\times$1 cases, which nearly overlay one another.  The bottom panels show the HRTS light bridge atlas spectrum (black) alongside the FALC (blue) and COX (red) radiation $\times$100 cases.  As in Figure \ref{fig:obs} the blue and red tic marks below the spectra indicate the wavelengths of previously identified atomic and \hh\ lines respectively.}
	\label{fig:obszoom2}
\end{figure*}

\begin{figure*}
	\begin{center}
		\includegraphics[height=3.75in]{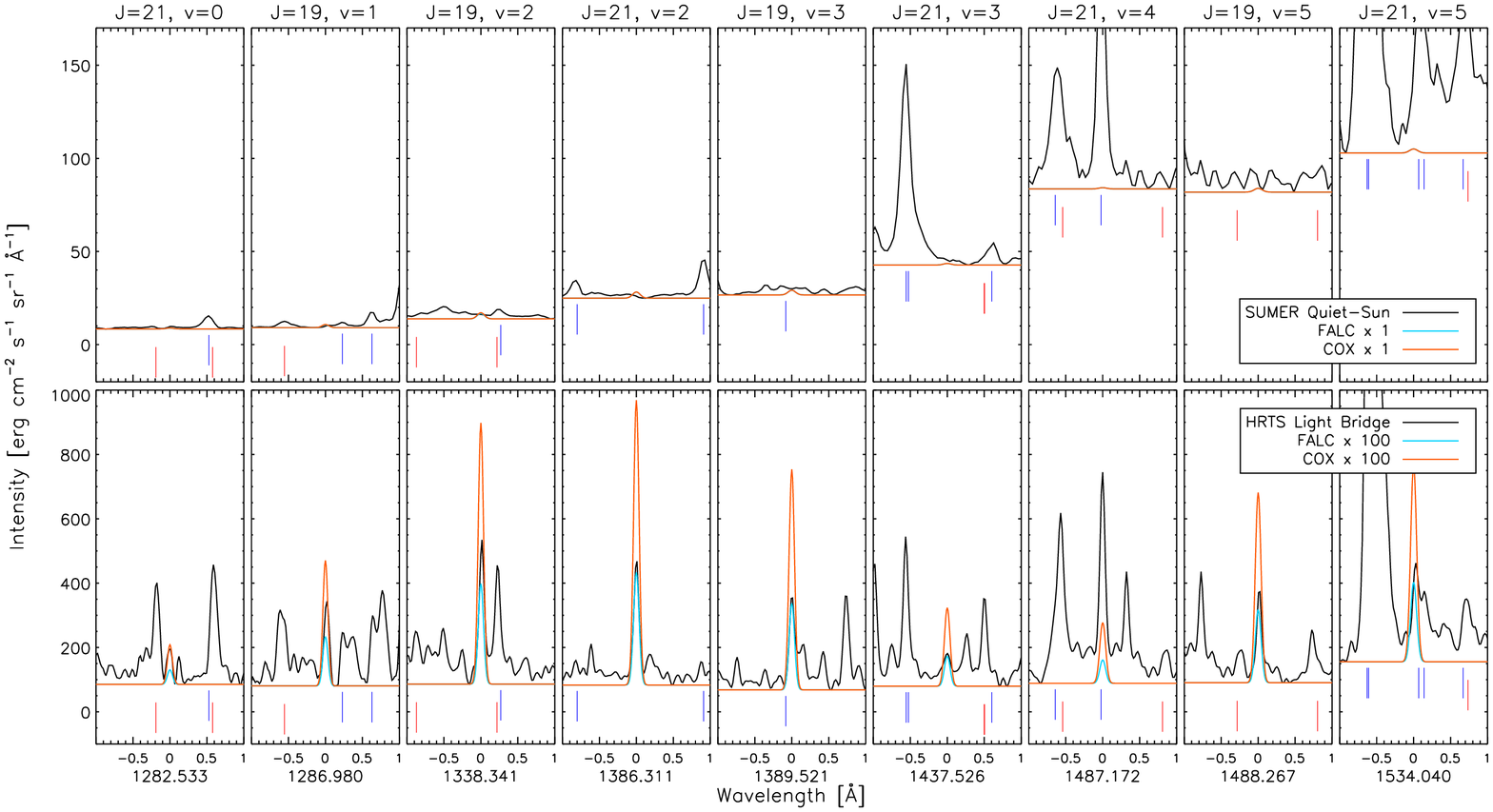}
	\end{center}
	\caption{Spectra from regions around selected \hh{} lines from the \level{20}{2}{u} upper level.  The top panels show the SUMER quiet-Sun atlas spectrum (black) alongside the \hh\ line profile predicted by FALC (blue) and COX (red) radiation $\times$1 cases, which nearly overlay one another.  The bottom panels show the HRTS light bridge atlas spectrum (black) alongside the FALC (blue) and COX (red) radiation $\times$100 cases.  As in Figure \ref{fig:obs} the blue and red tic marks below the spectra indicate the wavelengths of previously identified atomic and \hh\ lines respectively.}
	\label{fig:obszoom3}
\end{figure*}

Figures \ref{fig:obszoom1}, \ref{fig:obszoom2}, and \ref{fig:obszoom3} show a detailed view of the spectral regions where a selection of fluoresced \hh\ lines of the three upper levels, \level{0}{0}{u}; \level{2}{0}{u}; and \level{20}{2}{u} respectively, might be seen.  The top set of panels in each figure shows the SUMER quiet-Sun spectrum with the synthetic line profiles predicted by the FALC and COX models in the $\times1$ case.  The bottom panels show the HRTS light bridge spectrum and the FALC and COX models for the $\times100$ case which provide the best match to the \hh\ line strengths among the synthetic results.  The synthetic spectra shown in these figures only include the \hh\ radiation from the particular line without the background intensities from RH.  The minimum intensity level from the observation in each wavelength range has been added to the synthetic spectrum, and everything is plotted with a linear scaling so that the relative strengths can be easily compared.  As in Figure \ref{fig:obs}, the wavelengths of previously identified atomic and \hh\ lines are marked to resolve any possible blends (the blue and red tic marks respectively).

Lines from all three upper levels appear prominently in the HRTS light bridge spectrum, and a few \hh\ lines from the \level{0}{0}{u} and \level{2}{0}{u} can even be seen in the SUMER quiet-Sun spectrum (\re{1396.223} \AA\ from \level{0}{0}{u}, and \re{1333.797} and \re{1342.256} \AA\ from \level{2}{0}{u}).  The \re{1396.223} \AA\ line is blended with a \ion{S}{1} line at 1396.113 \AA, but can be clearly seen in the light bridge spectrum.  Knowing this we can attribute some of the skew seen in the quiet-Sun \ion{S}{1} profile to \hh.  The line at \re{1333.797} \AA\ might also be blended with another \ion{S}{1} line, but based on the presence and strength of the \re{1342.256} \AA\ line in the quiet-Sun spectrum, it seems likely that this is a mis-identification.  For other lines that cannot be cleanly identified, the predicted line strengths seem consistent with the noise level or the presence of strong atomic lines.  For lines with nearby or blended atomic lines, the presence of its comrades is an important piece of evidence as to its identity.

The nine lines from the \level{20}{2}{u} upper level in Figure \ref{fig:obszoom3} provide a convincing match to the observed HRTS spectrum.  While these lines were not previously identified by \citet{jordan77,jordan78} or \citet{bartoe79}, three of them are listed as unidentified features by \citet{sandlin86}.  All these works made use of the \hh\ line parameters from \citet{herzberg59}.  In the more recent work of \citet{abgrall93a}, the upper levels of the blended \re{1334.424} and \re{1334.426} \AA\ \hh\ lines are correctly identified as \upjv{20}{0} and \upjv{21}{3}.  Of the lines in Figure~\ref{fig:obszoom3}, the \re{1487.172} \AA\ line is blended with the stronger \hh\ \re{1487.136} \AA\ line from \upjv{6}{0}, and both the \re{1487.172} and \re{1534.040} \AA\ lines are possibly blended with \ion{Si}{1}, so neither is indicated here as a positive new identification.  Similarly, the \re{1533.628} line (not shown) is bended with \ion{Si}{1} and \ion{Si}{2}.  Thus, Figure \ref{fig:obszoom3} represents seven new line identifications for \upjv{20}{2}, including \re{1282.533} \AA , which did not meet the threshold for Table \ref{tbl:h2lines}.

\begin{figure*}
	\begin{center}
    	\includegraphics[height=4in]{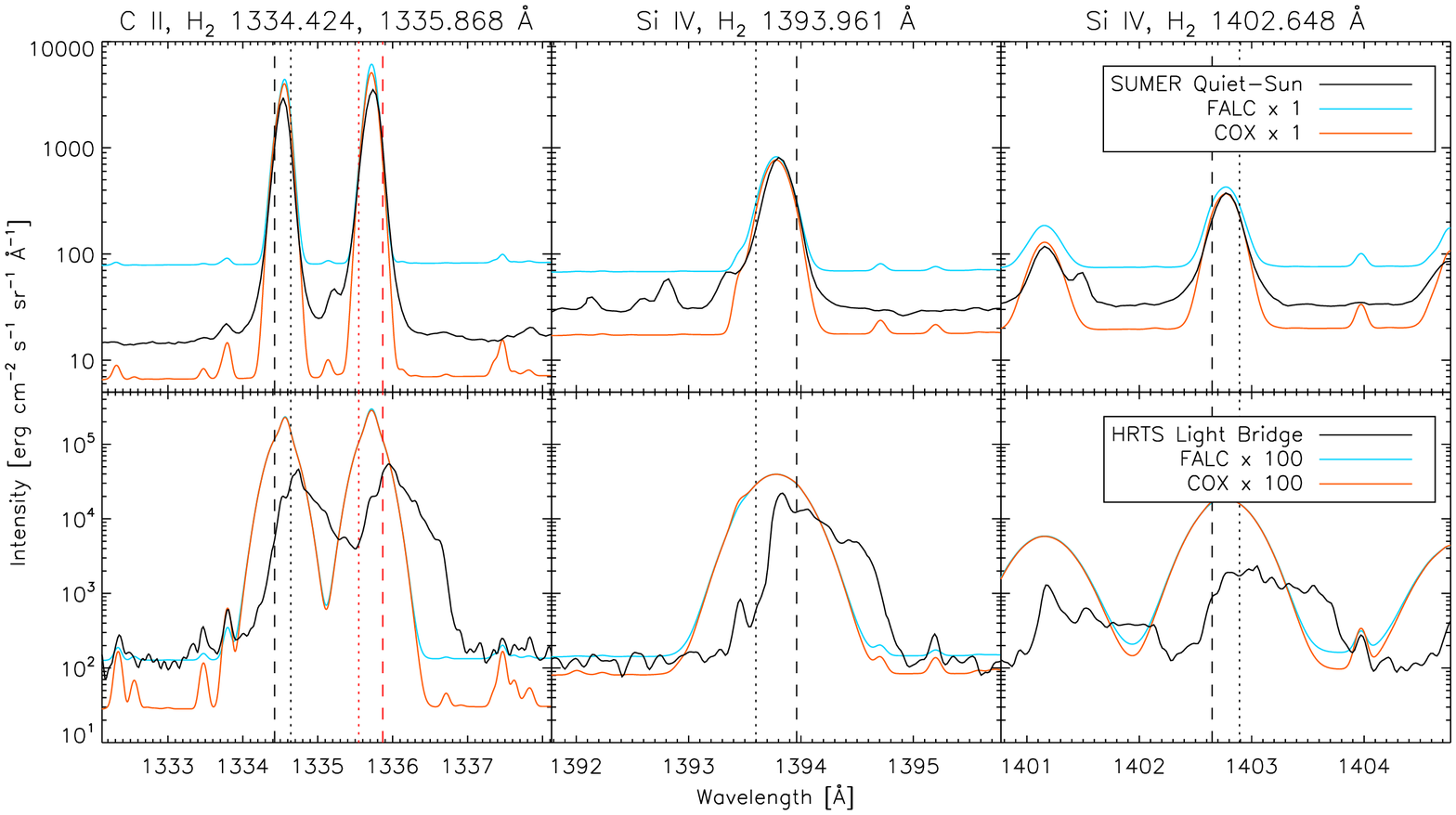}
        \caption{Emission near the excitation wavelength for the \hh\ upper levels \level{20}{2}{u} at 1334.431 \AA\ and \level{0}{0}{u} at 1335.865 \AA\ in the wings of the C II lines (left panel); and for the upper level \level{2}{0}{u} which gets significant contributions from both 1393.961 and 1402.648 \AA\ lines in the wings of both of the \ion{Si}{4} lines (right two panels).  Observed atlas spectra are shown in black, while the synthetic spectra are shown by the colored lines.  The SUMER quiet-Sun spectrum is compared to the FALC and COX $\times$ 1 multiplier (top panels), while the HRTS light bridge spectrum is compared to the FALC and COX $\times$ 100 multiplier for the down-going radiation (bottom panels).  The dashed vertical lines show the rest wavelength of the absorbing \hh\ transitions, but because \hh\ resides below the transition region, the velocity signature is reversed with respect to observations from above, the dotted line shows the \hh\ wavelength reversed with respect to the rest wavelength of the exciting transition region line.}
        \label{fig:obszoom4}
    \end{center}
\end{figure*}

To complete the picture we compare the intensity of the radiation that excites the \hh\ upper levels.  While we cannot sample the down-going radiation field at the height of \hh\ line formation in observations, as we can using the model atmospheres, the radiation exiting through the top of the atmosphere should be similar.  Figure \ref{fig:obszoom4} shows the intensities at the locations of the major pumped \hh{} lines, which are at \re{1335.868} \AA\ for the \level{0}{0}{u} upper level, at \re{1393.961} and \re{1402.646} \AA\ for the \level{2}{0}{u} upper level, and at \re{1334.424} \AA\ for the \level{20}{2}{u} upper level, in the atlas and synthetic spectra.  As in the previous two figures, the top set of panels show the SUMER quiet-Sun spectrum alongside the FALC and COX $\times1$ synthetic spectra, while the bottom panels show the HRTS light bridge spectrum with the FALC and COX $\times100$ spectra.  The dashed lines in each panel indicate the rest wavelength of the pumped \hh\ line.  Because \hh\ is situated in the low chromosphere, it will see line radiation from the transition region with a reversed velocity signature with respect to observations from above the transition region, assuming the line of sight corresponds with the local vertical direction.  Therefore we also show the conjugate wavelength of \hh\ reversed about the rest wavelength of the transition region line (dotted line).  The effect of the reversed velocity signature on \hh\ fluorescence can be significant when transition region lines show asymmetric profiles or large velocities.  The intensity values corresponding to the positions of the dotted lines in this plot are given in Table \ref{tbl:obspumps} for ease of reference.

\begin{deluxetable*}{rcrcccccccc}
\tablecaption{Primary Exciter Intensities for Selected Upper Levels \label{tbl:obspumps}}
\tablehead{& & & & & \multicolumn{6}{c}{Intensity [erg s$^{-1}$ cm$^{-2}$ str$^{-1}$ \AA$^{-1}$]} \\
\cmidrule(lr){6-11}
 & & & & & \multicolumn{2}{c}{Observed} & \multicolumn{2}{c}{FALC} & \multicolumn{2}{c}{COX} \\
\cmidrule(lr){6-7}\cmidrule(lr){8-9}\cmidrule(lr){10-11}
\colhead{$\mathcal{J}_u$} & \colhead{$v_u$} & \colhead{$\mathcal{J}_l$} & \colhead{$v_l$} & \colhead{$\lambda$ [\AA]} & \colhead{Quiet-Sun} & \colhead{Light Bridge} & \colhead{$\times1$} & \colhead{$\times100$} & \colhead{$\times1$} & \colhead{$\times100$} }
\startdata
0 & 0 & 1 & 4 & 1335.868 & 5.84e2 & 4.65e3 & 8.32e2 & 1.12e5 & 7.24e2 & 1.14e5 \\
2 & 0 & 1 & 5 & 1393.961 & 1.11e2 & 3.58e2 & 3.21e2 & 2.92e4 & 2.64e2 & 2.96e4 \\
2 & 0 & 3 & 5 & 1402.648 & 2.31e2 & 1.72e3 & 3.03e2 & 1.64e4 & 2.41e2 & 1.71e4 \\
20 & 2 & 21 & 1 & 1334.424 & 1.27e3 & 2.85e4 & 1.54e3 & 1.21e5 & 1.46e3 & 1.22e5
\enddata
\end{deluxetable*}

For the SUMER quiet-Sun spectrum, the transition region line radiation is well matched between the observed and synthetic spectra for the $\times$ 1 case for the FALC and COX models (by design), although the continuum radiation is not well matched by either model.  The hotter FALC atmosphere has a brighter continuum, and the cooler COX model has a dimmer continuum.  The \hh\ line intensities predicted by the FALC and COX $\times1$ models are small enough that only the brightest lines are able to be seen in the SUMER quiet-Sun spectrum.  The level of \hh\ emission predicted by our calculations seems consistent with observations of the real Sun for our quiet-Sun models.

The \hh\ line emission from the HRTS light bridge spectrum is very bright, and most of the \hh\ lines from the upper levels shown in Figures \ref{fig:obszoom1}, \ref{fig:obszoom2}, and \ref{fig:obszoom3} can be easily distinguished from the surrounding emission.  The line strengths are best matched by the FALC and COX $\times$ 100 cases, however, the transition region line radiation from \ion{C}{2} and \ion{Si}{4} in the light bridge atlas spectrum is considerably smaller than the emission in the synthetic spectra (see Figure \ref{fig:obszoom3} and Table \ref{tbl:obspumps}).  The intensities in the HRTS atlas spectra at the conjugate wavelength positions (reversed in terms of velocity) are 1-2 orders of magnitude smaller than in the synthetic spectra.  Although the continuum levels approximately match between the synthetic and observed spectra, we expect most of the pumping to come from line emission for the $\times$ 100 cases (refer back to Figure \ref{fig:pumps}).  Because it is red shifted, the level of transition region emission in the HRTS spectrum does not at first seem to be consistent with the large \hh\ line intensities that are observed, but there are a number of possible reasons for this, as discussed below.

\section{Discussion}
The agreement between the \hh\ intensities computed from the models and measured in the SUMER quiet-Sun spectrum is surprising.  While the reproduction of the relative \hh{} line strengths from a common upper level is to be expected from our calculations, the rough agreement between the \hh{} line strengths and the emitted transition region intensity, especially for the quiet-Sun, was not expected.  We had not anticipated that a typical quiet-Sun 1D model could generate \hh{} emission at about the level observed for two reasons.  Firstly, strong vibration-rotation lines of the CO molecule observed in the infrared imply that some regions of the thermal structure must be very cool \citep{solanki94}. Indeed this is why the COX model was first constructed.  Secondly, much work has emphasized the need for dynamical models to replace the 1D hydrostatic models which were developed in the 1960s-1980s.  The most relevant work to our discussion are the 1D calculations of \citet{carlsson95} and the 2D calculations of \citet{leenaarts11}.

Why do our calculations appear to be so successful?  We cannot answer this without performing more complex calculations, but we can nevertheless examine some of our assumptions in the light of the articles by \citet{carlsson95} and \citet{leenaarts11}.  The essential property of \hh{} molecules is that they are, in any model, only abundant below some 800 km above the solar photosphere.  As discussed earlier, this property is set by the strong gravitational stratification of the entire atmosphere, with some modification in the presence of a chromospheric average temperature rise.  Physical properties of the atmospheres above 700 km are, to a first approximation, generally irrelevant except for the irradiation.  Thus, although the atmosphere is very dynamic (as seen in simulations and indirectly through observed spectral signatures), the most extreme variations in temperature exhibited in the calculations of \citet{carlsson95} are only $\approx \pm 10\%$ below 700 km, dropping exponentially with height.  It seems that the \hh{} spectrum originates from deep enough in the atmosphere that the effects of dynamical molecular formation, radiative transfer and radiative excitation are captured in a 1D model with the upper boundary radiation specified at the transition region. 

The spectrum of \hh\ in the average quiet-Sun can be reproduced using standard atmospheric models without the need to invoke cool dense pockets of plasma (as in \citet{ayres86} or \citet{avrett95}), or regions where UV photons from above ``burn'' holes in the opacity through photo-ionization to allow penetration into cool regions with higher \hh\ abundance.  If we had found that the stratified models could not excite sufficient \hh\ we would have had to invoke a way to get the exciting UV photons to the places where \hh\ is abundant, but such efforts are not required to explain how some \hh\ emission is formed.  If our picture is correct, all that is needed is an overlying source of photons and stratification will take care of the fluorescent spectrum of \hh.

This presents us with some questions.  If we accept that bright UV emission and stratification are all that is needed for fluorescent \hh\ emission, then we should find it occurring everywhere beneath bright UV emission when we have sufficient instrumental sensitivity.  Figure \ref{fig:atmos} illustrates the delicate balance, in 1D, between the number of \hh\ molecules and the penetration of UV photons emitted from higher, tenuous plasma.  The same situation also occurs in a thermally inhomogeneous 3D atmosphere, because wherever \hh\ molecules exist, neutral atoms are also abundant, and the same balance between penetration of UV radiation into regions of abundant \hh\ molecules occurs.  In 3D however, UV radiation can penetrate cool dense pockets containing an abundance of \hh\ molecules when the exciting UV photons traverse distances $\lesssim \mathcal{H}$ in the atmosphere (The scale height $\mathcal{H}$ of the ambient atmosphere is the relevant distance to use, because the vertical optical depth $\tau_\kappa$ is $\approx \kappa\mathcal{H}$, and for a given opacity $\kappa$, $\tau_\kappa \gtrsim 1$ defines where the UV intensities begin to be strongly attenuated).

The fact that \hh{} lines \emph{can} be seen in the SUMER atlas of the quiet-Sun, as well as the light bridge spectrum from HRTS, means that our treatment of \hh{} fluorescence with a warm 1D, stratified atmospheric model is reasonable, at least for the families of lines which we have chosen to look at in detail.  In our calculations, we have included the brightest lines that we think are important.  However, other sources of pumping radiation may be \re{significant.  These sources could change based on the local conditions}, and the radiation may not scale isotropically with wavelength as we have assumed.  The transition region radiation which excites \hh{} emission is generated by energy transport from two reservoirs of energy - the chromosphere and corona - there is little mass in the transition region, and therefore it varies in time and space dramatically compared with both reservoirs.

The other major factor of importance to \hh{} emission is the continuum opacity which attenuates UV photon fluxes to deeper layers.  Radiation \re{longward} of 1520 \AA\ was explicitly excluded as a source of pumping radiation for the \hh{} lines.  However this region contains many \hh{} transitions.  Being a continuous source of photons, the continuum radiation at these wavelengths is unlikely to reproduce the selective excitation of the \hh{} line emission.  Also, a more careful discussion of continuum emission would require careful treatment of UV line blanketing as well as continua in non-LTE conditions, something beyond the scope of this paper.

While our calculations approximately predict the quiet-Sun \hh\ line intensities, large \hh\ line strengths seen in the HRTS light bridge spectrum seem inconsistent with the transitions region radiation observed.  Based on the observed \hh\ line intensities, 1-2 orders of magnitude more intense transition region line radiation is needed.  We do not have a definitive answer why this is, but speculate further on a few possibilities here.  The light bridge spectrum of \citet{brekke91} comes from the sunspot observed during second flight of HRTS on 13 February 1978.  This sunspot was the leading polarity in a $\beta\gamma$ sunspot group according to the Mt. Wilson sunspot drawing (\url{ftp://howard.astro.ucla.edu/pub/obs/drawings/1978/dr780215.jpg}) made two days later.  This region produced an X1 flare observed by the {\it SOLRAD 11} satellites around 3 UT prior to the HRTS rocket flight at 17:15 UT.  \citet{horan83} note that this flare had an extremely slow rise and decay phase, together lasting longer than 6 hours in certain wavebands, with several smaller flares superposed within the 9 hour time period they show.  This flare was also associated with a solar proton event (\url{https://umbra.nascom.nasa.gov/SEP/}) originating in the active region observed by HRTS.

We speculate that this ``light bridge spectrum'' might actually be the result of ongoing flare activity in this region following the large flare earlier in the day.  The H$\alpha$ image from the HRTS slit-jaw camera shown in \citet{rousseldupre84} shows a high level of emission overlying the region that \citet{brekke91} claims is the light bridge.  Another sign of unusual behavior in this spectrum is the non-standard ratio of the \ion{Si}{4} lines.  The \ion{Si}{4} 1393/1402 \AA\ line ratio is approximately 10:1 in this observation.  In optically thin situations found for most quiescent spectra, these lines show a 2:1 ratio \citep{mathioudakis99}.  However, flaring conditions are conducive to the alteration of this line ratio.  When optical depth effects become significant \re{the ratio becomes closer to 1:1, however} the ratio may decrease or increase when scattering out of or into the line of sight occurs, respectively depending on the atmospheric structure and viewing geometry \citep{kerr05}.  \re{Still, such a large difference in the line strengths is unexpected.}

There are many possibilities that might contribute to the very bright \hh\ emission seen in this spectrum.  Sunspot atmospheres are characterized by being much cooler, having higher molecular abundances, and lower opacity than the surrounding quiet-Sun.  This observation is not of a simple, quiescent light bridge, but shows signatures of dynamic flare activity.  The effects of pronounced 3D atmospheric structure during flares may allow UV photons to excite \hh\ more efficiently.  Another possibility is that the HRTS did not capture the full radiation source that was illuminating the \hh.  HRTS was a slit spectrograph and could easily miss bright emission from an adjacent region.  It is difficult to determine more from this observation so long after it was taken, however we hope that such dynamic spectra can be understood with the more comprehensive coverage provided by IRIS and other modern observatories.

Based on this and past work by \citet{jordan78} and \citet{bartoe79} we can provide some insight on more recent analysis of \hh\ observations.  \citet{kuhn06} suggested that the higher intensity of some \hh\ lines of the Werner band over a sunspot penumbra might indicate increased abundance due to the presence of ambipolar diffusion.  The dynamical effects of \hh{} formation in sunspots should be strongest where it has the highest partial pressure, close to the photosphere.  \hh{} fluorescent emission occurs some 500 km higher and may have little to do with the underlying photosphere.  The increased \hh\ intensity in this observation is more likely due to structure in the source of exciting photons.

The unusual event reported by \citet{schmit14} shows absorption of \hh{} on top of \ion{Si}{4} emission.  Based on our arguments here, it seems unlikely that such a spectrum could be produced by \hh{} high in the solar atmosphere.  Likewise, it is difficult for \ion{Si}{4} to produce emission very deep in the chromosphere \citep{judge15}.  Nevertheless, these results are not inconsistent, as events such as reported by \citet{schmit14} are indeed rare.  Our work provides a ``standard picture'' for the formation of \hh\ lines under normal circumstances, against which such observations can be pitted in future work.

\section{Conclusions}
We have carried out calculations of the fluorescent intensities of \hh\ lines of the Lyman and Werner bands using three semi-empirical 1D atmospheric models with different temperature stratifications.  \re{We have simulated the radiation conditions underneath transition region radiation, from quiet-Sun conditions to the intense emission that might occur during a major flare, by incorporating downward line radiation based on measured UV intensities in our radiative transfer calculations}.  This follows on the basic model of fluorescence put forth by \citet{jordan78} and \citet{bartoe79} following the initial discovery of the \hh\ lines, but we have carried out our calculations using a modern radiative transfer code with more detailed line intensities and opacities, updated molecular parameters\re{, and better measurements of the UV transition region lines.  We also} account for the optical depth of the \hh\ lines approximately using escape probabilities.

We have produced a variety of synthetic spectra which can be compared directly with new observations from the Interface Region Imaging Spectrograph and previous UV spectrographs.  The \hh\ emission in the three models originates near the UV continuum $\tau=1$ boundary for the ultraviolet, which occurs near 650 km, 700 km, and 450 km for the FALC, COX, and F2 atmospheres respectively.  Contrary to the work of \citet{jordan78} and \citet{bartoe79}, we find that chromospheric lines, including \ion{H}{1} Ly $\alpha$, have large opacities which make them inefficient pumps for \hh, therefore emission from optically thin transition region lines is primarily responsible for the brightest \hh\ lines.  The largest change in the appearance of the \hh\ spectrum is caused by the predominance of continuum radiation or line radiation as a pumping source for the upper level populations.  Complete tables of our results are available with the on-line version of this article.

We carry out a detailed comparison of our synthetic spectra from the FALC and COX models against the SUMER quiet-Sun and HRTS light bridge atlas spectra for three families of \hh\ lines in the Lyman band, where the downward transitions in each family share a single upper energy level populated through one or two upward transitions pumped by bright transition region lines (\ion{C}{2} and \ion{Si}{4}).  The average 1D atmospheric models for the quiet-Sun are able to reproduce the strength of a few \hh\ lines which can be seen in the SUMER quiet-Sun atlas spectrum, however all but the strongest \hh\ lines predicted are below the sensitivity limit in this spectrum.  The success of the simple, 1D models to explain the observed spectra is somewhat surprising, but indicates a simple fact:  fluorescent emission from \hh\ originates deep in the chromosphere where stratification is still the dominant effect.  In future work we can assess the impact of atmospheric variability on the intensity of the \hh\ emission by performing 1D radiative transfer calculations on a 3D model.

Based on comparison of our model spectra with the HRTS light bridge spectrum of \citet{brekke91}, we find very bright \hh\ emission occurs in all three of the selected line families, including the \level{20}{2}{u} upper level whose lines were not previously identified by \citet{jordan77}, \citet{jordan78}, or \citet{bartoe79}.  However the \hh\ emission that can be seen in the light bridge spectrum seems too bright compared with the observed radiation from the transition region lines, according to our models.  While sunspots may have significantly different atmospheric parameters than the quiet-Sun, this ``light bridge'' spectrum may be more than it seems.  Observations that occurred the same day HRTS rocket flight confirm that a high level of activity was present around this sunspot.  The spectrum itself shows large velocities and irregular profiles in the transition region lines, as well as a \ion{Si}{4} 1393/1402 line ratio that is not consistent with the optically thin 2:1 ratio, indicating that this spectrum may be the result of flare activity.  This observation reveals the possible limitations of slit spectrometers for characterizing the spatial structure of transition region radiation.  High cadence spectra and imaging from IRIS make it possible to further analyze similar dynamic events.

In addition to providing a set of synthetic reference spectra and solar identifications for 54 new lines and 8 new levels of \hh, this work also demonstrates an important lesson for future \re{observations of \hh:  the pumping radiation source cannot be neglected.}  The presence or absence of line and continuum radiation can drastically change the appearance of the \hh\ fluorescence spectrum.  Furthermore, in order to use \hh{} as a diagnostic, something must be known about the wavelength dependence and spatial structure of the radiation field responsible for pumping \hh.  The small segments of the spectrum observed in detail by IRIS do include some of the strongest pumping lines, but this alone may not be enough to characterize the most important sources of emission for the \hh\ lines contained within the FUV channels of that instrument.  Either the full UV spectrum must be observed, or some segments can be observed and others modeled or extrapolated.

\acknowledgements
The authors would like to thank Charles Kankelborg for his support on this project, and Carole Jordan for useful remarks on preparation of this manuscript.

The work of S. Jaeggli at the National Solar Observatory is sponsored by the National Science Foundation, and was supported by the NASA postdoctoral program at NASA's Goddard Space Flight Center, and at Montana State University under the NASA/IRIS subcontract from Lockheed Martin.  IRIS is a NASA small explorer mission developed and operated by LMSAL with mission operations executed at NASA Ames Research center and major contributions to downlink communications funded by ESA and the Norwegian Space Center.

Visits to MSU by P. Judge, which enabled this research, were supported by the Montana State University Physics Department, the Advanced Study Program of the High Altitude Observatory of the National Center for Atmospheric Research.


\clearpage
\LongTables
\begin{landscape}
\begin{deluxetable*}{ccrrrrrrrrrrrrrrrrl}
\tablecaption{Strongest \hh\ Lines 1205-1550 \AA\ \label{tbl:h2lines}}
\tablehead{
 & & & & & & \multicolumn{6}{c}{Line Intensity [erg s$^{-1}$ cm$^{-2}$ str$^{-1}$]} & \multicolumn{6}{c}{\% Intensity Due to Primary Pump} \\
\cmidrule(lr){7-12}\cmidrule(lr){13-18}
& \colhead{Wavelength} & & & & & \multicolumn{2}{c}{FALC} & \multicolumn{2}{c}{COX} & \multicolumn{2}{c}{F2} & \multicolumn{2}{c}{FALC} & \multicolumn{2}{c}{COX} & \multicolumn{2}{c}{F2} & \colhead{Wavelength of Primary Pump}\\
\cmidrule(lr){7-8}\cmidrule(lr){9-10}\cmidrule(lr){11-12}\cmidrule(lr){13-14}\cmidrule(lr){15-16}\cmidrule(lr){17-18}
\colhead{ID} &\colhead{[\AA]} & \colhead{$\mathcal{J}_{l}$} & \colhead{$v_{l}$}& \colhead{$\mathcal{J}_{u}$} & \colhead{$v_{u}$} & \colhead{$\times$1} & \colhead{$\times$100} & \colhead{$\times$1} & \colhead{$\times$100}  & \colhead{$\times$1} & \colhead{$\times$100} & \colhead{$\times$1} & \colhead{$\times$100} & \colhead{$\times$1} & \colhead{$\times$100}  & \colhead{$\times$1} & \colhead{$\times$100} & \colhead{[\AA]}\\ 
}

\footnotemark[1,2,3]

\startdata
 & 1231.010 &  15 &   1 &  16 &   2 &   0.07 &   4.01 &   0.09 &  15.40 &   5.47 &   6.29 & $^a$21 & $^a$95 & $^a$57 & $^a$97 & $^b$15 & $^a$18 & $^a$1335.845 $^b$1324.801 \\
 & 1246.194 &  10 &   2 &  11 &   1 &   0.33 &  19.01 &   0.58 &  52.23 &  29.14 &  34.08 & $^a$45 & $^a$97 & $^a$86 & $^a$98 & $^b$16 & $^a$26 & $^a$1334.493 $^b$1504.317 \\
 & 1248.338 &   8 &   2 &   9 &   0 &   0.37 &  23.78 &   0.75 &  71.75 &  24.65 &  29.77 & $^a$39 & $^a$88 & $^a$82 & $^a$92 & $^b$14 & $^a$18 & $^a$1334.498 $^b$1393.452 \\
 & 1258.771 &   9 &   9 &   9\footnotemark[*] &   5\footnote[*]{This is the only \cstate\ excited level in the table.  All others are \bstate.} &   0.04 &   0.90 &   0.07 &  45.37 &  21.39 &  21.67 & $^a$58 & $^b$85 & $^b$81 & $^b$99 & $^a$64 & $^a$63 & $^a$1258.771 $^b$ 977.135 \\
\ $n$\footnote{$n$ denotes a line identified by this work.} & 1271.774 &  17 &   1 &  16 &   2 &   0.27 &  15.04 &   0.20 &  37.09 &  51.13 &  56.97 & $^a$21 & $^a$95 & $^a$57 & $^a$97 & $^b$15 & $^a$18 & $^a$1335.845 $^b$1324.801 \\
 & 1274.921 &   1 &   3 &   2 &   0 &   0.82 &  14.78 &   0.48 &  35.45 & 111.07 & 115.76 & $^a$16 & $^b$51 & $^a$25 & $^b$55 & $^a$16 & $^a$18 & $^a$1402.648 $^b$1393.961 \\
$n$ & 1276.628 &  10 &   2 &   9 &   0 &   0.53 &  35.45 &   1.06 & 108.98 &  48.67 &  57.63 & $^a$39 & $^a$88 & $^a$82 & $^a$92 & $^b$14 & $^a$18 & $^a$1334.498 $^b$1393.452 \\
 & 1276.812 &   1 &   3 &   0 &   0 &   0.56 &  65.60 &   0.69 & 225.35 &  62.10 &  76.30 & $^a$40 & $^a$98 & $^a$74 & $^a$99 & $^a$27 & $^a$39 & $^a$1335.868 \\
$n$ & 1278.537 &  12 &   2 &  11 &   1 &   0.38 &  23.13 &   0.71 &  69.41 &  40.36 &  46.88 & $^a$45 & $^a$97 & $^a$86 & $^a$98 & $^b$16 & $^a$26 & $^a$1334.493 $^b$1504.317 \\
 & 1278.733 &   3 &   3 &   4 &   0 &   1.03 &   3.85 &   0.43 &   7.18 & 142.88 & 144.08 & $^a$16 & $^b$23 & $^a$17 & $^b$29 & $^a$17 & $^a$17 & $^a$1352.502 $^b$1397.419 \\
 & 1280.523 &   5 &   4 &   6 &   3 &   0.29 &   7.71 &   0.25 &  24.42 &  39.52 &  41.65 & $^a$25 & $^a$93 & $^a$74 & $^a$95 & $^b$15 & $^b$14 & $^a$1334.480 $^b$1462.581 \\
$n$ & 1282.533 &  21 &   0  &  20 &   2 &   0.05 &  3.88 &   0.06 &  10.57 &  8.53 &  9.96 & $^a$31 & $^a$96 & $^a$70 & $^a$98 & $^b$20 & $^a$22 & $^a$1334.424 $^b$1338.341 \\
\ $p$\footnote{$p$ denotes a previously identified line.} & 1283.110 &   3 &   3 &   2 &   0 &   0.99 &  17.65 &   0.58 &  43.38 & 128.57 & 134.09 & $^a$16 & $^b$51 & $^a$25 & $^b$55 & $^a$16 & $^a$18 & $^a$1402.648 $^b$1393.961 \\
$n$ & 1283.143 &  15 &   2 &  16 &   2 &   0.35 &  19.80 &   0.27 &  49.68 &  63.59 &  71.23 & $^a$21 & $^a$95 & $^a$57 & $^a$97 & $^b$15 & $^a$18 & $^a$1335.845 $^b$1324.801 \\
 & 1286.522 &   5 &   3 &   6 &   0 &   1.18 &   5.66 &   0.49 &   8.63 & 180.80 & 182.87 & $^a$15 & $^b$41 & $^b$16 & $^b$56 & $^a$15 & $^a$15 & $^a$1366.391 $^b$1404.747 \\
$n$ & 1286.980 &  19 &   1 &  20 &   2 &   0.18 &  13.19 &   0.20 &  33.56 &  30.05 &  35.39 & $^a$31 & $^a$96 & $^a$70 & $^a$98 & $^b$20 & $^a$22 & $^a$1334.424 $^b$1338.341 \\
$p$ & 1293.306 &   5 &   3 &   4 &   0 &   1.32 &   5.00 &   0.56 &   9.38 & 189.99 & 191.62 & $^a$16 & $^b$23 & $^a$17 & $^b$29 & $^a$17 & $^a$17 & $^a$1352.502 $^b$1397.419 \\
$n$ & 1298.138 &   7 &   3 &   8 &   0 &   1.19 &   4.03 &   0.52 &   4.56 & 192.98 & 194.61 & $^a$14 & $^a$14 & $^a$15 & $^a$14 & $^a$14 & $^a$14 & $^a$1383.653 \\
$p$ & 1299.818 &   7 &   4 &   6 &   3 &   0.36 &  10.01 &   0.31 &  31.81 &  53.70 &  56.57 & $^a$25 & $^a$93 & $^a$74 & $^a$95 & $^b$15 & $^b$14 & $^a$1334.480 $^b$1462.581 \\
\ $u$\footnote{$u$ denotes a line matching an unidentified feature in \citet{sandlin86}.} & 1301.278 &  10 &   3 &  11 &   1 &   0.44 &  27.90 &   0.84 &  86.85 &  50.02 &  57.85 & $^a$45 & $^a$97 & $^a$86 & $^a$98 & $^b$16 & $^a$26 & $^a$1334.493 $^b$1504.317 \\
$p$ & 1305.317 &   8 &   3 &   9 &   0 &   0.78 &  54.04 &   1.69 & 175.87 &  69.76 &  83.29 & $^a$39 & $^a$88 & $^a$82 & $^a$92 & $^b$14 & $^a$18 & $^a$1334.498 $^b$1393.452 \\
$p$ & 1307.211 &   7 &   3 &   6 &   0 &   1.51 &   7.42 &   0.67 &  11.79 & 234.75 & 237.50 & $^a$15 & $^b$41 & $^b$16 & $^b$56 & $^a$15 & $^a$15 & $^a$1366.391 $^b$1404.747 \\
 & 1313.369 &   9 &   3 &  10 &   0 &   1.09 &   5.59 &   0.51 &   8.59 & 180.14 & 182.35 & $^a$15 & $^a$41 & $^a$19 & $^a$57 & $^b$16 & $^b$16 & $^a$1371.414 $^b$1403.974 \\
 & 1324.589 &   9 &   3 &   8 &   0 &   1.51 &   5.34 &   0.71 &   6.34 & 242.15 & 244.29 & $^a$14 & $^a$14 & $^a$15 & $^a$14 & $^a$14 & $^a$14 & $^a$1383.653 \\
$n$ & 1324.801 &  17 &   2 &  16 &   2 &   0.36 &  21.43 &   0.29 &  58.68 &  70.50 &  78.42 & $^a$21 & $^a$95 & $^a$57 & $^a$97 & $^b$15 & $^a$18 & $^a$1335.845 $^b$1324.801 \\
 & 1331.954 &  11 &   3 &  12 &   0 &   0.96 &   3.45 &   0.47 &   3.77 & 162.57 & 164.10 & $^a$14 & $^b$15 & $^a$15 & $^b$16 & $^b$15 & $^b$15 & $^a$1389.583 $^b$1427.002 \\
$n$ & 1332.341  &  22 &   1 &  21 &   3 &   0.06 &   5.27 &   0.07 &  13.39 &  10.46 &  12.82 & $^a$42 & $^a$97 & $^a$79 & $^a$99 & $^a$18 & $^a$31 & $^a$1334.426 \\
$p$ & 1333.797 &   1 &   4 &   2 &   0 &   1.18 &  22.85 &   0.81 &  64.39 & 155.55 & 162.08 & $^a$16 & $^b$51 & $^a$25 & $^b$55 & $^a$16 & $^a$18 & $^a$1402.648 $^b$1393.961 \\
$n$ & 1334.424 &  21 &   1 &  20 &   2 &   0.24 &  18.13 &   0.27 &  47.99 &  44.31 &  51.69 & $^a$31 & $^a$96 & $^a$70 & $^a$98 & $^b$20 & $^a$22 & $^a$1334.424 $^b$1338.341 \\ 
$n$ & 1334.426 &  20 &   2 &  21 &   3 &   0.06 &   6.11 &   0.08 &  15.49 &  12.12 &  14.86 & $^a$42 & $^a$97 & $^a$79 & $^a$99 & $^a$18 & $^a$31 & $^a$1334.426 \\
$p$ & 1334.480 &   5 &   5 &   6 &   3 &   0.21 &   6.29 &   0.19 &  21.22 &  33.18 &  34.89 & $^a$25 & $^a$93 & $^a$74 & $^a$95 & $^b$15 & $^b$14 & $^a$1334.480 $^b$1462.581 \\
$n$ & 1334.493 &  12 &   3 &  11 &   1 &   0.41 &  27.17 &   0.80 &  90.23 &  48.63 &  55.90 & $^a$45 & $^a$97 & $^a$86 & $^a$98 & $^b$16 & $^a$26 & $^a$1334.493 $^b$1504.317 \\
$p$ & 1334.498 &  10 &   3 &   9 &   0 &   0.98 &  70.09 &   2.14 & 236.64 &  95.21 & 112.69 & $^a$39 & $^a$88 & $^a$82 & $^a$92 & $^b$14 & $^a$18 & $^a$1334.498 $^b$1393.452 \\
$n$ & 1335.845 &  15 &   3 &  16 &   2 &   0.27 &  15.97 &   0.22 &  45.45 &  52.79 &  58.49 & $^a$21 & $^a$95 & $^a$57 & $^a$97 & $^b$15 & $^a$18 & $^a$1335.845 $^b$1324.801 \\
$p$ & 1335.868 &   1 &   4 &   0 &   0 &   0.73 &  91.79 &   1.00 & 364.92 &  82.89 & 101.12 & $^a$40 & $^a$98 & $^a$74 & $^a$99 & $^a$27 & $^a$39 & $^a$1335.868 \\
$p$ & 1337.467 &   3 &   4 &   4 &   0 &   1.73 &   7.12 &   0.88 &  14.13 & 246.81 & 249.12 & $^a$16 & $^b$23 & $^a$17 & $^b$29 & $^a$17 & $^a$17 & $^a$1352.502 $^b$1397.419 \\
$n$ & 1338.341 &  19 &   2 &  20 &   2 &   0.36 &  27.92 &   0.41 &  72.62 &  67.04 &  78.63 & $^a$31 & $^a$96 & $^a$70 & $^a$98 & $^b$20 & $^a$22 & $^a$1334.424 $^b$1338.341 \\
$p$ & 1338.569 &   2 &   4 &   1 &   0 &   0.41 &   5.43 &   0.30 &  16.56 &  52.51 &  54.00 & $^a$18 & $^b$81 & $^b$34 & $^b$89 & $^a$19 & $^a$18 & $^a$1338.569 $^b$1393.719 \\
$p$ & 1342.256 &   3 &   4 &   2 &   0 &   1.55 &  30.15 &   1.05 &  84.01 & 209.27 & 218.14 & $^a$16 & $^b$51 & $^a$25 & $^b$55 & $^a$16 & $^a$18 & $^a$1402.648 $^b$1393.961 \\
$p$ & 1345.085 &   5 &   4 &   6 &   0 &   1.99 &  10.33 &   1.00 &  17.95 & 300.01 & 303.67 & $^a$15 & $^b$41 & $^b$16 & $^b$56 & $^a$15 & $^a$15 & $^a$1366.391 $^b$1404.747 \\
$n$ & 1345.170 &  11 &   3 &  10 &   0 &   1.39 &   7.49 &   0.72 &  12.35 & 227.26 & 230.13 & $^a$15 & $^a$41 & $^a$19 & $^a$57 & $^b$16 & $^b$16 & $^a$1371.414 $^b$1403.974 \\
$n$ & 1350.325 &   6 &   4 &   7 &   0 &   0.98 &   3.93 &   0.51 &   5.23 & 167.03 & 168.74 & $^a$14 & $^a$15 & $^a$15 & $^b$14 & $^a$15 & $^a$15 & $^a$1374.619 $^b$1234.789 \\
$p$ & 1352.502 &   5 &   4 &   4 &   0 &   2.56 &  11.03 &   1.26 &  19.61 & 427.97 & 432.38 & $^a$16 & $^b$23 & $^a$17 & $^b$29 & $^a$17 & $^a$17 & $^a$1352.502 $^b$1397.419 \\
 & 1353.591 &  13 &   3 &  14 &   0 &   0.92 &   3.63 &   0.46 &   3.45 & 191.82 & 193.93 & $^a$16 & $^a$17 & $^a$17 & $^a$17 & $^a$17 & $^a$17 & $^a$1452.342 \\
 & 1354.139 &   7 &   5 &   6 &   3 &   0.17 &   5.22 &   0.14 &  16.34 &  33.11 &  34.59 & $^a$25 & $^a$93 & $^a$74 & $^a$95 & $^b$15 & $^b$14 & $^a$1334.480 $^b$1462.581 \\
$p$ & 1356.482 &   7 &   4 &   8 &   0 &   2.50 &   9.66 &   1.22 &  10.88 & 462.69 & 467.36 & $^a$14 & $^a$14 & $^a$15 & $^a$14 & $^a$14 & $^a$14 & $^a$1383.653 \\
$u$ & 1357.273 &  10 &   4 &  11 &   1 &   0.28 &  18.92 &   0.48 &  60.71 &  41.51 &  46.43 & $^a$45 & $^a$97 & $^a$86 & $^a$98 & $^b$16 & $^a$26 & $^a$1334.493 $^b$1504.317 \\
$n$ & 1359.007 &   6 &   4 &   5 &   0 &   1.06 &   4.39 &   0.55 &   5.98 & 180.57 & 182.42 & $^a$15 & $^b$16 & $^a$17 & $^b$19 & $^a$16 & $^a$16 & $^a$1359.007 $^b$1400.609 \\
$p$ & 1363.524 &   8 &   4 &   9 &   0 &   1.50 & 109.80 &   2.96 & 353.16 & 181.79 & 209.07 & $^a$39 & $^a$88 & $^a$82 & $^a$92 & $^b$14 & $^a$18 & $^a$1334.498 $^b$1393.452 \\
$p$ & 1366.391 &   7 &   4 &   6 &   0 &   2.93 &  16.01 &   1.46 &  25.98 & 512.37 & 518.97 & $^a$15 & $^b$41 & $^b$16 & $^b$56 & $^a$15 & $^a$15 & $^a$1366.391 $^b$1404.747 \\
 & 1368.653 &  13 &   3 &  12 &   0 &   1.56 &   6.17 &   0.79 &   6.54 & 306.83 & 310.14 & $^a$14 & $^b$15 & $^a$15 & $^b$16 & $^b$15 & $^b$15 & $^a$1389.583 $^b$1427.002 \\
$p$ & 1371.414 &   9 &   4 &  10 &   0 &   2.23 &  12.57 &   1.14 &  19.50 & 419.79 & 425.37 & $^a$15 & $^a$41 & $^a$19 & $^a$57 & $^b$16 & $^b$16 & $^a$1371.414 $^b$1403.974 \\
$n$ & 1374.619 &   8 &   4 &   7 &   0 &   1.11 &   4.59 &   0.61 &   6.31 & 187.89 & 189.88 & $^a$14 & $^a$15 & $^a$15 & $^b$14 & $^a$15 & $^a$15 & $^a$1374.619 $^b$1234.789 \\
 & 1378.116 &  17 &   3 &  16 &   2 &   0.14 &   8.52 &   0.11 &  23.64 &  35.77 &  38.78 & $^a$21 & $^a$95 & $^a$57 & $^a$97 & $^b$15 & $^a$18 & $^a$1335.845 $^b$1324.801 \\
$n$ & 1382.283  &  22 &   2 &  21 &   3 &   0.04 &   4.30 &   0.05 &  10.32 &  11.08 &  13.04 & $^a$42 & $^a$97 & $^a$79 & $^a$99 & $^a$18 & $^a$31 & $^a$1334.426 \\
 & 1383.653 &   9 &   4 &   8 &   0 &   2.93 &  11.81 &   1.53 &  13.92 & 526.38 & 531.90 & $^a$14 & $^a$14 & $^a$15 & $^a$14 & $^a$14 & $^a$14 & $^a$1383.653 \\
$u$ & 1386.311 &  21 &   2 &  20 &   2 &   0.41 &  32.44 &   0.42 &  82.05 &  96.95 & 110.15 & $^a$31 & $^a$96 & $^a$70 & $^a$98 & $^b$20 & $^a$22 & $^a$1334.424 $^b$1338.341 \\ 
$n$ & 1389.521 &  19 &   3 &  20 &   2 &   0.32 &  25.02 &   0.33 &  63.68 &  78.03 &  88.22 & $^a$31 & $^a$96 & $^a$70 & $^a$98 & $^b$20 & $^a$22 & $^a$1334.424 $^b$1338.341 \\ 
 & 1389.583 &  11 &   4 &  12 &   0 &   1.98 &   7.99 &   1.03 &   8.68 & 387.09 & 391.31 & $^a$14 & $^b$15 & $^a$15 & $^b$16 & $^b$15 & $^b$15 & $^a$1389.583 $^b$1427.002 \\
$u$ & 1391.166 &  12 &   4 &  11 &   1 &   0.13 &   8.85 &   0.22 &  29.23 &  20.36 &  22.63 & $^a$45 & $^a$97 & $^a$86 & $^a$98 & $^b$16 & $^a$26 & $^a$1334.493 $^b$1504.317 \\
 & 1393.441 &   9 &   5 &   8 &   2 &   0.73 &   6.17 &   0.28 &  15.12 & 155.70 & 158.18 & $^a$15 & $^a$75 & $^a$19 & $^a$72 & $^a$19 & $^a$20 & $^a$1393.441 \\
$p$ & 1393.452 &  10 &   4 &   9 &   0 &   1.65 & 124.01 &   3.30 & 415.76 & 204.53 & 234.15 & $^a$39 & $^a$88 & $^a$82 & $^a$92 & $^b$14 & $^a$18 & $^a$1334.498 $^b$1393.452 \\
$p$ & 1393.961 &   1 &   5 &   2 &   0 &   1.60 &  33.06 &   1.14 &  93.91 & 254.97 & 264.58 & $^a$16 & $^b$51 & $^a$25 & $^b$55 & $^a$16 & $^a$18 & $^a$1402.648 $^b$1393.961 \\
 & 1394.706 &  15 &   3 &  14 &   0 &   1.21 &   5.03 &   0.64 &   5.16 & 250.66 & 253.53 & $^a$16 & $^a$17 & $^a$17 & $^a$17 & $^a$17 & $^a$17 & $^a$1452.342 \\
$p$ & 1396.223 &   1 &   5 &   0 &   0 &   0.87 & 113.30 &   1.11 & 449.62 & 126.20 & 147.24 & $^a$40 & $^a$98 & $^a$74 & $^a$99 & $^a$27 & $^a$39 & $^a$1335.868 \\
$p$ & 1397.419 &   3 &   5 &   4 &   0 &   2.60 &  12.16 &   1.49 &  22.29 & 429.11 & 433.90 & $^a$16 & $^b$23 & $^a$17 & $^b$29 & $^a$17 & $^a$17 & $^a$1352.502 $^b$1397.419 \\
$p$ & 1398.953 &   2 &   5 &   1 &   0 &   0.47 &   6.65 &   0.34 &  19.04 &  76.81 &  78.86 & $^a$18 & $^b$81 & $^b$34 & $^b$89 & $^a$19 & $^a$18 & $^a$1338.569 $^b$1393.719 \\
$p$ & 1400.609 &   4 &   5 &   5 &   0 &   0.98 &   4.35 &   0.57 &   6.27 & 169.17 & 171.02 & $^a$15 & $^b$16 & $^a$17 & $^b$19 & $^a$16 & $^a$16 & $^a$1359.007 $^b$1400.609 \\
$p$ & 1402.648 &   3 &   5 &   2 &   0 &   2.12 &  44.10 &   1.49 & 124.73 & 338.76 & 351.69 & $^a$16 & $^b$51 & $^a$25 & $^b$55 & $^a$16 & $^a$18 & $^a$1402.648 $^b$1393.961 \\
$p$ & 1403.974 &  11 &   4 &  10 &   0 &   2.71 &  15.88 &   1.49 &  25.92 & 512.52 & 519.45 & $^a$15 & $^a$41 & $^a$19 & $^a$57 & $^b$16 & $^b$16 & $^a$1371.414 $^b$1403.974 \\
 & 1404.747 &   5 &   5 &   6 &   0 &   3.14 &  18.19 &   1.77 &  31.61 & 546.67 & 553.96 & $^a$15 & $^b$41 & $^b$16 & $^b$56 & $^a$15 & $^a$15 & $^a$1366.391 $^b$1404.747 \\
$u$ & 1409.805 &   6 &   5 &   7 &   0 &   1.10 &   4.84 &   0.65 &   6.92 & 193.86 & 196.01 & $^a$14 & $^a$15 & $^a$15 & $^b$14 & $^a$15 & $^a$15 & $^a$1374.619 $^b$1234.789 \\
 & 1410.636 &  13 &   4 &  14 &   0 &   1.53 &   6.45 &   0.82 &   6.68 & 318.32 & 321.96 & $^a$16 & $^a$17 & $^a$17 & $^a$17 & $^a$17 & $^a$17 & $^a$1452.342 \\
 & 1412.838 &   5 &   5 &   4 &   0 &   3.00 &  14.24 &   1.74 &  26.39 & 497.72 & 503.32 & $^a$16 & $^b$23 & $^a$17 & $^b$29 & $^a$17 & $^a$17 & $^a$1352.502 $^b$1397.419 \\
 & 1415.750 &   7 &   5 &   8 &   0 &   3.21 &  13.75 &   1.82 &  17.02 & 582.40 & 588.79 & $^a$14 & $^a$14 & $^a$15 & $^a$14 & $^a$14 & $^a$14 & $^a$1383.653 \\
$p$ & 1419.271 &   6 &   5 &   5 &   0 &   1.07 &   4.86 &   0.63 &   7.10 & 187.50 & 189.58 & $^a$15 & $^b$16 & $^a$17 & $^b$19 & $^a$16 & $^a$16 & $^a$1359.007 $^b$1400.609 \\
$p$ & 1422.546 &   8 &   5 &   9 &   0 &   1.60 & 122.27 &   3.17 & 424.02 & 208.49 & 236.89 & $^a$39 & $^a$88 & $^a$82 & $^a$92 & $^b$14 & $^a$18 & $^a$1334.498 $^b$1393.452 \\
$u$ & 1422.957 &  17 &   3 &  16 &   0 &   0.89 &   3.89 &   0.49 &   4.13 & 199.28 & 201.65 & $^a$20 & $^a$21 & $^a$20 & $^a$21 & $^a$21 & $^a$21 & $^a$1479.539 \\
$p$ & 1426.548 &   7 &   5 &   6 &   0 &   3.45 &  20.50 &   2.02 &  36.56 & 598.99 & 607.09 & $^a$15 & $^b$41 & $^b$16 & $^b$56 & $^a$15 & $^a$15 & $^a$1366.391 $^b$1404.747 \\
$n$ & 1427.002 &  13 &   4 &  12 &   0 &   2.38 &  10.15 &   1.32 &  11.66 & 468.62 & 473.93 & $^a$14 & $^b$15 & $^a$15 & $^b$16 & $^b$15 & $^b$15 & $^a$1389.583 $^b$1427.002 \\
$p$ & 1430.151 &   9 &   5 &  10 &   0 &   2.86 &  17.33 &   1.67 &  29.73 & 539.65 & 547.09 & $^a$15 & $^a$41 & $^a$19 & $^a$57 & $^b$16 & $^b$16 & $^a$1371.414 $^b$1403.974 \\
$p$ & 1431.010 &   3 &   6 &   4 &   1 &   1.00 &   3.96 &   0.48 &   9.05 & 194.02 & 195.84 & $^a$15 & $^a$18 & $^a$17 & $^b$30 & $^a$19 & $^a$19 & $^a$1504.750 $^b$1202.449 \\
 & 1431.121 &  17 &   4 &  16 &   2 &   0.09 &   5.50 &   0.07 &  15.68 &  25.24 &  27.24 & $^a$21 & $^a$95 & $^a$57 & $^a$97 & $^b$15 & $^a$18 & $^a$1335.845 $^b$1324.801 \\
$p$ & 1434.068 &   4 &   6 &   5 &   1 &   0.36 &   5.61 &   0.22 &  18.40 &  69.01 &  70.59 & $^a$14 & $^b$65 & $^b$19 & $^b$75 & $^a$19 & $^a$18 & $^a$1510.687 $^b$1335.574 \\
 & 1434.166 &  15 &   4 &  16 &   0 &   1.13 &   4.90 &   0.62 &   5.30 & 246.18 & 249.08 & $^a$20 & $^a$21 & $^a$20 & $^a$21 & $^a$21 & $^a$21 & $^a$1479.539 \\
 & 1434.627 &   8 &   5 &   7 &   0 &   1.13 &   5.14 &   0.69 &   7.63 & 201.69 & 203.98 & $^a$14 & $^a$15 & $^a$15 & $^b$14 & $^a$15 & $^a$15 & $^a$1374.619 $^b$1234.789 \\
$u$ & 1437.526 &  21 &   3 &  20 &   2 &   0.11 &   8.93 &   0.11 &  23.44 &  32.36 &  36.08 & $^a$31 & $^a$96 & $^a$70 & $^a$98 & $^b$20 & $^a$22 & $^a$1334.424 $^b$1338.341 \\
$p$ & 1438.030 &   5 &   6 &   6 &   1 &   1.17 &   5.15 &   0.60 &  11.44 & 234.48 & 236.80 & $^a$15 & $^b$16 & $^b$16 & $^c$23 & $^b$18 & $^b$18 & $^a$1265.664 $^b$1517.338 $^c$1229.716 \\
$u$ & 1440.835 &  15 &   5 &  16 &   2 &   0.33 &  20.64 &   0.25 &  58.60 &  86.44 &  93.80 & $^a$21 & $^a$95 & $^a$57 & $^a$97 & $^b$15 & $^a$18 & $^a$1335.845 $^b$1324.801 \\
$p$ & 1442.741 &   5 &   7 &   6 &   3 &   0.42 &  13.76 &   0.36 &  44.39 &  87.36 &  91.35 & $^a$25 & $^a$93 & $^a$74 & $^a$95 & $^b$15 & $^b$14 & $^a$1334.480 $^b$1462.581 \\
$n$ & 1443.465 &   9 &   5 &   8 &   0 &   3.45 &  15.40 &   2.08 &  20.04 & 620.92 & 627.97 & $^a$14 & $^a$14 & $^a$15 & $^a$14 & $^a$14 & $^a$14 & $^a$1383.653 \\
$p$ & 1446.117 &   5 &   6 &   4 &   1 &   1.36 &   5.41 &   0.65 &  12.45 & 276.04 & 278.63 & $^a$15 & $^a$18 & $^a$17 & $^b$30 & $^a$19 & $^a$19 & $^a$1504.750 $^b$1202.449 \\
$n$ & 1447.602 &  11 &   5 &  12 &   0 &   2.51 &  11.04 &   1.44 &  13.12 & 523.13 & 529.22 & $^a$14 & $^b$15 & $^a$15 & $^b$16 & $^b$15 & $^b$15 & $^a$1389.583 $^b$1427.002 \\
$u$ & 1448.016 &  12 &   5 &  11 &   1 &   0.07 &   4.79 &   0.11 &  16.06 &  12.27 &  13.56 & $^a$45 & $^a$97 & $^a$86 & $^a$98 & $^b$16 & $^a$26 & $^a$1334.493 $^b$1504.317 \\
$p$ & 1448.520 &   7 &   6 &   8 &   1 &   1.04 &   4.43 &   0.50 &   7.39 & 239.93 & 242.31 & $^a$16 & $^b$17 & $^b$16 & $^c$20 & $^b$18 & $^b$18 & $^a$1246.850 $^b$1505.649 $^c$1222.028 \\
$p$ & 1452.332 &   6 &   6 &   5 &   1 &   0.49 &   7.76 &   0.30 &  25.33 &  99.78 & 102.05 & $^a$14 & $^b$65 & $^b$19 & $^b$75 & $^a$19 & $^a$18 & $^a$1510.687 $^b$1335.574 \\
 & 1452.342 &  15 &   4 &  14 &   0 &   1.79 &   7.96 &   1.01 &   8.85 & 394.61 & 399.32 & $^a$16 & $^a$17 & $^a$17 & $^a$17 & $^a$17 & $^a$17 & $^a$1452.342 \\
 & 1452.978 &  19 &   3 &  18 &   0 &   0.74 &   4.58 &   0.41 &   6.50 & 183.88 & 186.79 & $^a$20 & $^b$29 & $^a$20 & $^b$39 & $^a$21 & $^a$21 & $^a$1513.729 $^b$1404.600 \\
$p$ & 1453.015 &  10 &   5 &   9 &   0 &   1.59 & 122.78 &   3.13 & 438.08 & 226.16 & 255.01 & $^a$39 & $^a$88 & $^a$82 & $^a$92 & $^b$14 & $^a$18 & $^a$1334.498 $^b$1393.452 \\
$p$ & 1454.971 &   1 &   6 &   2 &   0 &   1.16 &  24.44 &   0.87 &  74.32 & 210.41 & 217.85 & $^a$16 & $^b$51 & $^a$25 & $^b$55 & $^a$16 & $^a$18 & $^a$1402.648 $^b$1393.961 \\
$p$ & 1457.435 &   1 &   6 &   0 &   0 &   0.55 &  72.11 &   0.69 & 301.01 &  90.97 & 104.39 & $^a$40 & $^a$98 & $^a$74 & $^a$99 & $^a$27 & $^a$39 & $^a$1335.868 \\
$p$ & 1458.131 &   3 &   6 &   4 &   0 &   2.12 &  10.84 &   1.36 &  20.42 & 390.24 & 395.00 & $^a$16 & $^b$23 & $^a$17 & $^b$29 & $^a$17 & $^a$17 & $^a$1352.502 $^b$1397.419 \\
$p$ & 1459.336 &   7 &   6 &   6 &   1 &   1.59 &   7.08 &   0.80 &  15.78 & 334.43 & 337.75 & $^a$15 & $^b$16 & $^b$16 & $^c$23 & $^b$18 & $^b$18 & $^a$1265.664 $^b$1517.338 $^c$1229.716 \\
$u$ & 1459.703 &  17 &   4 &  18 &   0 &   0.94 &   5.86 &   0.52 &   8.32 & 228.45 & 232.05 & $^a$20 & $^b$29 & $^a$20 & $^b$39 & $^a$21 & $^a$21 & $^a$1513.729 $^b$1404.600 \\
$n$ & 1462.142 &   9 &   6 &  10 &   1 &   0.92 &   4.00 &   0.46 &   7.19 & 221.68 & 223.96 & $^a$16 & $^b$16 & $^b$15 & $^c$33 & $^b$18 & $^b$18 & $^a$1267.221 $^b$1518.132 $^c$1237.292 \\
$p$ & 1462.581 &   7 &   7 &   6 &   3 &   0.47 &  15.72 &   0.41 &  50.46 & 105.12 & 109.85 & $^a$25 & $^a$93 & $^a$74 & $^a$95 & $^b$15 & $^b$14 & $^a$1334.480 $^b$1462.581 \\
 & 1463.226 &  11 &   5 &  10 &   0 &   3.04 &  19.03 &   1.85 &  33.40 & 611.37 & 619.99 & $^a$15 & $^a$41 & $^a$19 & $^a$57 & $^b$16 & $^b$16 & $^a$1371.414 $^b$1403.974 \\
$p$ & 1463.825 &   3 &   6 &   2 &   0 &   1.52 &  32.51 &   1.14 &  98.81 & 273.55 & 283.35 & $^a$16 & $^b$51 & $^a$25 & $^b$55 & $^a$16 & $^a$18 & $^a$1402.648 $^b$1393.961 \\
$p$ & 1465.017 &   5 &   6 &   6 &   0 &   2.69 &  16.79 &   1.71 &  31.34 & 514.96 & 522.20 & $^a$15 & $^b$41 & $^b$16 & $^b$56 & $^a$15 & $^a$15 & $^a$1366.391 $^b$1404.747 \\
 & 1467.688 &  13 &   5 &  14 &   0 &   1.87 &   8.46 &   1.07 &   9.56 & 414.17 & 419.18 & $^a$16 & $^a$17 & $^a$17 & $^a$17 & $^a$17 & $^a$17 & $^a$1452.342 \\
$u$ & 1470.000 &  10 &   6 &  11 &   1 &   0.33 &  24.33 &   0.58 &  81.41 &  59.15 &  65.67 & $^a$45 & $^a$97 & $^a$86 & $^a$98 & $^b$16 & $^a$26 & $^a$1334.493 $^b$1504.317 \\
$p$ & 1473.820 &   5 &   6 &   4 &   0 &   2.31 &  11.90 &   1.50 &  22.53 & 425.22 & 430.44 & $^a$16 & $^b$23 & $^a$17 & $^b$29 & $^a$17 & $^a$17 & $^a$1352.502 $^b$1397.419 \\
 & 1474.045 &  12 &   5 &  11 &   0 &   0.87 &   4.00 &   0.52 &   5.03 & 188.73 & 191.01 & $^a$15 & $^a$16 & $^b$15 & $^a$16 & $^a$16 & $^a$16 & $^a$1415.174 $^b$1380.115 \\
$p$ & 1475.405 &   7 &   6 &   8 &   0 &   2.80 &  13.30 &   1.77 &  17.82 & 562.66 & 569.47 & $^a$14 & $^a$14 & $^a$15 & $^a$14 & $^a$14 & $^a$14 & $^a$1383.653 \\
$u$ & 1475.508 &   9 &   6 &   8 &   1 &   1.46 &   6.27 &   0.70 &  10.55 & 334.12 & 337.46 & $^a$16 & $^b$17 & $^b$16 & $^c$20 & $^b$18 & $^b$18 & $^a$1246.850 $^b$1505.649 $^c$1222.028 \\
 & 1479.539 &  17 &   4 &  16 &   0 &   1.30 &   5.93 &   0.73 &   6.64 & 307.28 & 311.05 & $^a$20 & $^a$21 & $^a$20 & $^a$21 & $^a$21 & $^a$21 & $^a$1479.539 \\
 & 1481.816 &   8 &   6 &   9 &   0 &   1.14 &  88.74 &   2.20 & 321.22 & 174.53 & 195.27 & $^a$39 & $^a$88 & $^a$82 & $^a$92 & $^b$14 & $^a$18 & $^a$1334.498 $^b$1393.452 \\
 & 1483.015 &  17 &   5 &  16 &   2 &   0.42 &  26.81 &   0.32 &  75.81 & 119.17 & 129.05 & $^a$21 & $^a$95 & $^a$57 & $^a$97 & $^b$15 & $^a$18 & $^a$1335.845 $^b$1324.801 \\
$u$ & 1485.413 &  13 &   5 &  12 &   0 &   2.58 &  11.94 &   1.54 &  14.74 & 553.03 & 559.74 & $^a$14 & $^b$15 & $^a$15 & $^b$16 & $^b$15 & $^b$15 & $^a$1389.583 $^b$1427.002 \\
 & 1487.136 &   7 &   6 &   6 &   0 &   2.67 &  16.95 &   1.72 &  31.93 & 521.99 & 529.42 & $^a$15 & $^b$41 & $^b$16 & $^b$56 & $^a$15 & $^a$15 & $^a$1366.391 $^b$1404.747 \\
 & 1487.172 &  21 &   4 &  20 &   2 &   0.09 &   7.21 &   0.09 &  18.75 &  28.60 &  31.75 & $^a$31 & $^a$96 & $^a$70 & $^a$98 & $^b$20 & $^a$22 & $^a$1334.424 $^b$1338.341 \\ 
$u$ & 1488.267 &  19 &   5 &  20 &   2 &   0.28 &  22.60 &   0.28 &  58.83 &  83.58 &  93.27 & $^a$31 & $^a$96 & $^a$70 & $^a$98 & $^b$20 & $^a$22 & $^a$1334.424 $^b$1338.341 \\ 
$n$ & 1488.976 &   9 &   6 &  10 &   0 &   2.50 &  16.06 &   1.57 &  28.61 & 527.76 & 535.37 & $^a$15 & $^a$41 & $^a$19 & $^a$57 & $^b$16 & $^b$16 & $^a$1371.414 $^b$1403.974 \\
$p$ & 1489.072 &   1 &   7 &   0 &   1 &   0.45 &   1.59 &   0.61 &   3.08 &  82.18 &  82.96 & $^a$29 & $^a$34 & $^a$29 & $^a$27 & $^a$37 & $^a$37 & $^a$1489.072 \\
$p$ & 1489.562 &   3 &   7 &   4 &   1 &   1.53 &   6.32 &   0.74 &  14.41 & 320.39 & 323.49 & $^a$15 & $^a$18 & $^a$17 & $^b$30 & $^a$19 & $^a$19 & $^a$1504.750 $^b$1202.449 \\
 & 1489.923 &  15 &   5 &  16 &   0 &   1.35 &   6.22 &   0.77 &   7.02 & 321.02 & 324.98 & $^a$20 & $^a$21 & $^a$20 & $^a$21 & $^a$21 & $^a$21 & $^a$1479.539 \\
$n$ & 1491.602 &  15 &   6 &  16 &   2 &   0.32 &  20.64 &   0.25 &  58.37 &  94.64 & 102.31 & $^a$21 & $^a$95 & $^a$57 & $^a$97 & $^b$15 & $^a$18 & $^a$1335.845 $^b$1324.801 \\
$p$ & 1492.356 &   4 &   7 &   5 &   1 &   0.56 &   9.09 &   0.35 &  29.75 & 117.56 & 120.25 & $^a$14 & $^b$65 & $^b$19 & $^b$75 & $^a$19 & $^a$18 & $^a$1510.687 $^b$1335.574 \\
$n$ & 1494.191 &  11 &   6 &  10 &   1 &   1.37 &   6.04 &   0.68 &  10.95 & 327.85 & 331.23 & $^a$16 & $^b$16 & $^b$15 & $^c$33 & $^b$18 & $^b$18 & $^a$1267.221 $^b$1518.132 $^c$1237.292 \\
$p$ & 1495.217 &   3 &   7 &   2 &   1 &   1.16 &   5.07 &   0.58 &  11.03 & 237.49 & 239.85 & $^a$16 & $^a$19 & $^a$18 & $^b$29 & $^a$20 & $^a$20 & $^a$1495.217 $^b$1206.131 \\
$p$ & 1495.995 &   5 &   7 &   6 &   1 &   1.92 &   8.77 &   0.98 &  19.50 & 410.54 & 414.69 & $^a$15 & $^b$16 & $^b$16 & $^c$23 & $^b$18 & $^b$18 & $^a$1265.664 $^b$1517.338 $^c$1229.716 \\
$n$ & 1503.414 &   9 &   6 &   8 &   0 &   2.64 &  12.89 &   1.70 &  17.57 & 546.42 & 553.20 & $^a$14 & $^a$14 & $^a$15 & $^a$14 & $^a$14 & $^a$14 & $^a$1383.653 \\
$u$ & 1504.317 &  12 &   6 &  11 &   1 &   0.52 &  38.45 &   0.91 & 128.32 &  92.16 & 102.47 & $^a$45 & $^a$97 & $^a$86 & $^a$98 & $^b$16 & $^a$26 & $^a$1334.493 $^b$1504.317 \\
$p$ & 1504.750 &   5 &   7 &   4 &   1 &   1.83 &   7.55 &   0.88 &  17.27 & 379.64 & 383.32 & $^a$15 & $^a$18 & $^a$17 & $^b$30 & $^a$19 & $^a$19 & $^a$1504.750 $^b$1202.449 \\
$n$ & 1505.323 &  11 &   6 &  12 &   0 &   2.13 &  10.17 &   1.30 &  12.78 & 479.76 & 485.77 & $^a$14 & $^b$15 & $^a$15 & $^b$16 & $^b$15 & $^b$15 & $^a$1389.583 $^b$1427.002 \\
$p$ & 1505.649 &   7 &   7 &   8 &   1 &   1.84 &   8.00 &   0.89 &  13.48 & 420.13 & 424.35 & $^a$16 & $^b$17 & $^b$16 & $^c$20 & $^b$18 & $^b$18 & $^a$1246.850 $^b$1505.649 $^c$1222.028 \\
$n$ & 1508.057 &  19 &   4 &  18 &   0 &   1.06 &   6.80 &   0.60 &   9.79 & 267.81 & 272.09 & $^a$20 & $^b$29 & $^a$20 & $^b$39 & $^a$21 & $^a$21 & $^a$1513.729 $^b$1404.600 \\
 & 1509.549 &  15 &   5 &  14 &   0 &   1.86 &   8.80 &   1.10 &  10.32 & 430.59 & 436.01 & $^a$16 & $^a$17 & $^a$17 & $^a$17 & $^a$17 & $^a$17 & $^a$1452.342 \\
$p$ & 1510.687 &   6 &   7 &   5 &   1 &   0.64 &  10.44 &   0.40 &  34.14 & 134.68 & 137.77 & $^a$14 & $^b$65 & $^b$19 & $^b$75 & $^a$19 & $^a$18 & $^a$1510.687 $^b$1335.574 \\
$p$ & 1512.533 &  10 &   6 &   9 &   0 &   0.97 &  75.66 &   1.86 & 274.91 & 154.57 & 172.26 & $^a$39 & $^a$88 & $^a$82 & $^a$92 & $^b$14 & $^a$18 & $^a$1334.498 $^b$1393.452 \\
$n$ & 1513.729 &  17 &   5 &  18 &   0 &   1.10 &   7.04 &   0.62 &  10.14 & 276.66 & 281.09 & $^a$20 & $^b$29 & $^a$20 & $^b$39 & $^a$21 & $^a$21 & $^a$1513.729 $^b$1404.600 \\
$u$ & 1514.874 &  13 &   6 &  12 &   1 &   1.06 &   6.79 &   0.54 &  11.97 & 269.29 & 272.99 & $^a$16 & $^b$37 & $^c$14 & $^b$49 & $^a$15 & $^a$15 & $^a$1310.892 $^b$1402.854 $^c$1346.424 \\
$p$ & 1516.217 &   1 &   7 &   2 &   0 &   0.49 &  10.43 &   0.37 &  32.08 &  97.22 & 100.55 & $^a$16 & $^b$51 & $^a$25 & $^b$55 & $^a$16 & $^a$18 & $^a$1402.648 $^b$1393.961 \\
$p$ & 1517.338 &   7 &   7 &   6 &   1 &   2.15 &   9.89 &   1.10 &  21.93 & 460.37 & 465.06 & $^a$15 & $^b$16 & $^b$16 & $^c$23 & $^b$18 & $^b$18 & $^a$1265.664 $^b$1517.338 $^c$1229.716 \\
$p$ & 1518.132 &   9 &   7 &  10 &   1 &   1.76 &   7.77 &   0.87 &  14.12 & 414.95 & 419.22 & $^a$16 & $^b$16 & $^b$15 & $^c$33 & $^b$18 & $^b$18 & $^a$1267.221 $^b$1518.132 $^c$1237.292 \\
$n$ & 1518.892 &   1 &   7 &   0 &   0 &   0.21 &  28.11 &   0.26 & 117.83 &  37.28 &  42.52 & $^a$40 & $^a$98 & $^a$74 & $^a$99 & $^a$27 & $^a$39 & $^a$1335.868 \\
 & 1518.971 &   3 &   7 &   4 &   0 &   0.99 &   5.38 &   0.67 &  10.11 & 204.72 & 207.38 & $^a$16 & $^b$23 & $^a$17 & $^b$29 & $^a$17 & $^a$17 & $^a$1352.502 $^b$1397.419 \\
 & 1522.226 &  11 &   6 &  10 &   0 &   2.21 &  14.47 &   1.42 &  25.94 & 487.89 & 495.05 & $^a$15 & $^a$41 & $^a$19 & $^a$57 & $^b$16 & $^b$16 & $^a$1371.414 $^b$1403.974 \\
 & 1523.953 &  13 &   6 &  14 &   0 &   1.53 &   7.38 &   0.91 &   8.77 & 369.81 & 374.56 & $^a$16 & $^a$17 & $^a$17 & $^a$17 & $^a$17 & $^a$17 & $^a$1452.342 \\
$p$ & 1525.152 &   3 &   7 &   2 &   0 &   0.63 &  13.60 &   0.48 &  41.79 & 124.98 & 129.28 & $^a$16 & $^b$51 & $^a$25 & $^b$55 & $^a$16 & $^a$18 & $^a$1402.648 $^b$1393.961 \\
$p$ & 1525.224 &   5 &   7 &   6 &   0 &   1.34 &   8.73 &   0.89 &  16.52 & 288.66 & 292.89 & $^a$15 & $^b$41 & $^b$16 & $^b$56 & $^a$15 & $^a$15 & $^a$1366.391 $^b$1404.747 \\
 & 1525.284 &  10 &   7 &  11 &   1 &   0.67 &  49.76 &   1.18 & 165.88 & 117.65 & 130.93 & $^a$45 & $^a$97 & $^a$86 & $^a$98 & $^b$16 & $^a$26 & $^a$1334.493 $^b$1504.317 \\
$p$ & 1532.031 &   7 &   8 &   8 &   2 &   0.71 &   6.51 &   0.28 &  15.58 & 176.99 & 179.93 & $^a$15 & $^a$75 & $^a$19 & $^a$72 & $^a$19 & $^a$20 & $^a$1393.441 \\
$n$ & 1532.546 &   9 &   7 &   8 &   1 &   1.97 &   8.64 &   0.94 &  13.96 & 448.50 & 453.06 & $^a$16 & $^b$17 & $^b$16 & $^c$20 & $^b$18 & $^b$18 & $^a$1246.850 $^b$1505.649 $^c$1222.028 \\
 & 1532.951 &  11 &   7 &  12 &   1 &   1.34 &   8.62 &   0.67 &  14.57 & 332.05 & 336.61 & $^a$16 & $^b$37 & $^c$14 & $^b$49 & $^a$15 & $^a$15 & $^a$1310.892 $^b$1402.854 $^c$1346.424 \\
 & 1533.628 &  19 &   6 &  20 &   2 &   0.18 &  14.80 &   0.18 &  36.67 &  56.57 &  62.97 & $^a$31 & $^a$96 & $^a$70 & $^a$98 & $^b$20 & $^a$22 & $^a$1334.424 $^b$1338.341 \\ 
 & 1534.040 &  21 &   5 &  20 &   2 &   0.31 &  25.39 &   0.30 &  62.89 &  93.49 & 104.37 & $^a$31 & $^a$96 & $^a$70 & $^a$98 & $^b$20 & $^a$22 & $^a$1334.424 $^b$1338.341 \\
 & 1534.718 &   7 &   7 &   8 &   0 &   1.42 &   7.16 &   0.92 &   9.41 & 317.69 & 321.80 & $^a$14 & $^a$14 & $^a$15 & $^a$14 & $^a$14 & $^a$14 & $^a$1383.653 \\
$p$ & 1534.768 &   5 &   7 &   4 &   0 &   0.97 &   5.27 &   0.64 &   9.46 & 199.08 & 201.68 & $^a$16 & $^b$23 & $^a$17 & $^b$29 & $^a$17 & $^a$17 & $^a$1352.502 $^b$1397.419 \\
 & 1535.071 &  17 &   5 &  16 &   0 &   1.27 &   6.04 &   0.72 &   6.61 & 313.91 & 317.92 & $^a$20 & $^a$21 & $^a$20 & $^a$21 & $^a$21 & $^a$21 & $^a$1479.539 \\
 & 1536.972 &  15 &   6 &  14 &   1 &   0.78 &   3.53 &   0.37 &   4.16 & 217.27 & 219.66 & $^a$18 & $^b$17 & $^b$17 & $^b$17 & $^a$18 & $^a$18 & $^a$1332.307 $^b$1372.120 \\
$n$ & 1540.569 &   8 &   7 &   9 &   0 &   0.49 &  38.15 &   0.92 & 132.44 &  81.21 &  90.17 & $^a$39 & $^a$88 & $^a$82 & $^a$92 & $^b$14 & $^a$18 & $^a$1334.498 $^b$1393.452 \\
 & 1543.056 &   9 &   8 &  10 &   2 &   0.68 &   2.69 &   0.27 &   4.87 & 190.55 & 192.32 & $^a$18 & $^b$17 & $^b$18 & $^b$14 & $^b$18 & $^b$18 & $^a$1248.144 $^b$1412.812 \\
$n$ & 1543.072 &  13 &   6 &  12 &   0 &   1.75 &   8.59 &   1.06 &  10.38 & 414.53 & 419.88 & $^a$14 & $^b$15 & $^a$15 & $^b$16 & $^b$15 & $^b$15 & $^a$1389.583 $^b$1427.002 \\
$n$ & 1543.652 &   7 &   8 &   6 &   2 &   0.89 &   3.39 &   0.35 &   7.73 & 216.76 & 218.71 & $^a$16 & $^a$20 & $^a$22 & $^b$17 & $^a$19 & $^a$19 & $^a$1376.848 $^b$1089.510 \\
 & 1544.273 &  15 &   6 &  16 &   0 &   1.04 &   5.02 &   0.59 &   5.50 & 269.57 & 273.07 & $^a$20 & $^a$21 & $^a$20 & $^a$21 & $^a$21 & $^a$21 & $^a$1479.539 \\
$p$ & 1546.422 &   5 &   9 &   6 &   3 &   0.20 &   6.66 &   0.17 &  20.33 &  46.89 &  48.94 & $^a$25 & $^a$93 & $^a$74 & $^a$95 & $^b$15 & $^b$14 & $^a$1334.480 $^b$1462.581 \\
$n$ & 1547.081 &   9 &   7 &  10 &   0 &   1.24 &   8.23 &   0.79 &  13.96 & 294.90 & 299.31 & $^a$15 & $^a$41 & $^a$19 & $^a$57 & $^b$16 & $^b$16 & $^a$1371.414 $^b$1403.974 \\
$p$ & 1547.332 &   3 &   8 &   4 &   1 &   1.09 &   4.63 &   0.53 &   9.92 & 238.63 & 241.06 & $^a$15 & $^a$18 & $^a$17 & $^b$30 & $^a$19 & $^a$19 & $^a$1504.750 $^b$1202.449 \\
$p$ & 1547.415 &   7 &   7 &   6 &   0 &   1.13 &   7.38 &   0.73 &  13.25 & 245.12 & 248.74 & $^a$15 & $^b$41 & $^b$16 & $^b$56 & $^a$15 & $^a$15 & $^a$1366.391 $^b$1404.747 \\
 & 1549.510 &  13 &   7 &  14 &   1 &   0.99 &   4.47 &   0.46 &   5.25 & 269.41 & 272.36 & $^a$18 & $^b$17 & $^b$17 & $^b$17 & $^a$18 & $^a$18 & $^a$1332.307 $^b$1372.120 \\
$p$ & 1549.760 &   4 &   8 &   5 &   1 &   0.39 &   6.33 &   0.24 &  19.70 &  82.70 &  84.60 & $^a$14 & $^b$65 & $^b$19 & $^b$75 & $^a$19 & $^a$18 & $^a$1510.687 $^b$1335.574 \\
$n$ & 1549.851 &  11 &   7 &  10 &   1 &   1.83 &   8.17 &   0.90 &  14.15 & 432.25 & 436.77 & $^a$16 & $^b$16 & $^b$15 & $^c$33 & $^b$18 & $^b$18 & $^a$1267.221 $^b$1518.132 $^c$1237.292

\enddata
\end{deluxetable*}

\clearpage
\end{landscape}


\begin{thebibliography}{16}

\bibitem[{Abgrall, Roueff, \& Drira}(2000)]{abgrall00}
{Abgrall}, H., {Roueff}, E., \& {Drira}, I. 2000, A \& A S, 141, 297

\bibitem[{Abgrall et al.}(1997)]{abgrall97}
{Abgrall}, H., {Roueff}, E., {Liu}, X., \& {Shemansky}, D.~E. 1997, \apj, 481, 557


\bibitem[{Abgrall et~al.}(1993a)]{abgrall93a}
{Abgrall}, H., {Roueff}, E., {Launay}, F., {Roncin}, J.-Y., \& {Subtil}, J.~L. 1993, A \& A S, 101, 273

\bibitem[{Abgrall et~al.}(1993b)]{abgrall93b}
{Abgrall}, H., {Roueff}, E., {Launay}, F., {Roncin}, J.-Y., \& {Subtil}, J.~L. 1993, A \& A S, 101, 323

\bibitem[{Allen}(1973)]{allen1973}
Allen, C.~W.: 1973,
{\em Astrophysical Quantities},
 Athlone Press, Univ.\ London

\bibitem[{Avrett}(1995)]{avrett95}
{Avrett}, E.~H. 1995, Infrared tools for solar astrophysics: What's next?, ed. J. R. Kuhn \& M. J. Penn, (Singapore: World Scientific), 303

\bibitem[{Ayres, Testerman, \& Brault}(1986)]{ayres86}
{Ayres}, T.~R., {Testerman}, L., \& {Brault}, J.~W. 1986, \apj, 304, 542

\bibitem[{Ayres \& Wiedemann}(1989)]{ayres89}
{Ayres}, T.~R. \& Wiedemann, G.~R. 1989 \apj, 338, 1033

\bibitem[{Barklem \& Collet}(2016)]{barklem16}
{Barkelm}, P.~S. \& {Collet}, R. 2016, A \& A, 588, 96

\bibitem[{Bartoe \& Brueckner}(1975)]{bartoe75}
{Bartoe}, J.-D.~F., {Brueckner}, G.~E. 1975, J. Opt. Soc. Am., 65, 13

\bibitem[{Bartoe et~al.}(1979)]{bartoe79}
{Bartoe}, J.-D.~F., {Brueckner}, G.~E., \& {Jordan}, C. 1979, MNRAS, 187, 463

\bibitem[{Brekke et al.}(1991)]{brekke91}	
{Brekke}, P., {Kjeldseth-Moe}, O., {Bartoe}, J.-D. F., \& {Brueckner}, G.~E. 1991, \apjs, 75, 1337 

\bibitem[{Carlsson et al.}(2016)]{carlsson16}
{Carlsson}, M., {Hansteen}, V.~H., {Gudiksen}, B.~V., {Leenaarts}, J. \& {De Pontieu}, B. 2016, A \& A, 585, 4

\bibitem[{Carlsson, Leenaarts, \& De Pontieu}(2015)]{carlsson15}
{Carlsson}, M., {Leenaarts}, J., \& {De Pontieu}, B. 2015, \apjl, 809, L30

\bibitem[{Carlsson \& Stein}(1995)]{carlsson95}
{Carlsson}, M., \& {Stein}, R. F. 1995, \apjl, 440, L29

\bibitem[{Curdt et~al.}(2001)]{curdt01}
{Curdt}, W., {Brekke}, P., {Feldman}, U. et al. 2001, A \& A, 375, 591


\bibitem[{De Pontieu et~al.}(2014)]{depontieu14}
{De Pontieu}, B., {Title}, A.~M., {Lemen}, J.~R. et al. 2014, Sol. Phys., 289, 2733

\bibitem[{Fontenla, Avrett, \& Loeser}(1993)]{fontenla93}
Fontenla, J. M., Avrett, E. H., \& Loeser, R. 1993, \apj, 406, 319

\bibitem[{Gudiksen et al.}(2011)]{Gudiksen2011} Gudiksen, B.~V., Carlsson, M., Hansteen, V.~H., et al.\ 2011, A \& A, 531, A154 

\bibitem[{Heays, Bosman, \& van Dishoeck}(2017)]{heays17}
{Heays}, A.~N., {Bosman}, A.~D., \& {van Dishoeck}, E.~F. 2017, A \& A, 602, 105

\bibitem[{Herzberg \& Howe}(1959)]{herzberg59}
{Herzberg}, G. \& {Howe}, L.~L. 1959, Can. J. Phys., 37, 636

\bibitem[{Horan, Kreplin, \& Dere}(1983)]{horan83}
{Horan}, D.~M., {Kreplin}, R.~W., \& {Dere}, K.~P. 1983, Sol. Phys., 85, 303

\bibitem[{Innes}(2008)]{innes08}
{Innes}, D.~E. 2008, A \& A L, 481, 41

\bibitem[{Irwin}(1987)]{irwin87}
{Irwin}, A.~W. 1987, A\&A, 182, 348

\bibitem[{Jordan et al.}(1977)]{jordan77}
{Jordan}, C., {Brueckner}, G.~E., {Bartoe}, J.-D.~F., {Sandlin}, G.~D., \& {VanHoosier}, M.~E. 1977, Natur., 270, 326L

\bibitem[{Jordan et al.}(1978)]{jordan78}
{Jordan}, C., {Brueckner}, G.~E., {Bartoe}, J.-D.~F., {Sandlin}, G.~D., \& {VanHoosier}, M.~E. 1978, ApJ, 226, 687

\bibitem[{Judge}(2015)]{judge15}
{Judge}, P.~G. 2015, \apj, 808, 116

\bibitem[{Kerr et al.}(2005)]{kerr05}
{Kerr}, F.~M., {Rose}, S.~J., {Wark}, J.~S., \& {Keenan}, F.~P. 2005, \apj, 629, 1091

\bibitem[{Krishna Swamy}(1975)]{krishna75}
{Krishna Swamy}, K.~S. 1975, Sol. Phys., 41, 301

\bibitem[{Kuhn \& Morgan}(2006)]{kuhn06}
{Kuhn}, J.~R. \& {Morgan}, H. 2006, ASPC, 354, 230

\bibitem[{Kurucz}(2016)]{kurucz}
{Kurucz}, R.~L. http://kurucz.harvard.edu/


\bibitem[{Leenaarts et al.}(2011)]{leenaarts11}
{Leenaarts}, J., {Carlsson}, M., {Hansteen}, V., \& {Gudiksen}, B.~V. 2011, A \& A, 530, A124

\bibitem[{Lemaire et al.}(1981)]{lemaire81}
{Lemaire}, P., {Gouttebroze}, P., {Vial}, J.-C., \& {Artzner}, G.~E. 1981, A \& A, 103, 160

\bibitem[{Liu}(2009)]{liu09}
{Liu}, J., {Salumbides}, E.~J., {Hollenstein}, U. et al. 2009, J. Chem. Phys., 130, 174306

\bibitem[{Machado et al.}(1980)]{machado80}
{Machado}, M.~E., {Avrett}, E.~H., {Vernazza}, J.~E., \& {Noyes}, R.~W. 1980, \apj, 242, 336

\bibitem[{Machado, Emslie, \& Avrett}(1989)]{machado89}
{Machado}, M.~E., {Emslie}, A.~G., \& {Avrett}, E.~H. 1989, Sol. Phys., 124, 303

\bibitem[{Maltby et al.}(1986)]{maltby86}
{Maltby}, P., {Avrett}, E.~H., {Carlsson}, M., {Kjeldseth-Moe}, O., \& {Kurucz}, R.~L. 1986, \apj, 306, 284

\bibitem[{Mathioudakis et al.}(1999)]{mathioudakis99}
{Mathioudakis}, M., {McKenny}, J., {Keenan}, F.~P., {Williams}, D.~R., \& {Phillips}, K.~J.~H. 1999, A \& A L, 351, 23


\bibitem[{Nahar}(2000)]{nahar00}
\re{{Nahar}, S.~N. 2000, \apjs, 126, 537}

\bibitem[{Roussel-Dupr\'e et al.}(1984)]{rousseldupre84}
{Roussel-Dupr\'e}, R., {Wrathall}, J., {Nicolas}, K.~R., {Bartoe}, J.~D.~F., \& {Brueckner}, G.~E. 1984, \apj, 278, 428

\bibitem[{Rybicki}(1984)]{rybicki84}
{Rybicki}, G.~B. 1984, in Methods in Radiative Transfer, ed. W. Kalkofen (Cambridge University), 21

\bibitem[{Sandlin et~al.}(1986)]{sandlin86}
{Sandlin}, G.~D., {Bartoe}, J.D.~F., {Brueckner}, G.~E., {Tousey}, R., \& {VanHoosier}, M.~E. 1986, ApJS, 61, 801S

\bibitem[{Schmit et~al.}(2014)]{schmit14}
{Schmit}, D.~J., {Innes}, D., {Ayres}, T. et al. 2014, A \& A L, 569, 7

\bibitem[{Sch\"uhle et~al.}(1999)]{schuhle99}
{Sch\"uhle}, U., {Brown}, C.~M., {Curdt}, W., \& {Feldman}, U. 1999, in Plasma Dynamics and Diagnostics in the Solar Transition Region and Corona: Eighth SOHO Workshop, ed. J.-C. Vial \& B. Kaldeich-Sch\"urmann (ESA SP-446; Noordwijk: ESA), 617

\bibitem[{Shaw et al.}(2005)]{shaw05}
{Shaw}, G., {Ferland}, G.~J., {Abel}, N.~P., {Stancil}, P.~C., \& {van Hoof}, P.~A.~M. 2005, 624, 794

\bibitem[{Socas-Navarro et al.}(2015)]{socas15}
{Socas-Navarro}, H., {de la Cruz Rodr\'iguez}, J., {Asensio Ramos}, A., {Trujillo Bueno}, J., \& {Ruiz Cobo}, B., A \& A, 577, 7

\bibitem[{Solanki, Livingston, \& Ayres}(1994)]{solanki94}
{Solanki}, S., {Livingston}, W., \& {Ayres}, T. 1994, Sci., 263, 64

\bibitem[{Uitenbroek}(2000)]{uitenbroek00}
{Uitenbroek}, H. 2000, \apj, 531, 571

\bibitem[{van Dishoeck \& Visser}(2011)]{vandishoeck11}
{van Dishoeck}, E. \& {Visser}, R. 2015, in Laboratory Astrochemistry: From Molecules through Nanoparticles to Grains, eds. S. Schlemmer, T. Giesen, H. Mutschke, and C. J\"ager (Weinheim: Wiley-VCH), 229

\bibitem[{Vernazza, Avrett, \& Loeser}(1981)]{vernazza81}
{Vernazza}, J., {Avrett}, E., \& {Loeser}, R. 1981, \apjs, 45, 635

\bibitem[{Vernazza, Avrett, \& Loeser}(1976)]{vernazza76}
{Vernazza}, J., {Avrett}, E., \& {Loeser}, R. 1976, \apjs, 30, 1

\end{thebibliography}
\end{document}